\def\DIRvalue{Marino}
\def\IDvalue{MA}
\def\titlevalue{Localization at large $N$ \\ in Chern--Simons--matter theories}
\def\authorvalue{Marcos Mari\~no}
\def\shortauthorvalue{\authorvalue}
\def\addressvalue{D\'epartement de Physique Th\'eorique et Section de Math\'ematiques\\
Universit\'e de Gen\`eve, Gen\`eve, CH-1211 Switzerland\\
\tt Marcos.Marino@unige.ch}
\def\abstractvalue{We review some exact results for the matrix models appearing in the localization 
of Chern--Simons--matter theories, focusing on the structure of non-perturbative effects and on the M-theory expansion of ABJM theory. 
We also summarize some of the results obtained for other Chern--Simons--matter theories, as well as recent applications to topological strings.}
\def\preprintvalue{}

\ifx\ifLONG\undefined
\documentclass[a4paper,11pt]{article}
\pdfoutput=1 
\newcommand{\chapterauthor}[1]{
\begin{center}
{\bf \normalsize  #1}
\end{center}
}

\newcommand{\chapteraddress}[1]{
\begin{center}
{ \small \it \addressvalue}
\end{center}
}

\newcommand{\chapterabstract}[1]{
\vspace{\baselineskip}
\begin{center}
\textbf{\small Abstract}
\end{center}
#1}

\newcommand{\chapterheader}{

\chapter[\titlevalue{}  (by \shortauthorvalue)]{\titlevalue}
\label{Chapter\IDvalue}
\chapterauthor{\authorvalue}
\chapteraddress{\addressvalue}
\chapterabstract{\abstractvalue}
\tightmtctrue
\minitoc
}

\newcommand{\documentheader}{
\begin{flushright} \small
  \preprintvalue
 \end{flushright}

\begin{center}
{\bf \Large \titlevalue}
\end{center}

\chapterauthor{\authorvalue}
\chapteraddress{\addressvalue}
\chapterabstract{\abstractvalue}

\medskip

This is a contribution to the review volume ``Localization techniques
in quantum field theories'' (eds. V.~Pestun and M.~Zabzine) which
contains 17 Chapters available at \cite{ContributionSummary}

\tableofcontents
}

\newcommand{\ifvolume}[2]{\ifx\ifLONG\undefined#2\else#1\fi}

\newcommand{\documentfinish}{
\ifx\ifLONG\undefined
\bibliographystyle{bibreview} 
\bibliography{\IDvalue,review}  
\end{document}
\else
\addcontentsline{toc}{section}{References}
\input{\DIRvalue/\IDvalue.bbl}
\fi
}

\newcommand{\documentfinishBBL}{
\addcontentsline{toc}{section}{References}
\ifx\ifLONG\undefined

\end{document}
\else

\fi
}

\ifx\ifLONG\undefined
\def\volcite#1{Contribution \cite{Contribution#1}}
\else 
\def\volcite#1{Chapter \ref{Chapter#1}}
\fi

\newcommand{\ContributionSummaryBibItemReference}
{
\bibitem{ContributionSummary}
V.~Pestun and M.~Zabzine, eds., {\em Localization techniques in quantum field
  theory}, vol.~xx.
\newblock Journal of Physics A, 2016.
\newblock \href{http://arxiv.org/abs/1608.02952}{{\tt 1608.02952}}.
\newblock \url{https://arxiv.org/src/1608.02952/anc/LocQFT.pdf},
  \url{http://pestun.ihes.fr/pages/LocalizationReview/LocQFT.pdf}.
}

\usepackage{amsmath,amssymb,amsthm,amscd,graphicx}
\usepackage{hyperref}
\usepackage{fullpage}

\numberwithin{equation}{section}

\begin{document}
\thispagestyle{empty}
\documentheader\else
\chapterheader\fi

\newcommand{\CA}{{\cal A}}
\newcommand{\CB}{{\cal B}}
\newcommand{\CC}{{\cal C}}
\newcommand{\CE}{{\cal E}}
\newcommand{\CF}{{\cal F}}
\newcommand{\CG}{{\cal G}}
\newcommand{\CH}{{\cal H}}
\newcommand{\CI}{{\cal I}}
\newcommand{\CJ}{{\cal J}}
\newcommand{\CK}{{\cal K}}
\newcommand{\CL}{{\cal L}}
\newcommand{\CM}{{\cal M}}
\newcommand{\CN}{{\cal N}}
\newcommand{\CO}{{\cal O}}
\newcommand{\CQ}{{\cal Q}}
\newcommand{\CR}{{\cal R}}
\newcommand{\CS}{{\cal S}}
\newcommand{\CT}{{\cal T}}
\newcommand{\CU}{{\cal U}}
\newcommand{\CV}{{\cal V}}
\newcommand{\CW}{{\cal W}}
\newcommand{\CX}{{\cal X}}
\newcommand{\CY}{{\cal Y}}
\newcommand{\CZ}{{\cal Z}}

\newcommand{\IZ}{\mathbb{Z}}
\newcommand{\IR}{\mathbb{R}}
\newcommand{\IC}{\mathbb{C}}
\newcommand{\IP}{\mathbb{P}}
\newcommand{\IT}{\mathbb{T}}
\newcommand{\IS}{\mathbb{S}}
\newcommand{\IN}{\mathbb{N}}
\newcommand{\IH}{\mathbb{H}}
\newcommand{\MAtr}{{\rm Tr}}
\newcommand{\MAre}{{\rm e}}
\newcommand{\ri}{{\rm i}}
\newcommand{\rd}{{\rm d}}

\newcommand{\refb}[1]{(\ref{MA#1})}

\section{Introduction}

The use of localization techniques in superconformal gauge theories, pioneered in \cite{MApestun}, 
has led to many new exact results in QFT. Typically, these techniques give expressions for 
the partition functions or correlation functions of the theory in terms of a matrix integral, 
and the number of variables of this integral scales as the rank of the gauge group $N$. Although the resulting expressions 
are relatively explicit, it is not so easy to evaluate them analytically for small $N$, and it is even harder to determine 
their behavior as $N$ grows large. However, this is precisely the 
regime that one wants to study in applications of localization to the AdS/CFT correspondence. 

In this paper we will review some aspects of the large $N$ solution to the matrix integrals appearing 
in the localization of Chern--Simons--matter theories \cite{MAkwy}. The first exact results for 
their large $N$ limits were found in \cite{MAmpabjm,MAdmp}, where the planar Wilson loop vevs and the planar free energy of 
ABJM theory \cite{MAabjm} were calculated explicitly. Many subsequent works, 
following \cite{MAhkpt}, have studied the strict large $N$ limit of these theories and compared them with their gravity counterparts\footnote{By strict large $N$ limit, we mean the dominant 
term in the large $N$ expansion. This contains less information than the exact planar limit, since it is given by its leading term at strong 't Hooft coupling.}. However, in this review we will focus on 
the rich structures appearing {\it beyond} the strict large $N$ limit: the determination of exact planar free energies, 
the 't Hooft $1/N$ expansion beyond the planar limit, and specially the structure of non-perturbative effects at large $N$, 
which are invisible in the 't Hooft expansion. We will also focus on the results for the partition functions on the three-sphere. There have been 
studies of these theories on other three-manifolds, and also of other observables, such as Wilson loops, but we will not consider these extensions here. 

Going beyond the strict large $N$ limit is not easy, and so far the only theory for which we have a 
rather complete picture is ABJM theory (and its close cousin, ABJ theory \cite{MAabj}.) 
 After considerable effort, a detailed expression for the full $1/N$ expansion of the partition function of ABJM theory, including non-perturbative 
corrections, is now available. This is arguably the most complete result obtained so far for a gauge theory observable 
in the general framework of the $1/N$ expansion (of course, simpler models 
have been solved with the same level of detail, but they do not have the same level of complexity.) Therefore, section 2 (which comprises most of this review), is devoted 
to a relatively self-contained explanation of this result, since many of its ingredients are scattered across the literature. 
In section 3, we summarize what is known beyond the strict large $N$ limit in other Chern--Simons--matter theories. We also comment on some 
related developments in topological string theory. Finally, in section 4 we list some conclusions and open problems. 

While preparing this review for publication, another review paper on this subject appeared \cite{MAhmo-rev}.
 
\section{The ABJM matrix model}
 
\subsection{A short review of ABJM theory}

\begin{center}
 \begin{figure}
\begin{center}
\includegraphics[height=4cm]{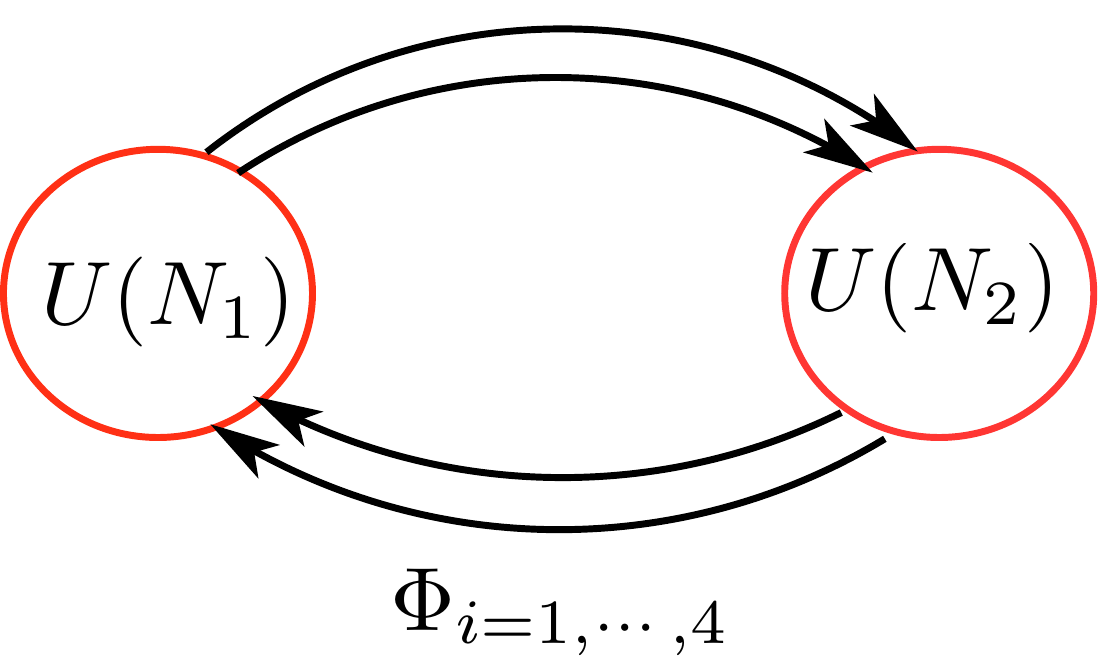}
\caption{The quiver for ABJ(M) theory. The two nodes represent the $U(N_{1,2})$ Chern--Simons theories (with opposite levels) and the arrows between 
the nodes represent the four matter multiplets in the bifundamental representation.}
\label{MAquiver}
\end{center}
\end{figure}  
\end{center}
ABJM theory and its generalization, also called ABJ theory, were proposed in \cite{MAabjm,MAabj} 
to describe $N$ M2 branes on $\IC^4/\IZ_k$. They are particular examples of supersymmetric Chern--Simons--matter theories 
and their basic ingredient is a pair of vector multiplets with gauge groups $U(N_1)$, $U(N_2)$, described by two supersymmetric Chern--Simons theories 
with opposite levels $k$, $-k$. In addition, we have four matter supermultiplets $\Phi_i$, $i=1,\cdots, 4$, in the bifundamental representation of the gauge group 
$U(N_1) \times U(N_2)$. This theory can be represented as a quiver with two nodes, which stand for the two 
supersymmetric Chern--Simons theories, and four edges between the nodes representing the matter supermultiplets (see Fig. \ref{MAquiver}). In addition, there is a 
superpotential involving the matter fields, which after integrating out the auxiliary fields in the Chern--Simons--matter system, reads (on $\IR^3$)
\begin{equation}
\label{MAsuperpot}
W={4 \pi \over k} \MAtr\left(\Phi_1 \Phi_2^{\dagger} \Phi_3 \Phi_4^{\dagger} - \Phi_1 \Phi_4^{\dagger} \Phi_3 \Phi_2^{\dagger}\right). 
\end{equation}
In this expression we have used the standard superspace notation for $\CN=1$ supermultiplets in 4d. 
When the two gauge groups have identical rank, i.e. $N_1=N_2=N$, 
the theory is called ABJM theory. The generalization in which $N_1 \not= N_2$ is called ABJ theory. More details on the construction of these theories can be 
found in \cite{MAblmp,MAmmlec}. In most of this review we will focus on ABJM theory, which has two parameters: $N$, the common rank of the gauge group, and $k$, 
the Chern--Simons level. Note that, in this theory, all the fields are in the adjoint representation of $U(N)$ or in the bifundamental representation of 
$U(N) \times U(N)$. Therefore, they have two color indices and one can use 
the standard 't Hooft rules \cite{MAthooft-paper} to perform a $1/N$ expansion. Since $k$ plays the r\^ole of the inverse gauge coupling 
$1/g^2$, the natural 't Hooft parameter is given by 
\begin{equation}
\label{MAth-abjm}
\lambda={N \over k}. 
\end{equation}

One of the most important aspects of ABJM theory is that, at large $N$, it describes a non-trivial background of $M$ theory, 
as it was already postulated in \cite{MAmalda}. In the large distance limit in which M-theory can be described by 
supergravity, this is nothing but the Freund--Rubin background 
\begin{equation}
\label{MAads4}
X_{11}={\rm AdS}_4 \times \IS^7/\IZ_k.
\end{equation}
If we represent $\IS^7$ inside $\IC^4$ as
\begin{equation}
\sum_{i=1}^4 |z_i|^2=1,
\end{equation}
the action of $\IZ_k$ in (\ref{MAads4}) is simply given by 
\begin{equation}
\label{MAmod-action}
z_i \rightarrow \MAre^{2 \pi \ri \over k} z_i. 
\end{equation}
The metric on ${\rm AdS}_4 \times \IS^7$ depends on a single parameter, the radius $L$, and by using metrics on ${\rm AdS}_4$ and $\IS^7$ of unit radius, we have
\begin{equation}
\rd s^2= L^2 \left( {1\over 4} \rd s^2_{{\rm AdS}_4}+ \rd s^2_{\IS^7} \right). 
\end{equation}
As it is well-known, the Freund--Rubin background also involves a non-zero flux for the four-form field strength $G$ 
of 11d SUGRA, see for example \cite{MAdnp} for an early review of eleven-dimensional 
supergravity on this background. 

The AdS/CFT correspondence between ABJM theory and M-theory in the above Freund--Rubin background 
comes with a dictionary between the gauge theory parameters and the M-theory parameters. 
The parameter $k$ in the gauge theory has a purely geometric interpretations and 
it is the same $k$ appearing in the modding out by $\IZ_k$ in (\ref{MAads4}) and (\ref{MAmod-action}). The parameter $N$ 
corresponds to the number of M2 branes, which lead to the non-zero flux of $G$, and also determines the radius of the background. One finds, 
\begin{equation}
\label{MALNdic}
\left( {L \over \ell_p}\right)^6= 32 \pi^2 k N, 
\end{equation}
where $\ell_p$ is the eleven-dimensional Planck constant. It should be emphasized that 
the above relation is in principle only valid in the large $N$ limit, and it has been 
argued that it is corrected due to a shift in the M2 charge \cite{MAbh,MAahho}. According to this argument, the physical charge determining the radius is not $N$, but rather
\begin{equation}
\label{MAqshift}
Q= N-{1\over 24} \left( k-{1\over k} \right).   
\end{equation}
The geometric, M-theory description in terms of the background (\ref{MAads4}) emerges when
\begin{equation}
\label{MAmtregime}
N \rightarrow \infty, \qquad \text{$k$ fixed}. 
\end{equation}
The corresponding regime in the dual gauge theory will be called the {\it M-theory regime}. In this regime, one looks for 
asymptotic expansions of the observables at large $N$ but $k$ fixed. This is the so-called {\it M-theory expansion} of the gauge theory. 

It has been known for a while that the above Freund--Rubin background of M-theory can be used to find a background of type IIA superstring theory of the form 
\begin{equation}
\label{MAx10}
X_{10}={\rm AdS}_4 \times \IC\IP^3.  
\end{equation}
This is due to the existence of the Hopf fibration, 
\begin{equation}
\begin{array}{ccc}
\IS^1 & \rightarrow & \IS^7 \\
  & & \downarrow\\
  & & \IC\IP^3
  \end{array}
  \end{equation}
  and the circle of this fibration can be used to perform a non-trivial reduction from M-theory to type IIA theory \cite{MAnp}. 
  In order to have a perturbative regime for the type IIA superstring, we need 
  the circle to be small, and this is achieved when $k$ is large. Indeed, by using the standard dictionary relating M-theory and type IIA theory, we find that 
  the string coupling constant $g_{\rm st}$ is given by 
  \begin{equation}
g_{\rm st}^{2} ={1\over k^2} \left( {L \over \ell_s}\right)^2,
\end{equation}
where $\ell_s$ is the string length. On the other hand, we also have from this dictionary that
\begin{equation}
 \lambda ={1\over 32 \pi^2 } \left( {L \over \ell_s}\right)^4, 
\end{equation}
where $\lambda$ is the 't Hooft parameter (\ref{MAth-abjm}). 
 We conclude that the perturbative regime of the type IIA superstring corresponds to the 't Hooft $1/N$ expansion, in which 
 \begin{equation}
 \label{MAthooft-regime}
N \rightarrow \infty, \qquad \lambda={N\over k} \qquad \text{fixed}, 
\end{equation}
i.e. the genus expansion in the 't Hooft regime of the 
 gauge theory corresponds to the perturbative genus expansion of the superstring. In addition, 
 the regime of strong 't Hooft coupling corresponds to the point-particle limit of the superstring, in which $\alpha'$ corrections are suppressed.  
 
 A very important aspect of ABJM theory is that there are two different regimes to consider: the M-theory regime (\ref{MAmtregime}), and the standard 't Hooft regime (\ref{MAthooft-regime}). 
 The existence of a well-defined M-theory limit is somewhat surprising from the gauge theory point of view. This limit is more like a thermodynamic limit of the theory, in which 
 the number of degrees of freedom goes to infinity but the coupling constant remains fixed. General aspects of this limit have been discussed in \cite{MAhanada-m}. 
 
 One of the consequences of the AdS/CFT correspondence is that the partition function of the Euclidean ABJM theory  
on $\IS^3$ should be equal to the partition function of the Euclidean version of M-theory/string theory on the dual AdS backgrounds \cite{MAwittenads}, i.e. 
\begin{equation}
\label{MAzz}
Z(\IS^3)=Z(X), 
\end{equation}
where $X$ is the eleven-dimensional background (\ref{MAads4}) or the ten-dimensional background (\ref{MAx10}), 
appropriate for the M-theory regime or the 't Hooft regime, 
respectively. In the M-theory limit we can use the 
supergravity approximation to compute the M-theory partition function, which is just given by the classical action of eleven-dimensional 
supergravity evaluated on-shell, i.e. on the metric of (\ref{MAads4}). 
This requires a regularization of IR divergences but eventually leads to a finite result, which gives a prediction for the behavior of the partition function of 
ABJM theory at large $N$ and fixed $k$ (see \cite{MAmmlec} for a review of these isssues). If we define the free energy of the theory as the logarithm of the partition function, 
 \begin{equation}
 F(N,k)= \log Z(N,k), 
 \end{equation}
one finds, from the supergravity approximation to M-theory \cite{MAejm},
\begin{equation}
\label{MAmsugra}
F(N,k) \approx -{\pi \sqrt{2} \over 3} k^{1/2} N^{3/2}, \qquad  N\gg 1. 
\end{equation}
The $N^{3/2}$ behavior of the free energy is a famous prediction of AdS/CFT \cite{MAkt} for the large 
$N$ behavior of a theory of M2 branes. 

The AdS/CFT prediction (\ref{MAzz}) can be also studied in the 't Hooft regime. The free energy of the gauge theory on the sphere has a large $N$ 
expansion which we will write as
\begin{equation}
\label{MAgenusg}
F(N, k)= \sum_{g\ge 0} F_g(\lambda) g_s^{2g-2}, 
\end{equation}
where
\begin{equation}
g_s={2 \pi \ri \over k}. 
\end{equation}
This expansion is of course equivalent to a $1/N$ expansion, since $g_s= 2 \pi \ri \lambda/N$, and the 't Hooft parameter is kept fixed. In the string theory side, it corresponds 
to the genus expansion of the free energy. In particular, the planar free energy $F_0(\lambda)$ of the gauge theory should agree with the superstring free energy at tree level, i.e. 
at genus zero. When $\lambda$ is large, the string is small as compared to the AdS radius, and we can use the point-particle approximation to string theory, i.e. we can 
approximate the genus zero free energy by the type IIA supergravity result. One obtains in this way a prediction for the planar free energy of ABJM theory 
at strong 't Hooft coupling, of the form 
\begin{equation}
\label{MAssugra}
F_0(\lambda) \approx {4 \pi^3 {\sqrt{2}} \over 3} \lambda^{3/2} ,  \qquad \lambda \gg 1.
\end{equation}
Interestingly, both predictions are equivalent, in the sense that one can obtain (\ref{MAssugra}) from (\ref{MAmsugra}) by setting $k =N/\lambda$, and viceversa. This is not 
completely obvious from the point of view of the gauge theory, since it could happen that higher genus corrections in the 't Hooft expansion contribute to the M-theory limit. 
That this is not the case has been conjectured in \cite{MAhanada-m} 
to be a general fact and it has been called ``planar dominance." 
It seems to be a general property 
of Chern--Simons--matter theories with both an M-theory expansion and a 't Hooft expansion. 
Note as well that, as explained in detail in \cite{MAmmlec}, the behavior of the planar free 
energy at weak 't Hooft coupling is very different from the prediction (\ref{MAssugra}). Therefore, 
the planar free energy should be a non-trivial interpolating function 
between the weakly coupled regime and the strongly coupled regime. 

 In order to analyze in detail the implications of the large $N$ duality between ABJM theory and M-theory on the AdS background (\ref{MAads4}), it is extremely useful to 
 be able to perform reliable computations on the gauge theory side. The techniques of localization pioneered in \cite{MApestun} have led to a wonderful result for the 
 partition function of ABJM theory on the three-sphere $\IS^3$, due to \cite{MAkwy}. This result expresses this partition function, which {\it a priori} is given by 
 a complicated path integral, in terms of a matrix model (i.e. a path integral in zero dimensions). We will refer to it as the ABJM matrix model, and it takes the following form 
 (the derivation of this and similar expressions can be found in \volcite{WI}):
 \begin{equation}
\label{MAabjm-mm}
\begin{aligned}
&Z(N,k)\\
&={1\over N!^2} \int {\rd ^N \mu \over (2\pi)^N} {\rd ^N \nu \over (2\pi)^N} {\prod_{i<j} \left[ 2 \sinh \left( {\mu_i -\mu_j \over 2} \right)\right]^2
  \left[ 2 \sinh \left( {\nu_i -\nu_j \over 2} \right)\right]^2 \over \prod_{i,j} \left[ 2 \cosh \left( {\mu_i -\nu_j \over 2} \right)\right]^2 } 
  \exp \left[ {\ri k \over 4 \pi} \sum_{i=1}^N (\mu_i^2 -\nu_i^2) \right].
  \end{aligned}
  \end{equation}
 In the remaining of this review, we will analyze this matrix model in detail. We will study it in different regimes and we will try to extract lessons and consequences for the 
 AdS/CFT correspondence.
   
\subsection{The 't Hooft expansion}

\subsubsection{The planar limit}

In order to test the prediction (\ref{MAssugra}), it would be useful to have an explicit expression for $F_0(\lambda)$, and eventually for the full series of 
genus $g$ free energies $F_g(\lambda)$. This is in principle a formidable problem, involving the 
resummation of double-line diagrams with a fixed genus in the perturbative expansion of the total free energy. 
However, since the partition function is given by the matrix integral (\ref{MAabjm-mm}), 
we can try to obtain the $1/N$ expansion directly in the matrix model. The large $N$ expansion of matrix models has been extensively studied since the seminal work of 
Br\'ezin, Itzykson, Parisi and Zuber \cite{MAbipz}, and there are by now many different techniques to solve this problem. The first step in this calculation is of course to obtain 
the planar free energy $F_0(\lambda)$, which is the dominant term at large $N$. 

A detailed review of the calculation of the planar free energy of the ABJM matrix model can be found in \cite{MAmmlec}, and we won't 
repeat it here. We will just summarize the most important aspects of the solution. As usual, at large $N$,  the eigenvalues of the matrix 
model ``condense" around cuts in the complex plane. This means that the equilibrium values of the eigenvalues $\mu_i$, $\nu_i$, $i=1, \cdots, N$, 
fall into two arcs in the complex plane as $N$ becomes large. 
The equilibrium conditions for the eigenvalues $\mu_i$, $\nu_i$ can be found immediately from the integrand of the matrix integral: 
\begin{equation}
\begin{aligned}
{\ri k \over 2 \pi}\mu_i  =&- \sum_{j\ne i}^{N}\coth\frac{\mu_i-\mu_j}2+\sum_{j=1}^{N}\tanh\frac{\mu_i-\nu_j}2, \\
{\ri k \over 2 \pi}\nu_i =& \sum_{j\ne i}^{N}\coth\frac{\nu_i-\nu_j}2-\sum_{j=1}^{N}\tanh\frac{\nu_i-\mu_j}2.
 \label{MAsaddleabjm}
\end{aligned}
\end{equation}
In standard matrix models, it is useful to think about the equilibrium values of the eigenvalues as the result of a competition between 
a confining one-body potential and a repulsive two-body potential. Here we can not do that, since the one-body potential is imaginary. 
One way to go around this is to use analytic continuation: we rotate $k$ to an imaginary value, and then at the end of the calculation we rotate it back. 
This is the procedure followed originally in \cite{MAdmp}. We consider then the saddle-point equations
\begin{equation}
\begin{aligned}
\mu_i =&{t_1 \over N_1} \sum_{j\ne i}^{N_1}\coth\frac{\mu_i-\mu_j}2+{t_2 \over N_2} \sum_{j=1}^{N_2}\tanh\frac{\mu_i-\nu_j}2, \\
\nu_i =&{t_2 \over N_2}  \sum_{j\ne i}^{N_2}\coth\frac{\nu_i-\nu_j}2+{t_1 \over N_1}\sum_{j=1}^{N_1}\tanh\frac{\nu_i-\mu_j}2. 
 \label{MAsaddlealt}
\end{aligned}
\end{equation}
where 
\begin{equation}
t_i= g_s N_i, \quad i=1,2. 
\end{equation}
The planar free energy obtained from these equations will be a function only of $t_1$ and $t_2$, and to recover the planar free energy of the original ABJM matrix model we have to set 
\begin{equation}
\label{MAabjm-limit}
t_1=-t_2= {2 \pi \ri \over k} N. 
\end{equation}
The equations (\ref{MAsaddlealt}), for real $g_s$, are equivalent to the original ones (\ref{MAsaddleabjm}) after rotating $k$ to the imaginary axis, and then performing an analytic continuation 
$N_2 \rightarrow -N_2$. At large $N_i$, and for real $g_s$, $t_i$, the eigenvalues $\mu_i$, $i=1, \cdots, N_1$ and $\nu_j$, $j=1, \cdots N_2$, condense around 
two cuts in the real axis, $\CC_{1,2}$ (respectively.) Due to the symmetries of the problem, these cuts are symmetric around the origin. We will denote by $[-A, A]$, $[-B, B]$, respectively. 

It turns out that the equations (\ref{MAsaddlealt}) are the saddle-point equations for the so-called lens space matrix model 
studied in \cite{MAmmcs,MAakmv,MAhy}, whose planar solution 
is well-known. To write down the solution, one introduces a resolvent
$\omega(z)$, as defined in \cite{MAhy}:
\begin{equation}
\label{MAresolv}
\omega(z)
=g_s \left\langle \sum_{i=1}^{N_1} \coth \left( {z-\mu_i \over 2} \right) \right\rangle +g_s \left\langle\sum_{a=1}^{N_2}  \tanh \left( {z-\nu_a \over 2} \right)\right\rangle.
\end{equation}
In terms of the variable $Z=\MAre^z$ , it is given by
\begin{equation}
\omega(z)\rd z =-t {\rd Z  \over Z}  +2  g_s\left\langle \sum_{i=1}^{N_1}{\rd Z \over Z-\MAre^{\mu_i}}\right\rangle+ 2 g_s \left\langle \sum_{a=1}^{N_2}{\rd Z \over Z+\MAre^{\nu_a}}\right\rangle, 
\end{equation}
where 
\begin{equation}
t=t_1 +t_2.
\end{equation}
 In the planar approximation, the sum over eigenvalues can be replaced by an integration involving their densities, and we have that
\begin{equation}
\label{MAplanarres}
\omega_0(z) =-t +2  t_1 \int_{\CC_1} \rho_1(\mu) {Z \over Z-\MAre^{\mu}}\rd \mu + 2  t_2 \int_{\CC_2} \rho_2 (\nu) {Z \over Z+\MAre^{\nu}}\rd \nu, 
\end{equation}
where $\rho_1(\mu)$, $\rho_2(\nu)$ are the large $N$ densities of eigenvalues on the cuts $\CC_1$, $\CC_2$, respectively, normalized as
\begin{equation}
\label{MAnorma}
\int_{\CC_1} \rho_1(\mu) \rd \mu=\int_{\CC_2} \rho_2 (\nu) \rd \nu=1.
\end{equation}
A standard discontinuity argument tells us that 
\begin{equation}
\label{MAdensities}
\begin{aligned}
\rho_1(X) \rd X &=-{1\over 4 \pi \ri t_1}{\rd X \over X} \left( \omega_0(X+\ri \epsilon) -\omega_0(X-\ri \epsilon)\right), \qquad X\in \CC_1,\\
\rho_2(Y) \rd Y &={1\over 4 \pi \ri t_2 } {\rd Y \over Y}\left( \omega_0(Y+\ri \epsilon) -\omega_0(Y-\ri \epsilon)\right), \qquad Y\in \CC_2.
\end{aligned}
\end{equation}
The planar resolvent turns out to have the explicit expression \cite{MAakmv,MAhy}
\begin{equation} \label{MAw/2}
\omega_0(Z)   = \log \left( {\MAre^{-t} \over 2  } \left[ f(Z) -{\sqrt{ f^2(Z) -4 \MAre^{2t} Z^2}}  \right]\right).
\end{equation}
Notice that $\MAre^{\omega_0}$ has a square root branch cut involving the function
\begin{equation}
\label{MAsigmaz}
\sigma(Z)= f^2(Z) -4 \MAre^{2t} Z^2 =\left(Z-a\right) \left(Z-1/a\right) \left(Z+b\right) \left(Z+1/b\right), 
\end{equation}
where $a^{\pm 1}, b^{\pm1}$ are the endpoints of the cuts in the $Z=\MAre^z$ plane (i.e. $A=\log a$, $B=\log b$). They are determined, 
in terms of the parameters $t_1$, $t_2$ 
by the normalization conditions for the densities (\ref{MAnorma}). We will state the final results in ABJM theory. A detailed derivation can be found in the original 
papers \cite{MAmpabjm,MAdmp} and in the review \cite{MAmmlec}. 

In the ABJM case we have to consider the special case or ``slice" given in (\ref{MAabjm-limit}), therefore $t=0$. One can parametrize the endpoints 
of the cut in terms of a single parameter $\kappa$, as 
\begin{equation}
\label{MAendpoints}
a+{1\over a}=2+\ri \kappa, \qquad b+{1\over b}=2-\ri \kappa. 
\end{equation}
The 't Hooft coupling $\lambda$ turns out to be a non-trivial function of $\kappa$, determined by the normalization of the density. 
In order for $\lambda$ to be real and well-defined, $\kappa$ has to be real as well, and one finds the equation \cite{MAmpabjm}
 \begin{equation}
 \label{MAlamkap}
 \lambda(\kappa)={\kappa \over 8 \pi}   {~}_3F_2\left(\frac{1}{2},\frac{1}{2},\frac{1}{2};1,\frac{3}{2};-\frac{\kappa^2
   }{16}\right). 
   \end{equation}
Notice that the endpoints of the cuts are in general complex, i.e. the cuts $\CC_1$, $\CC_2$ are arcs in the complex plane. This is a consequence of the 
analytic continuation and it has been 
verified in numerical simulations of the original saddle-point equations (\ref{MAsaddleabjm}) \cite{MAhkpt}. 
Using similar techniques (see again \cite{MAmmlec}), one finds a very explicit expression for the derivative of the planar free energy,
\begin{equation}
\label{MAcomf}
 \partial_\lambda F_0 (\lambda)={\kappa \over 4} G^{2,3}_{3,3} \left( \begin{array}{ccc} {1\over 2}, & {1\over 2},& {1\over 2} \\ 0, & 0,&-{1\over 2} \end{array} \biggl| -{\kappa^2\over 16}\right)+{ \pi^2 \ri \kappa \over 2} 
  {~}_3F_2\left(\frac{1}{2},\frac{1}{2},\frac{1}{2};1,\frac{3}{2};-\frac{\kappa^2
   }{16}\right).
\end{equation}
This is written in terms of the auxiliary variable $\kappa$, but by using the explicit map (\ref{MAlamkap}), one can re-express it in terms of the 't Hooft coupling, and one finds the following expansion 
around $\lambda=0$, 
\begin{equation}
\label{MAGcomp}
 \partial_\lambda F_0 (\lambda)=-8 \pi ^2 \lambda  \left(\log \left(\frac{\pi  \lambda }{2}\right)-1\right)+\frac{16 \pi ^4 \lambda ^3}{9} +\CO\left(\lambda^5\right). 
 \end{equation}
It is easy to see that this reproduces the perturbative, weak coupling expansion of the matrix integral. This also fixes the integration constant, and one can write 
 \begin{equation}
 \label{MAcorrectweak}
  F_0(\lambda)=\int_0^{\lambda} \rd \lambda' \,  \partial_{\lambda'} F_0 (\lambda').
\end{equation}
To study the strong 't Hooft coupling behavior, we notice from (\ref{MAlamkap}) that large $\lambda \gg1$ requires $\kappa \gg 1$. More concretely, we find the following expansion at large $\kappa$:
\begin{equation}
 \label{MAlamas}
 \lambda(\kappa) ={\log ^2(\kappa)\over 2 \pi
   ^2}+\frac{1}{24}+\CO\left( {1\over \kappa^2}\right), \qquad \kappa \gg 1. 
   \end{equation}
 This suggests to define the shifted coupling
\begin{equation}
\label{MAshift-lam}
\hat \lambda=\lambda-{1\over 24}.
\end{equation}
Notice from (\ref{MAqshift}) that this shift is precisely the one needed in order for $\hat\lambda$ to be identified with $Q/k$, at leading order in the string coupling constant. 
The relationship (\ref{MAlamas}) is immediately inverted to 
\begin{equation}
\kappa \approx \MAre^{\pi {\sqrt{2\hat \lambda}}}, \qquad \lambda\gg 1. 
 \end{equation}
  To compute the planar free energy, we have to analytically continue the r.h.s. of (\ref{MAcomf}) to $\kappa=\infty$, 
and we obtain
\begin{equation}
\label{MAslice}
\partial_\lambda F_0 (\lambda)=2\pi^2 \log \kappa +{4 \pi^2 \over \kappa^2} \, {}_4 F_3 \left( 1, 1, {3\over 2}, {3\over 2}; 2,2,2; -{16 \over \kappa^2} \right). 
\end{equation}
After integrating w.r.t. $\lambda$, we find, 
\begin{equation}
\label{MAprepotf}
F_0 (\hat \lambda)= {4\pi^3 \sqrt{2}  \over 3} \hat \lambda^{3/2}+{\zeta(3) \over 2}
+\sum_{\ell\ge1}  \MAre^{- 2\pi \ell  {\sqrt{2\hat \lambda}}} f_{\ell}\left({1\over \pi {\sqrt{2 \hat \lambda}}} \right),
\end{equation}
where $f_{\ell}(x)$ is a polynomial in $x$ of degree $2 \ell-3$ (for $\ell\ge 2$). The leading term in (\ref{MAprepotf}) agrees precisely with the 
prediction from the AdS dual in (\ref{MAssugra}). The series of exponentially small corrections in (\ref{MAprepotf}) were interpreted in \cite{MAdmp} as coming from 
worldsheet instantons of type IIA theory wrapping the $\IC\IP^1$ cycle in $\IC\IP^3$. This is a novel type of correction in AdS$_4$ dualities which is not present in 
the large $N$ dual to $\CN=4$ super Yang--Mills theory, see \cite{MAsorokin} for a preliminary investigation of these effects. 

An important aspect of the above planar solution is the following. 
As we explained above, in finding this solution it is useful to take into account the relationship to the lens space matrix model 
of \cite{MAmmcs,MAakmv} discovered in \cite{MAmpabjm}. On the other hand, this matrix model computes, in the $1/N$ expansion, the partition function 
of topological string theory on a non-compact Calabi--Yau (CY) known as local $\IP^1 
\times \IP^1$, and in particular its planar free energy is given by the genus zero free energy or 
prepotential of this topological string theory. Local $\IP^1 \times \IP^1$ has two complexified 
K\"ahler parameters $T_{1,2}$. It turns out that the ABJM slice in which $N_1=N_2$ corresponds to the ``diagonal" geometry in which $T_1=T_2$. The relationship of ABJM 
theory to this topological string theory has been extremely useful in deriving exact answers for many of these quantities, and we will find it again in the sections to follow. For example, 
the constant term involving $\zeta(3)$ in (\ref{MAprepotf}) is well-known in topological string theory and it gives the constant map contribution to the genus 
zero free energy. The series of worldsheet instantons appearing in (\ref{MAprepotf}) is 
related to the worldsheet instantons of genus zero in topological string theory. There is however one subtlety: the 
genus zero free energy in (\ref{MAprepotf}) is the one appropriate to the so-called ``orbifold frame" studied in \cite{MAakmv}, 
and then it is analytically continued to large $\lambda$, which in topological string theory 
corresponds to the so-called large radius regime. This is not a natural procedure to follow from the point of view of topological strings on local $\IP^1 \times \IP^1$, 
where quantities in the orbifold frame are typically expanded around the orbifold point.

\subsubsection{Higher genus corrections}

The analysis of the previous subsection gives us the leading term in the $1/N$ expansion, but it is of course an important and interesting problem to compute the higher genus free energies with 
$g\ge 1$. This involves computing subleading $1/N$ corrections to the free energy of the ABJM matrix model. The computation of such 
corrections in Hermitian matrix models has a long history, and a general algorithm solving the problem was found in \cite{MAeo}. 
However, this algorithm is difficult to implement in practice. In some examples, one can use 
a more efficient method, developed in the context of topological string theory, which is known as the direct integration of the holomorphic anomaly equations. 
This method was introduced in \cite{MAhk}, and applied to the ABJM matrix model in \cite{MAdmp}. The $F_g(\lambda)$ obtained by this method 
are written in terms of modular forms. The modular parameter is given by 
\begin{equation}
\label{MAtauex}
\tau=\ri  {K'\left({\ri \kappa \over 4}\right)\over K \left({\ri \kappa \over 4}\right)}. 
\end{equation}
In this equation, $K(k)$ is the elliptic integral of the first kind, $K'(k)=K(k')$, and 
\begin{equation}
(k')^2=1-k^2
\end{equation}
is the complementary modulus. $\tau$ is related to the second derivative of the planar free energy by
\begin{equation}
\label{MAtauABJM}
{\ri \over 4 \pi^3} \partial_\lambda^2 F_0(\lambda)=\tau-1,
\end{equation}
which is a standard relation in special geometry. The genus one free energy is given by 
\begin{equation}
\label{MAgenus-one}
F_1(\lambda)= -\log \eta(\tau-1)-{1\over2} \log(2), 
\end{equation}
where $\eta(\tau)$ is Dedekind's eta function. The higher genus free energies are expressed in terms of $E_2(\tau)$, 
the standard Eisenstein series, and the Jacobi theta functions
\begin{equation}
b=\vartheta_2^4(\tau), \qquad d=\vartheta_4^4(\tau). 
\end{equation} 
They have the general structure
\begin{equation}
F_g(\lambda)={1\over \left( b d^2 \right)^{g-1}} \sum_{k=0}^{3g-3}
E_2^{k}(\tau) p^{(g)}_k(b,d), \qquad g\ge 2, 
\end{equation}
where $p^{(g)}_k(b,d)$ are polynomials in $b, d$ of modular weight $6g-6-2k$. For example, for the genus two free energy one finds the explicit expression 
\begin{equation}
F_2(\lambda)=\frac{1}{432 b d^2}\left(-\frac{5}{3}\,E_2^3+3bE_2^2-2E_4E_2\right)+\frac{16 b^3+15 d b^2-15 d^2 b+2 d^3}{12960 b d^2}.  
\end{equation}
The higher genus $F_g$ can be found recursively, although there is no known closed form expression or generating functional for them. A detailed analysis 
for the very first $g$ shows that they have the following structure, in terms of the auxiliary variable $\kappa$ \cite{MAdmpnp}\footnote{The constant contribution $c_g$ was not 
originally included 
in \cite{MAdmp,MAdmpnp}, but this omission was corrected in \cite{MAhanada}.}:
\begin{equation}
\label{MAfgl}
F_g=c_g+  f_g\left( {1\over \log \, \kappa} \right)+ \CO\left( {1\over  \kappa^2} \right), \qquad g\ge2,
\end{equation}
where 
\begin{equation}
\label{MAcg}
c_g=- {4^{g-1} |B_{2g} B_{2g-2}| \over g (2g-2) (2g-2)!}
\end{equation}
involves the Bernoulli numbers $B_{2g}$, and 
\begin{equation}
f_g(x)=\sum_{j=0}^g c_j^{(g)}x^{2g-3+j}
\end{equation}
is a polynomial. Physically, the equation (\ref{MAfgl}) tells us that the higher genus free energy has a constant contribution, 
a polynomial contribution 
in inverse powers of $\lambda^{1/2}$, going like 
\begin{equation}
 F_g(\lambda)-c_g\approx \lambda^{{3\over 2}-g}, \qquad \lambda \gg 1, \quad g\ge2, 
 \end{equation}
 and an infinite series of corrections due to worldsheet instantons of genus $g$. The quantities appearing here have a natural 
 interpretation in the context of topological string theory, since the $F_g(\lambda)$ are simply the orbifold higher genus free energies of local $\IP^1 \times \IP^1$. 
 The constants (\ref{MAcg}) are the well-known constant map contributions to the higher genus free energies, and the worldhseet instantons 
 of type IIA superstring theory appearing in $F_g(\lambda)$ come from the worldsheet instantons of the topological string. 
 
In \cite{MAdmpnp} it was noted that, if we drop the worldsheet instanton corrections in the $F_g(\lambda)$, the expansion of the free energy 
 has a simple expression in terms of a variable $\zeta$ defined by 
 \begin{equation}
\zeta=32 \pi^2 k \left( N- B(k) \right),
\end{equation}
where
\begin{equation}
\label{MAbk}
B(k)={k \over 24} +{1\over 3 k}. 
\end{equation}
The free energy truncated in this way, which we will denote by $F^{(\rm p)}(N,k)$ (where the superscript means perturbative), has the following 
expansion, 
\begin{equation}
\label{MAfp}
F^{(\rm p)}(N,k)=-{1\over 384\pi^2 k} \zeta^{3/2}+{1\over 6} \log \left[ {\pi^3 k^3 \over  \zeta^{3/2}} \right] +A(k)\\
+ \sum_{n=1}^{\infty} d_{n+1} \pi^{2n} k^n  \zeta^{-3n/2},
\end{equation}
where the coefficients $d_n$ are just rational numbers, 
\begin{equation}
\label{MAdjs}
d_2=-{80\over 3} , \qquad d_3=5120, \qquad d_4=-{18104320 \over 9}, \qquad d_5=1184890880, \qquad \cdots
\end{equation}
and the constant term $A(k)$ is an appropriate resummation at all genera of the contribution from the constant maps. Its explicit expression was first 
found in \cite{MAhanada} and it was slightly simplified 
in \cite{MAho} to the form, 
\begin{equation}
\label{MAak}
A(k)= \frac{2\zeta(3)}{\pi^2 k}\left(1-\frac{k^3}{16}\right)
+\frac{k^2}{\pi^2} \int_0^\infty \rd x \frac{x}{\MAre^{k x}-1}\log(1-\MAre^{-2x}).
\end{equation}
It can be expanded, around $k =\infty$, as
\begin{equation}
\label{MAlargeka}
A(k)=-{k^2\over 8 \pi^2} \zeta(3)+ {1\over 2} \log(2) + 2 \zeta'(-1) +{1\over 6} \log \left({\pi \over 2k}\right) + \sum_{g\ge 2} \left({2 \pi \over k}\right)^{2g-2} (-1)^{g} c_g, 
\end{equation}
where the $c_g$ are given in (\ref{MAcg}). 
The expansion (\ref{MAfp}) is remarkable, both physically and mathematically. First of all, it was shown in \cite{MAfhm} 
that it can be resummed in terms of the well-known Airy function: after exponentiation, one finds that 
the partition function has the form
\begin{equation}
\label{MAzpnk}
Z^{\rm (p)}(N,k)=\MAre^{A(k)} C^{-1/3}(k) {\rm Ai} \left[ C^{-1/3}(k) \left( N- B(k) \right) \right], 
\end{equation}
where
\begin{equation}
\label{MAck}
C(k)= {2 \over  \pi^2 k}. 
\end{equation}
The expression (\ref{MAzpnk}) gives an excellent approximation to the integral (\ref{MAabjm-mm}) for large $N$ and fixed $k$ \cite{MAhanada}. On the other hand, 
from a physical point of view, if we assume that the parameter $\zeta$ gives the right ``renormalized" dictionary between the gauge theory data and the geometry, i.e., if
\begin{equation}
\left( {L \over \ell_p} \right)^6= \zeta, 
\end{equation}
then (\ref{MAfp}) is the expected expansion for a free energy in a theory of quantum gravity in 
eleven dimensions. Indeed, an $\ell$-loop term for a vacuum diagram in gravity in $d$ dimensions 
goes like (see for example \cite{MAburgess,MAsen-s})
\begin{equation}
\left( {\ell_p \over L}\right)^{(d-2)(\ell-1)}, 
\end{equation}
which for $d=11$ agrees with the expansion parameter $\zeta^{-3/2}$ appearing in (\ref{MAfp}). The log term in (\ref{MAfp}) should correspond to a one-loop correction in supergravity, and this was checked 
by a direct computation in \cite{MAbgms}, providing in this way a test of the AdS/CFT correspondence beyond the planar limit (in type IIA, this correction comes from the genus one free energy). 

The 't Hooft expansion (\ref{MAgenusg}) gives an asymptotic series for the free energy, at fixed 't Hooft parameter. General arguments 
(see \cite{MAshenker} for an early statement and 
\cite{MAmmlargen} for a recent review) suggest that this series diverges factorially. The divergence of the series is controlled by a large $N$ 
instanton with action $A_{\rm st}(\lambda)$. The correction due to such an instanton is proportional to the exponentially suppressed factor, 
\begin{equation}
\label{MAnp-effects}
\exp(-A_{\rm st}(\lambda)/g_s). 
\end{equation}
%
An explicit expression for the instanton action $A_{\rm st}(\lambda)$ was conjectured 
in \cite{MAdmpnp}. When $\lambda$ is real and sufficiently large, it is given by 
\begin{equation}
A_{\rm st}(\lambda)=\frac{\ri \kappa}{4\pi} G^{2,3}_{3,3} \left(\left. 
  \begin{array}{ccc} 
    \frac{1}{2}, & \frac{1}{2},& \frac{1}{2} \\ 
    0, & 0,&-\frac{1}{2} 
  \end{array} \right| -\frac{\kappa^2}{16}\right)-\frac{ \pi \kappa}{2} 
  {~}_3F_2\left(\frac{1}{2},\frac{1}{2},\frac{1}{2};1,\frac{3}{2};-\frac{\kappa^2
   }{16}\right)
- \pi^2, 
\end{equation}
and it is essentially proportional to the derivative of the free energy (\ref{MAcomf}). 
The function $A_{\rm st}(\lambda)$ is complex, and at strong coupling it behaves like, 
\begin{equation}
\label{MAiaction}
-\ri A_{\rm st}(\lambda)=2 \pi^2 {\sqrt{2 \lambda}} + \pi^2 \ri + \CO\left(\MAre^{-2 \pi {\sqrt{2\lambda}}}\right), \qquad \lambda \gg 1. 
\end{equation}
Since the genus $g$ amplitudes are real, the complex 
instanton governing the large order behavior of the $1/N$ expansion must appear together with its complex conjugate, 
and it leads to an oscillatory asymptotics. If we write
\begin{equation}
A_{\rm st}(\lambda)= \left|A_{\rm st}(\lambda)\right| \MAre^{\ri \theta(\lambda)}.
\end{equation}
we have the behavior, 
\begin{equation}
\label{MAoscil}
F_g(\lambda)-c_g  \sim (2g)! \left| A_{\rm st}(\lambda)\right| ^{-2g} \cos\left( 2g\theta (\lambda) +\delta (\lambda) \right), \qquad  g\gg 1, 
\end{equation}
where $\delta(\lambda)$ is a function of the 't Hooft coupling, which in simple cases is determined by the one-loop corrections around the instanton. 
The oscillatory asymptotics in (\ref{MAoscil}) suggests that the 't Hooft expansion is Borel summable. This  was tested in \cite{MAgmz} by 
detailed numerical calculations. However, the Borel resummation of the expansion does not 
reproduce the correct values of the free energy at finite $N$ and $k$. The contribution of the complex instanton, which is of order (\ref{MAnp-effects}), 
should be added in an appropriate way to the Borel-resummed 't Hooft expansion 
in order to reconstruct the exact answer for the free energy. In practice, this means that one should consider ``trans-series" incorporating these exponentially 
small effects (see for example \cite{MAmmlargen} for an introduction to trans-series.)

The resummation of the perturbative free energies in (\ref{MAzpnk}) in terms of an Airy function suggests another approach to the problem. Conceptually, the resummation of the 
genus expansion in type IIA superstring theory should be achieved by going to M-theory. The non-perturbative effects appearing in (\ref{MAnp-effects}) should also appear naturally in 
an M-theory approach: by using (\ref{MAiaction}), we see that they have the form, for $\lambda\gg 1$, 
\begin{equation}
\label{MAg-membrane}
\exp\left(-{\sqrt{2}} \pi k^{1/2} N^{1/2} \right). 
\end{equation}
In view of the AdS/CFT dictionary (\ref{MALNdic}), the exponent in (\ref{MAg-membrane}) goes like
\begin{equation}
k^{1/2} N^{1/2}  \sim \left( {L \over \ell_p} \right)^3. 
\end{equation}
This is the expected dependence on $L$ for the action of a membrane instanton in M-theory, 
which corresponds to a D2-brane in type IIA theory. In \cite{MAdmpnp}, it was shown that a D2 brane 
wrapping the $\IR\IP^3$ cycle inside $\IC \IP^3$ would lead to the correct strong coupling limit of the action (\ref{MAiaction}). Therefore, 
by going to M-theory, we could in principle incorporate not only the 
worldsheet instantons which were not taken into account in (\ref{MAzpnk}), 
but also the non-perturbative effects due to membrane instantons. In fact, it is well-known that in M-theory membrane and 
worldsheet instantons appear on equal footing \cite{MAbbs}.

\subsection{The M-theory expansion}

In the M-theory expansion, $N$ is large and $k$ is fixed, corresponding to the regime (\ref{MAmtregime}). The original 
study of the ABJM matrix model (\ref{MAabjm-mm}) in \cite{MAmpabjm,MAdmp} was done in the 't Hooft regime (\ref{MAthooft-regime}). It is now time to see if we can 
understand the matrix model directly in the M-theory regime and solve the problems raised at the end of the previous section: can 
we resum the genus expansion in some way? Can we incorporate the non-perturbative effects due to membrane instantons?

\subsubsection{The strict large $N$ limit}
\label{MAhkpt-review}
The first direct study of the M-theory regime of the matrix model (\ref{MAabjm-mm}) was performed in \cite{MAhkpt}. 
What should we expect in this regime, based on the results from the 't Hooft expansion? First of all, note that, in this regime, $\lambda$ scales with $N$, 
therefore the M-theory regime corresponds to strong 't Hooft coupling. If we analyze 
the planar solution at strong coupling, we find that the endpoints of the cuts for the eigenvalues $\mu_i$, $\nu_i$, given in (\ref{MAendpoints}), behave like, 
\begin{equation}
\label{MAplanar-strong}
\begin{aligned}
A & \approx \pi {\sqrt{ {2N \over k}-{1\over 12}}} + {\ri \pi \over 2}, \\
B & \approx \pi {\sqrt{ {2N \over k}-{1\over 12}}} - {\ri \pi \over 2}. 
\end{aligned}
\end{equation}
Therefore, the equilibrium positions for the eigenvalues occur around arcs in the complex plane, and the real part of their endpoints 
grows like $N^{1/2}$ at fixed $k$. Note that $A$ and $B$ are related by complex conjugation, 
in agreement with the symmetry of the equations (\ref{MAsaddleabjm}) under $\mu_i \rightarrow \nu_i^*$. 
Although the above result is obtained by looking at the strong coupling behavior of the planar limit, it was verified in \cite{MAhkpt} by a numerical analysis of 
the equations (\ref{MAsaddleabjm}) at large $N$ 
and fixed $k$. It suggests the following ansatz for the M-theory limit of the distribution of the eigenvalues, 
\begin{equation}
 \label{MAscaling}
  \mu_j=N^{1/2} x_j+ \ri y_j, \qquad \nu_j=N^{1/2} x_j-\ri y_j,  \qquad j=1, \cdots, N,
  \end{equation}
  where $x_j$, $y_j$ are of order one at large $N$. If we assume that the values of $x_j$, $y_j$ become dense at large $N$, as suggested both by the planar 
  limit and the numerical analysis, we should introduce a continuous parameter in the standard way, 
  \begin{equation}
  {j\over N} \rightarrow \xi \in [0,1], 
  \end{equation}
so that the limiting distributions are described by functions $x(\xi)$, $y(\xi)$. We also introduce the density of eigenvalues
\begin{equation}
\rho(x) ={\rd \xi \over \rd x}. 
\end{equation}
A detailed analysis performed in \cite{MAhkpt} shows that, when $N$ is large, the free energy of the matrix model can be written as a 
functional of $\rho(x)$ and $y(x) =y \left( \xi(x) \right)$, 
\begin{equation}
\label{MAhkfree}
-F(N,k)=N^{3/2} \left[ {k \over \pi} \int \rd x \, x \rho(x) y(x) + \int \rd x \, \rho^2(x) f\left( 2 y(x)\right) -{m \over 2 \pi} \left( \int \rd x \, \rho(x)-1\right)\right]. 
\end{equation}
Here, $f(t)$ is a periodic function of $t$, with period $2\pi$, and given by
\begin{equation}
f(t) =\pi^2 -t^2, \qquad t\in [-\pi, \pi]. 
\end{equation}
The last term in (\ref{MAhkfree}) involves, as usual, a Lagrange multiplier $m$ imposing the normalization of $\rho(x)$. As stressed in \cite{MAhkpt}, 
the above functional is local in the functions $\rho(x)$, $y(x)$, in 
contrast to the standard functional for the planar limit of matrix model, which involves an interaction 
between $\rho(x)$ and $\rho(x')$ at different points $x,x'$. The reason is that 
the non-local part of the interaction between the eigenvalues cancels due to the 
presence of the $\cosh$ term in the denominator of (\ref{MAabjm-mm}). 
Varying the functional (\ref{MAhkfree}) w.r.t. $\rho(x)$ 
 and $y(x)$, one obtains the two equations
 \begin{equation}
 \begin{aligned}
 2 \pi \rho(x) f'\left(2 y(x)\right)&=-kx, \\
 4 \pi \rho(x) f\left(2 y(x)\right)&=m-2 k x y(x),
 \end{aligned}
 \end{equation}
 which are solved by 
 \begin{equation}
 \label{MAnomsol}
 \rho(x)={m \over 4 \pi^3}, \qquad y(x)={\pi^2 k x \over 2 m}. 
 \end{equation}
The support of $\rho(x)$, $y(x)$ is the interval $[-x_*, x_*]$. One fixes $x_*$ and $m$ from the normalization of $\rho$ and by minimizing $-F$. This 
gives
\begin{equation}
\label{MAxmABJM}
x_*=\pi {\sqrt{2 \over k}}, \qquad m={2 \pi^3 \over x_*}. 
\end{equation}
Therefore, 
\begin{equation}
y(x_*)= {\pi \over 2}, 
\end{equation}
in agreement with the planar solution at strong coupling (\ref{MAplanar-strong}). Evaluating the free energy for the functions (\ref{MAnomsol}) and the values (\ref{MAxmABJM}) of 
$x_*$, $m$, one finds
\begin{equation}
\label{MAlead-m}
-F(N,k) \approx {\pi {\sqrt{2}} \over 3} k^{1/2} N^{3/2}, \qquad N \gg 1, 
\end{equation}
which agrees with the prediction of M-theory (\ref{MAmsugra}). Note that the density of eigenvalues $\rho(x)$ in (\ref{MAnomsol}) is a constant. This 
agrees again with the strong coupling limit 
of the planar densities of eigenvalues $\rho_1(X)$, $\rho_2(Y)$ in (\ref{MAdensities}), as shown in \cite{MAcmp}. 

The result (\ref{MAfp}), obtained by a partial resummation of the 't Hooft expansion, shows that the M-theory expansion of the 
ABJM free energy has subleading corrections at large $N$, as well as non-perturbative 
corrections coming from worldsheet instantons. Can we derive these corrections directly from a study of the ABJM matrix model? In particular, 
we would like to obtain in the M-theory expansion a quantitative understanding of the non-perturbative corrections of the form (\ref{MAg-membrane}), 
which are invisible in the 't Hooft expansion. As shown in \cite{MAm-pufu}, the next-to-leading 
correction to (\ref{MAlead-m}) at large $N$ and fixed $k$ can be computed by extending the analysis of \cite{MAhkpt} that we 
have just reviewed. However, this just captures the leading effect due to the shift of
$\lambda$ incorporated in the variable $\hat \lambda$ of (\ref{MAshift-lam}), which is the variable appearing naturally in the 
planar expression (\ref{MAprepotf}). Including further corrections seems 
difficult to do in the approach of \cite{MAhkpt}. This motivates another approach which was started in \cite{MAmp} and has been very useful in understanding the corrections to the strict large $N$ limit. 

\subsubsection{The Fermi gas approach}

There is a long tradition relating matrix integrals to fermionic theories. One reason for this is that the Vandermonde determinant 
\begin{equation}
\Delta(\mu)= \prod_{i<j} (\mu_i - \mu_j)
\end{equation}
appearing in these integrals can be regarded, roughly speaking, as 
a Slater determinant in a theory of $N$ one-dimensional fermions with positions $\mu_i$. For example, the fact that 
this factor vanishes whenever two particles are at the same point can be regarded as a manifestation of Pauli's exclusion principle. 

The rewriting of the ABJM matrix integral in terms of fermionic quantities can be regarded as a variant of this idea. It should be remarked however that the 
Fermi gas approach that we will explain in this section is not a universal technique which can be applied to any matrix integral with a Vandermonde-like 
interaction. It rather requires a specific type of eigenvalue interaction, which turns out to be typical of many matrix integrals appearing in the localization 
of Chern--Simons--matter theories. 

The starting point for the Fermi gas approach is the observation that the interaction term in 
the matrix integral (\ref{MAabjm-mm}) can be rewritten by using the Cauchy identity, 
 \begin{equation}
 \label{MAcauchy}
 \begin{aligned}
  {\prod_{i<j}  \left[ 2 \sinh \left( {\mu_i -\mu_j \over 2}  \right)\right]
\left[ 2 \sinh \left( {\nu_i -\nu_j   \over 2} \right) \right] \over \prod_{i,j} 2 \cosh \left( {\mu_i -\nu_j \over 2} \right)}  
 & ={\rm det}_{ij} \, {1\over 2 \cosh\left( {\mu_i - \nu_j \over 2} \right)}\\
 &=\sum_{\sigma \in S_N} (-1)^{\epsilon(\sigma)} \prod_i {1\over 2 \cosh\left( {\mu_i - \nu_{\sigma(i)} \over 2} \right)}.
 \end{aligned}
  \end{equation}
  In this equation, $S_N$ is the permutation group of $N$ elements, and $\epsilon(\sigma)$ is the signature of the permutation $\sigma$. 
  After some manipulations spelled out in detail in \cite{MAkwytwo}, one obtains \cite{MAkwytwo,MAmp}
\begin{equation}
\label{MAfgasform}
Z (N,k)={1 \over N!} \sum_{\sigma  \in S_N} (-1)^{\epsilon(\sigma)}  \int  {\rd ^N x \over (2 \pi k)^N}  \prod_{i=1}^N  \rho\left(x_i, x_{\sigma(i)}\right), 
\end{equation}
where
\begin{equation}
\label{MAdensitymat}
\rho(x_1, x_2)={1\over 2 \pi k} {1\over \left( 2 \cosh  {x_1 \over 2}  \right)^{1/2} }  {1\over \left( 2 \cosh {x_2  \over 2} \right)^{1/2} } {1\over 
2 \cosh\left( {x_1 - x_2\over 2 k} \right)}. 
\end{equation}
The expression (\ref{MAfgasform}) can be immediately identified \cite{MAfeynman} as the canonical partition function of a one-dimensional ideal Fermi 
gas of $N$ particles, where (\ref{MAdensitymat}) is the canonical density matrix. Notice that, by using the Cauchy identity again, 
with $\mu_i=\nu_i$, we can rewrite (\ref{MAfgasform}) as a matrix integral involving one single set of $N$ eigenvalues, 
\begin{equation}
\label{MAtanh-form}
Z (N,k)={1\over N!}  \int \prod_{i=1}^N {\rd x_i \over 4 \pi k}  {1\over 2 \cosh {x_i \over 2} } \prod_{i<j} \left( \tanh \left( {x_i - x_j \over 2 k } \right) \right)^2. 
\end{equation}
The canonical density matrix (\ref{MAdensitymat}) is related to the Hamiltonian operator $\mathsf{H}$ in the usual way, 
\begin{equation}
\label{MAhamil}
\rho(x_1, x_2)= \langle x_1| \mathsf{\rho} | x_2\rangle,  \qquad   \mathsf{\rho}=\MAre^{-\mathsf{H}},
\end{equation}
where the inverse temperature $beta=1$ is fixed. We will come back to the construction of the Hamiltonian shortly. 

Since ideal quantum gases are better studied 
in the grand canonical ensemble, the above representation suggests to look at the grand canonical partition function, defined by 
\begin{equation}
\label{MAxips}
\Xi(\mu, k)=1+\sum_{N\ge 1} Z(N,k) \MAre^{N \mu}. 
\end{equation}
Here, $\mu$ is the chemical potential. The grand canonical potential is 
\begin{equation}
\CJ(\mu,k) =\log \Xi (\mu,k). 
\end{equation}
A standard argument (presented for example in \cite{MAfeynman}) tells us that
\begin{equation}
\label{MAjsum}
\CJ(\mu,k)=-\sum_{\ell \ge 1}{(-\kappa)^\ell \over \ell} Z_\ell, 
\end{equation}
where 
\begin{equation}
\label{fuga}
\kappa= \MAre^\mu
\end{equation}
is the fugacity, and 
\begin{equation}
\label{MAtrzl}
Z_\ell= \MAtr \mathsf{\rho}^\ell =\int \rd x_1 \cdots \rd x_\ell   \, \rho(x_1, x_2)\rho(x_2, x_3)\cdots \rho(x_{\ell-1}, x_\ell) \rho(x_{\ell}, x_1).
\end{equation}
As is well-known, the canonical and the grand-canonical formulations are equivalent, and the canonical partition function is recovered from the grand canonical one by integration, 
\begin{equation}
\label{MAexactinverse}
Z(N,k) =\oint {\rd \kappa \over 2 \pi \ri } {\Xi (\mu, k) \over \kappa^{N+1}}.
\end{equation}
Since we are dealing with an ideal gas, all the physics is in principle encoded in the spectrum of the Hamiltonian $\mathsf{H}$. This spectrum is defined by, 
\begin{equation}
\label{MAspec}
\mathsf{\rho} |\varphi_n \rangle =\MAre^{-E_n} |\varphi_n \rangle, \qquad n=0, 1, \dots, 
\end{equation}
or, equivalently, by the integral equation associated to the kernel (\ref{MAdensitymat}),
\begin{equation}
\label{MAspec-2}
\int  \rho(x, x') \varphi_n (x') \rd x'= \MAre^{-E_n} \varphi_n (x),  \qquad n=0, 1, \dots. 
\end{equation}
It can be verified that this spectrum is indeed discrete and the energies are real. This is because, as it can be easily checked, (\ref{MAdensitymat}) defines a positive, trace-class operator 
on $L^2(\IR)$, and the above properties of the spectrum follow from standard results in the theory of such operators (see, for example, \cite{MAsimon-course}). The thermodynamics is 
completely determined by the spectrum: the grand canonical partition function is given by the Fredholm determinant associated to the integral operator (\ref{MAdensitymat}), 
\begin{equation}
\label{MAfred}
\Xi(\mu,k)= {\rm det}\left(1+ \kappa \mathsf{\rho} \right)=\prod_{n\ge 0} \left(1+ \kappa \MAre^{-E_n}\right). 
\end{equation}
In terms of the density of eigenvalues
\begin{equation}
\rho(E)= \sum_{n \ge 0} \delta(E-E_n),
\end{equation}
we also have the standard formula
\begin{equation}
\label{MAst-form}
\CJ (\mu, k)=\int_0^{\infty} \rd E \, \rho(E) \, \log\left(1+ \kappa \MAre^{-E} \right). 
\end{equation}

What can we learn from the ABJM partition function in the Fermi gas formalism? The first thing we can do is to derive the strict large $N$ limit of the free energy, including the correct 
coefficient. To do this, we have to be more precise about the Hamiltonian of the theory, which is defined implicitly by (\ref{MAhamil}) and (\ref{MAdensitymat}). Let us
 first write the density matrix (\ref{MAdensitymat}) as 
\begin{equation}
\label{MArhopq}
\mathsf{\rho}=\MAre^{-{1\over 2} U(\mathsf{x})} \MAre^{-T(\mathsf{p})} \MAre^{-{1\over 2} U(\mathsf{x})}.
\end{equation}
In this equation, $\mathsf{x}, \mathsf{p}$ are canonically conjugate operators, 
\begin{equation}
\label{MAcc-operators}
[\mathsf{x}, \mathsf{p}]=\ri \hbar, 
\end{equation}
and
\begin{equation}
\label{MAhache}
\hbar = 2 \pi k. 
\end{equation}
Note that $\hbar$ is the inverse coupling constant of the gauge theory/string theory, therefore 
semiclassical or WKB expansions in the Fermi gas correspond to strong coupling expansions in gauge theory/string theory. 
Finally, the potential $U(x)$ in (\ref{MArhopq}) is given by
\begin{equation}
\label{MAupotential}
U(x)=\log \left( 2 \cosh {x\over 2} \right), 
\end{equation}
and the kinetic term $T(p)$ is given by the same function,
\begin{equation}
\label{MAkinabjm}
T(p)=\log\left( 2\cosh  {p \over 2} \right). 
\end{equation}
Indeed, we have 
\begin{equation}
\begin{aligned}
\langle x'| \mathsf{\rho} |x\rangle&=\MAre^{-{1\over 2}U(x') -{1\over 2} U(x)} \int \rd p \rd p' \, \langle x'| p\rangle \langle p|\MAre^{-T(p)} |p'\rangle \langle p'| x\rangle\\ 
&=\MAre^{-{1\over 2}U(x') -{1\over 2} U(x)}\int {\rd p \over 2\pi \hbar } {\MAre^{{\ri \over \hbar} p (x'-x)} \over 2\cosh\left( {p \over 2} \right)}={1\over 2\pi k} \MAre^{-{1\over 2}U(x') -{1\over 2} U(x)} {1\over 2 \cosh \left( {x-x' \over 2 k} \right) },
\end{aligned}
\end{equation}
which is (\ref{MAdensitymat}). The resulting Hamiltonian is not standard. 
First of all, the kinetic term leads to an operator involving an infinite number of derivatives (by expanding 
it around $p=0$), and it should be regarded as a difference operator, as we will see later. 
Second, the ordering of the operators in (\ref{MArhopq}) 
shows that the Hamiltonian we are dealing with is not the sum of the kinetic term plus the potential, 
but it includes $\hbar$ corrections due to non-trivial commutators. This is for example what happens when one considers 
quantum theories on the lattice: the standard Hamiltonian is only recovered in the continuum limit, 
which sets the commutators to zero. All these complications can be treated appropriately, and we will 
address some of them in this expository article, but let us first try to understand what happens when $N$ is large.

\begin{center}
 \begin{figure}
\begin{center}
\includegraphics[height=6cm]{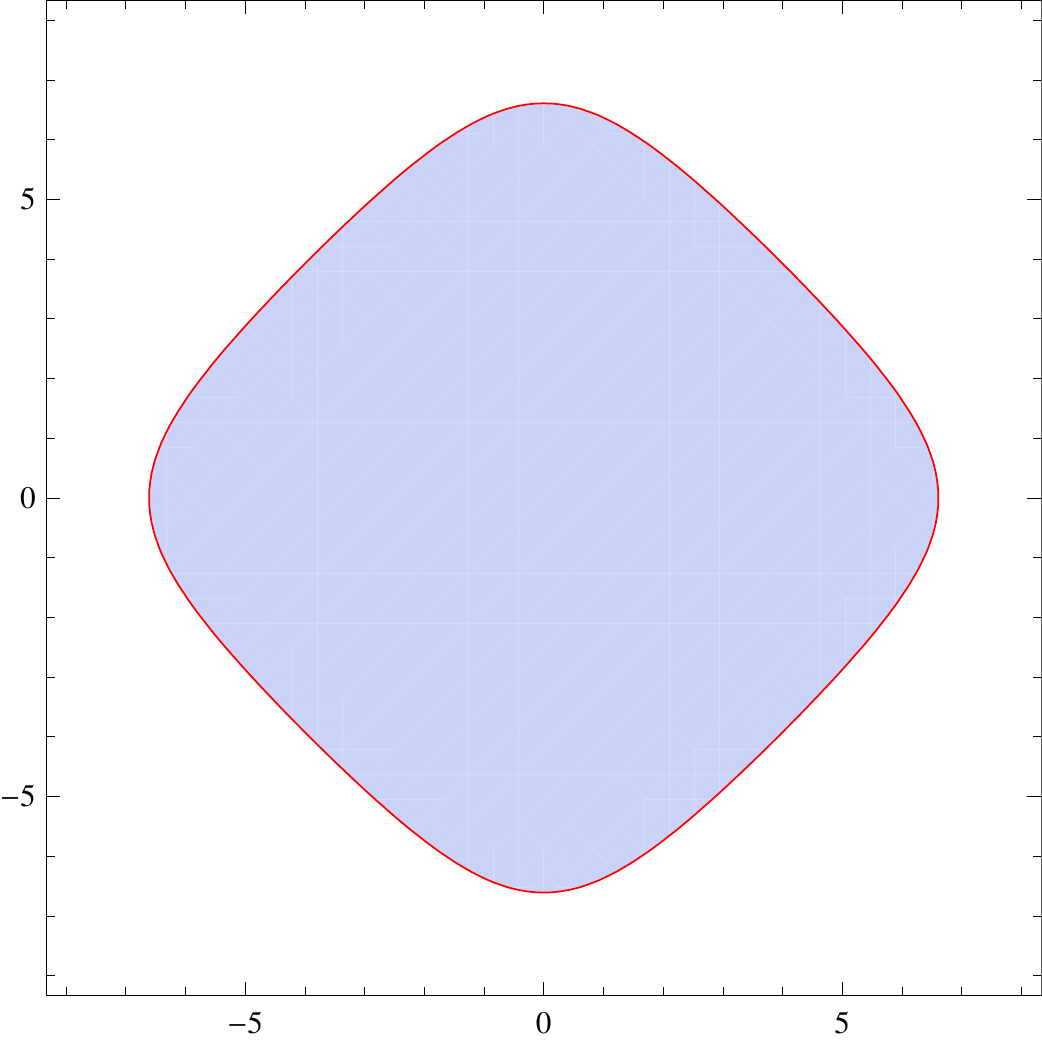} \qquad 
\includegraphics[height=6cm]{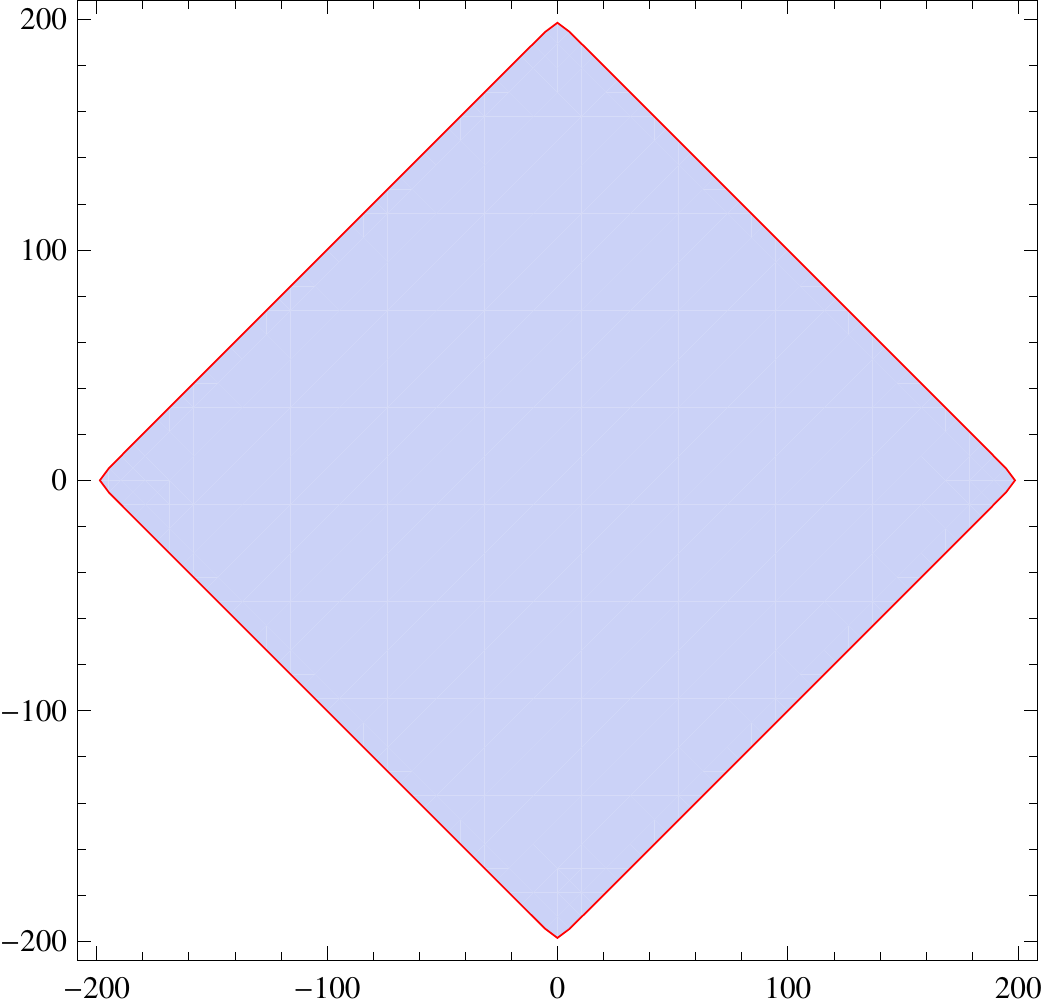}
\caption{The Fermi surface (\ref{MAclfermi}) for ABJM theory in the $q=x$-$p$ plane, for $E=4$ (left) and $E=100$ (right). When the energy is large, the Fermi surface approaches the polygon (\ref{MApolygon}).}
\label{MAfermisurface}
\end{center}
\end{figure}  
\end{center}
The potential in (\ref{MAupotential}) is a confining one, and at large $x$ it behaves linearly, 
\begin{equation}
U(x) \approx {|x|\over 2} , \qquad |x| \rightarrow \infty. 
\end{equation}
When the number of particles in the gas, $N$, is large, the typical energies are large, and we are in the semiclassical regime. In that case, 
we can ignore the quantum corrections to the Hamiltonian and 
take its classical limit
\begin{equation}
\label{MAclassical-hamil}
H_{\rm cl}(x,p)= U(x)+ T(p).
\end{equation}
Standard semiclassical considerations indicate that the number of particles $N$ is given by the area of the Fermi surface, defined by 
\begin{equation}
\label{MAclfermi}
H_{\rm cl}(x,p)=E, 
\end{equation}
divided by $2\pi \hbar$, the volume of an elementary cell. However, for large $E$, we can replace $U(x)$ and $T(p)$ by their leading behaviors at large argument, so that the 
Fermi surface is well approximated by the polygon, 
\begin{equation}
\label{MApolygon}
{|x|\over 2}+{|p| \over 2}=E.
\end{equation} 
 This can be seen in Fig. \ref{MAfermisurface}, where we show the Fermi surface computed from (\ref{MAclfermi}) for two values of the energies, a moderate one and a large one. For the large one, 
 the Fermi surface is very well approximated by the polygon of (\ref{MApolygon}). The area of this polygon is $8 E^2$. Therefore, by using the relation between the grand potential and the 
 average number of particles, 
 \begin{equation} 
 \label{MAn-aver}
 {\partial \CJ (\mu, k) \over \partial \mu} =\langle N(\mu, k) \rangle  \approx {8  \mu^2 \over 2 \pi \hbar}, 
 \end{equation}
 we obtain immediately
 \begin{equation}
 \label{MAleading-J}
 \CJ(\mu, k)  \approx {2 \mu^3 \over 3 \pi^2 k}.
 \end{equation}
To compute the free energy, we note that, at large $N$, the contour integral (\ref{MAexactinverse}) can be computed by a saddle--point approximation, which leads to the standard Legendre 
transform, 
\begin{equation}
\label{MAfreethermo}
F(N,k) \approx \CJ(\mu_*,k) - \mu_*N, 
\end{equation}
where $\mu_*$ is the function of $N$ and $k$ defined by (\ref{MAn-aver}), i.e. 
\begin{equation}
\label{MAmu-saddle}
\mu_* \approx {\sqrt{2} \over 2} \pi k^{1/2} N^{1/2}.
\end{equation}
In this way, we immediately recover the result (\ref{MAlead-m}) from (\ref{MAfreethermo}). In particular, the scaling $3/2$ is a 
simple consequence from the analysis: 
it is the expected scaling for a Fermi gas in one dimension with a linearly confining potential and an ultra-relativistic dispersion relation $T(p)\propto |p|$ at large $p$. This is arguably the 
simplest derivation of the result (\ref{MAlead-m}), as it uses only elementary notions in Statistical Mechanics. Note that in this derivation we have considered the M-theory regime in 
which $N$ is large and $k$ is fixed, and we have focused on the strict large $N$ limit considered in 
\cite{MAhkpt} and reviewed in the last section. The main questions is now: can we use the Fermi gas formulation to obtain explicit results for the corrections to the strict large $N$ limit? In the next 
sections we will address this question. 

\subsubsection{The WKB expansion of the Fermi gas}

In the Fermi gas approach, the physics of the partition function is encoded in a quantum ideal gas. Although the gas is non-interacting, 
its one-particle Hamiltonian is complicated, and the energy levels $E_n$ in (\ref{MAspec}) are not known in closed form. What can we do in this situation? 
As we have seen in (\ref{MAhache}), the parameter $k$ corresponds to the Planck constant of the quantum Fermi gas. Therefore, we can try a systematic 
development around $k=0$, i.e. a semiclassical WKB approximation. Such an approach should give a way of computing corrections to (\ref{MAleading-J}) and (\ref{MAlead-m}). 
Of course, we are not {\it a priori} interested in the physics at small $k$, but rather at {\it finite} $k$, and in particular at integer $k$. However, the expansion at small $k$ gives important clues 
about the problem at finite $k$ and it can be treated systematically. 

There are two ways of working out the WKB expansion: we can work directly at the level of the grand potential, or we can work at the level of the energy spectrum. Let us first 
consider the problem at the level of the grand potential. It turns out that, in order to perform a systematic semiclassical expansion, the most useful approach 
is Wigner's phase space formulation of Quantum Mechanics (in fact, this formulation was originally introduced by Wigner in order to understand the semiclassical expansion of 
thermodynamic quantities.) A detailed application of this formalism to the ABJM Fermi gas can be found in \cite{MAmp,MAfermi-wilson}. The main idea of the method is 
to map quantum-mechanical operators to functions in classical phase space through the Wigner transform. Under this map, 
the product of operators famously becomes the $\star$ or Moyal product 
(see for example \cite{MAmoyal-review} for a review, and \cite{MAgvoros} for an elegant summary with applications). 
This approach is particularly useful in view of the nature of our Hamiltonian $\mathsf{H}$, which includes 
quantum corrections. The Wigner transform of $\mathsf{H}$ has the structure 
\begin{equation}
H_{\rm W}(x,p)= H_{\rm cl}(x,p)+ \sum_{n\ge 1} \hbar^{2n} H_{\rm W}^{(n)}(x,p), 
\end{equation}
where $H_{\rm cl}(x,p)$ is the classical Hamiltonian introduced in (\ref{MAclassical-hamil}). Proceeding in this way, we obtain a systematic $\hbar$ 
expansion of all the quantities of the theory. The WKB expansion of the grand potential reads, 
\begin{equation}
\label{MAj-wkb}
\CJ^{\rm WKB} (\mu, k)= \sum_{n \ge 0} \CJ_n(\mu) k^{2n-1}. 
\end{equation}
Note that this is principle an approximation to the full function $\CJ(\mu, k)$, since it does not take into account non-perturbative effects in $\hbar$. 
The functions $\CJ_n(\mu)$ in this expansion can be in principle computed in closed form, 
although their calculation becomes more and more cumbersome as $n$ grows. 
The leading term $n=0$ is however 
relatively easy to compute \cite{MAmp}. We first notice that the traces (\ref{MAtrzl}) have a simple semiclassical limit, 
\begin{equation}
Z_\ell \approx \int {\rd x \rd p \over 2 \pi \hbar} \MAre^{-\ell H_{\rm cl}(x,p)}, \qquad \hbar \rightarrow 0, 
\end{equation}
which is just the classical average, with an appropriate measure which includes the volume of the elementary quantum cell $2 \pi \hbar$. By using the integral 
\begin{equation}
\label{MAcoshint}
\int_{-\infty}^\infty\frac{\rd\xi}{\left(2\cosh{\xi \over 2} \right)^\ell}=\frac{\Gamma^2(\ell/2)}{\Gamma(\ell)},
\end{equation}
we find
\begin{equation}
 k Z_\ell \approx \frac{1}{2\pi}\frac{\Gamma^4(\ell/2)}{\Gamma^2(\ell)}, \qquad \hbar \rightarrow 0, 
\end{equation}
and
\begin{equation}
\label{MAexactJ0}
 \CJ_0(\mu)=-\sum_{\ell=1}^\infty\frac{(-\kappa)^{\ell}}{4\pi^2}\frac{\Gamma^4(\ell/2)}{\ell \Gamma^2(\ell)}.
\end{equation}
This expression is convenient when $\kappa$ is small, i.e. for $\mu \rightarrow -\infty$. 
To make contact with the large $N$ limit, we need to consider the limit of large, positive 
chemical potential, $\mu \rightarrow +\infty$. As we will see in the next section, this can be done by using a Mellin--Barnes integral, and one finds 
\begin{equation}
\label{MAj0mu}
 \CJ_0(\mu)=\frac{2 \mu ^3}{3 \pi ^2}+\frac{\mu }{3}+\frac{2 \zeta (3)}{\pi ^2}+J_0^{\rm M2}(\mu), 
\end{equation}
 where 
 \begin{equation}
 \label{MAjm2}
 J^{\rm M2}_0(\mu)=\sum_{\ell=1}^{\infty} \left(a_{0,\ell} \mu^2 + b_{0,\ell} \mu + c_{0,\ell} \right) \MAre^{-2 \ell \mu},
 \end{equation}
and $a_{0,\ell}$, $b_{0,\ell}$ and $c_{0,\ell}$ are computable coefficients. 
The leading, cubic term in $\mu$ in (\ref{MAj0mu}) is the one we found in (\ref{MAleading-J}). The subleading term in $\mu$ gives a correction of order $N^{1/2}$ to the leading 
behavior (\ref{MAlead-m}). The function $J_0^{\rm M2}(\mu)$ involves an infinite series of exponentially small corrections in $\mu$. Note that, 
although this result for $\CJ_0(\mu)$ is semiclassical in $\hbar$, it goes beyond the leading result at large $N$ in (\ref{MAlead-m}). This is because it 
takes into account the exact classical Fermi surface (\ref{MAclfermi}), rather than its polygonal approximation (\ref{MApolygon}). Therefore, we see that, already at this level, 
the Fermi gas approach makes it possible to go beyond the strict large $N$ limit.

Of particular interest are the exponentially small terms in $\mu$ in (\ref{MAjm2}). What is their meaning? By taking into account that, at large $N$, 
$\mu$ is given in (\ref{MAmu-saddle}), one finds that these corrections to $\CJ(\mu, k)$ lead to corrections in $Z(N,k)$ precisely of the form (\ref{MAg-membrane}). 
We recall that these were found originally in the matrix model as non-perturbative effects in the 't Hooft expansion. We conclude that the exponentially small 
corrections in $\mu$ in (\ref{MAjm2}), which in the Fermi gas approach appear already in the semi-classical approximation, 
correspond to non-perturbative corrections to the genus expansion, and should be identified as membrane instanton contributions.

It is possible to go beyond the leading order of the WKB expansion of the grand potential and compute the corrections appearing in (\ref{MAj-wkb}). 
The function $\CJ_1(\mu)$ was also derived in \cite{MAmp} and 
its large $\mu$ expansion has the following form, 
\begin{equation}
\label{MAj1}
\CJ_1(\mu)= {\mu \over 24} -{1\over 12} + \CO \left(\mu^2 \MAre^{-2 \mu}\right). 
\end{equation}
Moreover, the following non-renormalization theorem can be proved \cite{MAmp}: for $n\ge 2$, 
the $n$-th order correction to the WKB expansion is given by a $\mu$-independent constant $A_n$, and a function which is exponentially suppressed as $\mu \rightarrow \infty$, i.e. 
\begin{equation}
\CJ_n(\mu) = A_n+\CO \left(\mu^2 \MAre^{-2 \mu}\right), \qquad n \ge 2. 
\end{equation}
The exponentially small terms appearing in the functions $\CJ_n(\mu)$ with $n \ge 1$ have the same structure as for $n=0$. 
We then conclude that, in the WKB approximation, i.e. as a power series in $k$ around $k=0$, the grand potential has the structure
\begin{equation}
\label{MApert-gp}
\CJ^{\rm WKB} (\mu, k)= J^{(\rm p)}(\mu)+ J^{\rm M2}(\mu, k). 
\end{equation}
In this equation, the perturbative piece $J^{(\rm p)}(\mu)$ is given by 
\begin{equation}
\label{MAjpmu}
J^{(\rm p)}(\mu)= {C(k) \over 3} \mu^3 + B(k) \mu  + A(k), 
\end{equation}
where $B(k)$ and $C(k)$ were defined in (\ref{MAbk}), and (\ref{MAck}), respectively, and $A(k)$ is given by the formal power series expansion 
\begin{equation}
A(k)=\sum_{n\ge 0} A_n k^{2n-1}, 
\end{equation}
where 
\begin{equation}
A_0= {2 \zeta(3) \over \pi^2}, \qquad A_1=-{1\over 12}, 
\end{equation}
as one finds from (\ref{MAj0mu}) and (\ref{MAj1}). This series turns out to be the asymptotic expansion around $k=0$ of the 
function defined in (\ref{MAak}) (and this is why we have used the same notation for both). Therefore, the function $A(k)$ has two different asymptotic expansions: 
one of them gives the constants $A_n$ appearing in the WKB analysis of the grand potential, as we have just seen. The other one gives the genus $g$, constant map contributions 
to the free energy $c_g$ which appear in the 't Hooft expansion, as we saw in (\ref{MAlargeka}). The second term in the r.h.s. of 
(\ref{MApert-gp}) has the structure, 
\begin{equation}
\label{MAjm2k}
J^{\rm M2}(\mu)=\sum_{\ell=1}^{\infty} \left(a_{\ell}(k) \mu^2 + b_{\ell}(k) \mu + c_{\ell}(k) \right) \MAre^{-2 \ell \mu}, 
\end{equation}
where the coefficients have the WKB expansion, 
\begin{equation}
a_\ell (k)= \sum_{n \ge 0} a_{n, \ell} k^{2n-1}.  
\end{equation}
Similar expansions hold for $b_\ell(k)$, $c_\ell (k)$.

\begin{figure}
\center
\includegraphics[height=6cm]{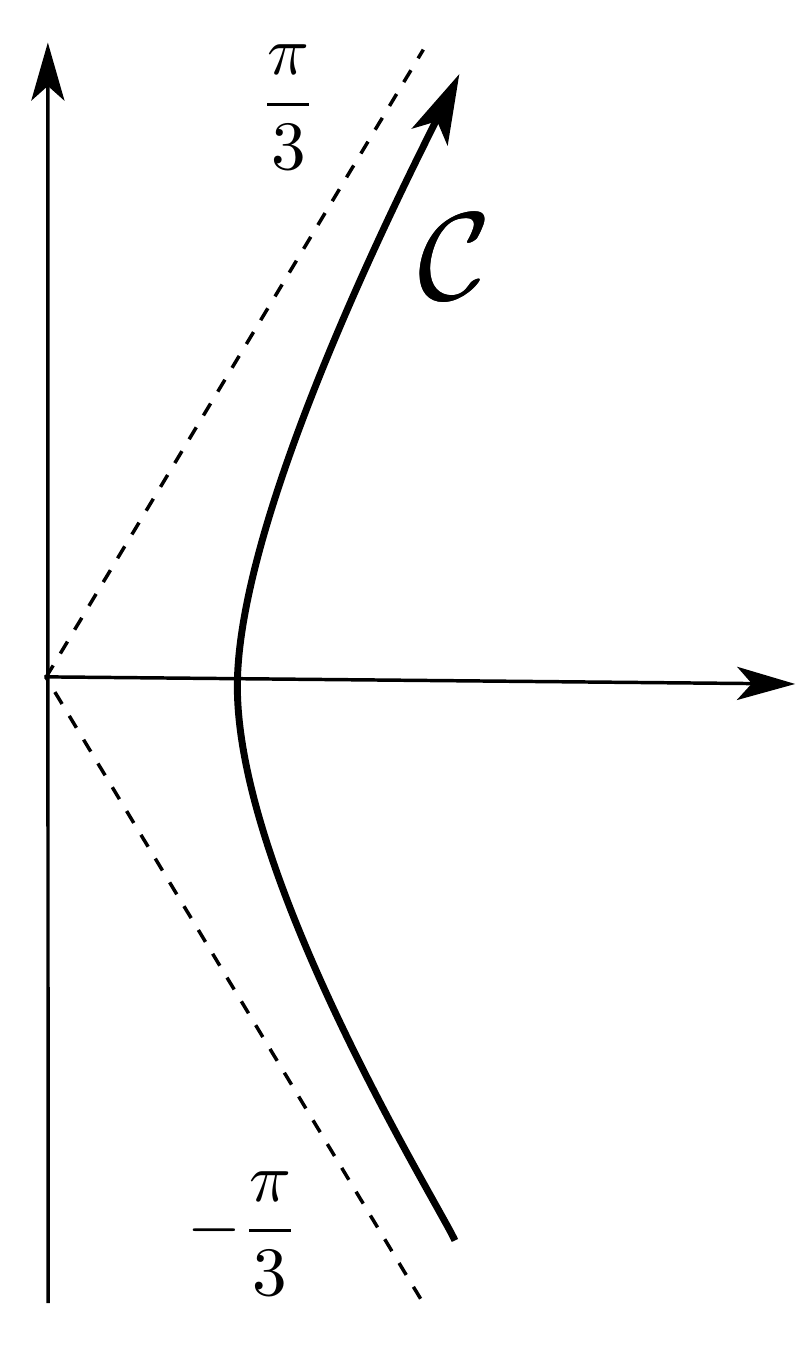}  
\caption{The contour $\CC$ in the complex plane of the chemical potential.}
\label{MAairy-c}
\end{figure}
We can now plug the result (\ref{MApert-gp}) in (\ref{MAexactinverse}). In the $\mu$-plane, this is an integral from $-\pi \ri$ to $\pi \ri$:
\begin{equation}
\label{MAfinite-n}
Z(N,k)={1\over 2 \pi \ri} \int_{-\pi \ri}^{\pi \ri}{\rd \mu \over 2 \pi \ri} \MAre^{\CJ(\mu,k) - N \mu}. 
\end{equation}
If we neglect exponentially small corrections in $N$, 
we can deform the contour $[-\pi \ri, \pi \ri]$ to the contour $\CC$ shown in Fig. \ref{MAairy-c}. Therefore, we find that, up to these corrections, 
\begin{equation}
\label{MAairy}
Z(N,k) \approx {1 \over 2 \pi \ri} \int_\CC \exp \left(J^{(\rm p)}(\mu)  - \mu N\right) \rd \mu. 
\end{equation}
The above integral is given by an Airy function, and we immediately recover the result (\ref{MAzpnk}) for the perturbative $1/N$ expansion 
at fixed $k$. As explained before, 
this includes all the $1/N$ corrections to the partition function in a single strike. 

We see that the Fermi gas approach leads to a powerful derivation of the Airy function behavior of the partition function. In this approach, such a derivation 
just requires computing the grand potential at next-to-leading order 
in the WKB expansion. Although this is a one-loop result, it is exact in $k$ if we neglect exponentially small corrections in $\mu$. 
Therefore, a one-loop calculation in the grand-canonical ensemble leads to an all-orders result in the canonical ensemble.

\subsubsection{From the WKB expansion to the refined topological string}

In order to complete our understanding of the WKB expansion of the Fermi gas, we should determine the coefficients appearing in 
the expansion (\ref{MAjm2k}). We can in principle compute them order by order in powers of $k$ by using standard semiclassical techniques, 
but it would be much better to know the full expansion explicitly. As first noted in \cite{MAhmmo}, it turns out that there is an elegant and powerful answer for the 
all-orders WKB grand potential, which involves the {\it refined} topological string of local $\IP^1 \times \IP^1$ in the so-called Nekrasov--Shatashvili (NS) limit \cite{MAns}. 
A good starting point for understanding this connection is to 
formulate the problem of computing the semiclassical limit $\CJ_0(\mu)$ in a way which makes contact with the theory of periods of CY manifolds. 
\begin{figure}
\center
\includegraphics[height=5cm]{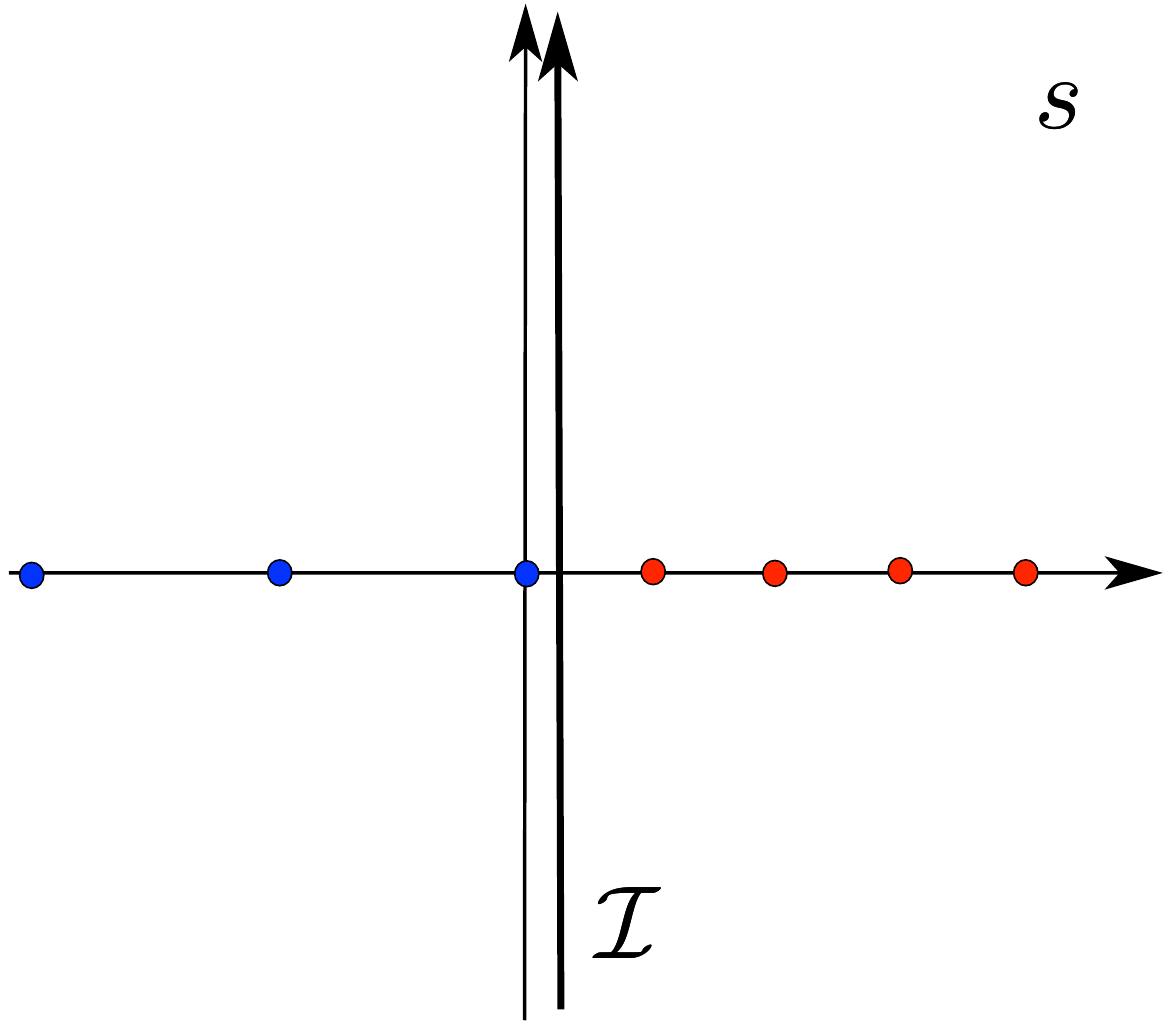}  
\caption{The contour $\CI$ in the complex $s$ plane. By closing the contour to the right, we encircle the poles at $s=n$, $n\in \IZ_{>0}$. By closing the contour 
to the left, we encircle the poles at $s=-2n$, $n\in \IZ_{\ge 0}$.}
\label{MAmb-c}
\end{figure}

Let us consider again the expression (\ref{MAexactJ0}) for the semiclassical grand potential, and let us write this infinite sum as 
a Mellin--Barnes integral, 
\begin{equation}
\CJ_0(\kappa)=-{1\over 4 \pi^2} \int_{\CI} {\rd s \over 2 \pi \ri} {\Gamma(-s) \Gamma(s/2)^4 \over \Gamma(s)} \kappa^s, 
\end{equation}
where the contour $\CI$ runs parallel to the imaginary axis, see Fig. \ref{MAmb-c}\footnote{The Mellin--Barnes technique to study the grand potential 
was independently developed in \cite{MAhatsuda-spectral}.}. It can be deformed so that the integral 
encloses the poles of $\Gamma(-s)$ at $s=n$, $n=1, \cdots$ (in the clockwise direction). The residues at these poles give back the infinite series in (\ref{MAexactJ0}). 
We can however deform the contour in the opposite direction, so that it encloses the poles at
\begin{equation}
s=-2 m, \qquad m=0, 1, 2, \cdots, 
\end{equation}
and we find
\begin{equation}
\CJ_0(\kappa)=-{1\over 4 \pi^2} \sum_{n=0}^\infty {\rm Res}_{s=-2n}\left[ {\Gamma(-s) \Gamma(s/2)^4 \over \Gamma(s)} \kappa^s \right]. 
\end{equation}
The pole at $s=0$ gives
\begin{equation}
{2\over 3 \pi^2} \mu^3+{1 \over 3} \mu + {2 \zeta(3) \over  \pi^2}, 
\end{equation}
which is precisely the leading part of (\ref{MAj0mu}) as $\mu \rightarrow \infty$. 
To understand the contribution of the rest of the poles, let us consider the following differential operator, 
\begin{equation}
\label{MApfop}
 \CL=\theta^3 - 4z \theta (2 \theta+1)^2, 
\end{equation}
where 
\begin{equation}
\theta= z {\rd \over \rd z}. 
\end{equation}
A basis of solutions to the equation 
\begin{equation}
\label{MAPFeq}
\CL \Pi=0
\end{equation}
can be obtained by using the Frobenius method. One first considers the so-called fundamental period or solution, 
\begin{equation}
\varpi_0(z, \rho)= \sum_{n\ge 0} a_n(\rho) z^{n+\rho},
\end{equation}
where
\begin{equation}
\label{MAanrho}
a_n(\rho)=16^n { \Gamma^2\left( n+ \rho+{1\over 2} \right) \Gamma(n+\rho) \over \Gamma^3( n+\rho+1)} {\Gamma^3(\rho+1) \over \Gamma^2(\rho+{1\over 2}) \Gamma(\rho)}.
\end{equation}
The Frobenius method instruct us to look at the functions, 
\begin{equation}
\varpi_k(z)= { \rd^k  \varpi_0 (z, \rho)  \over \rd \rho^k} \biggl|_{\rho=0}. 
\end{equation}
For $k=1,2,3$, they have the following structure, 
\begin{equation}
\begin{aligned}
\varpi_1(z)&= \log z + \widetilde \varpi_1(z), \\
\varpi_2(z)&= \left( \log z  \right)^2+ 2 \log z \widetilde \varpi_1(z) + \widetilde \varpi_2(z),\\
\varpi_3(z)&= \left( \log z  \right)^3+  3 \left( \log z  \right)^2 \widetilde \varpi_1(z) +3 \log z \,  \widetilde \varpi_2(z) +  \widetilde \varpi_3(z), 
\end{aligned}
\end{equation}
where the $ \widetilde \varpi_k(z)$ are power series in $z$, 
\begin{equation}
 \widetilde \varpi_k(z)=\sum_{n=1}^\infty  \left({ \rd^k  a_n (\rho)  \over \rd \rho^k} \right)_{\rho=0} z^n.
 \end{equation}
 We have, for example, 
\begin{equation}
\label{MAom-pers}
\begin{aligned}
 \widetilde \varpi_1(z)&=\sum_{n=1}^\infty {1\over  n} \left( {\Gamma \left( n+{1\over 2} \right) \over 
\Gamma({1\over 2}) n!} \right)^2 (16 z)^n, \\
 \widetilde \varpi_2(z)&=\sum_{n=1}^\infty {4 \over  n} \left( {\Gamma \left( n+{1\over 2} \right) \over 
\Gamma({1\over 2}) n!} \right)^2 \left[ \psi\left(n+{1\over 2} \right) -\psi(n+1)+ 2 \log 2-{1\over 2n} \right] (16 z)^n. 
\end{aligned}
\end{equation}
When $k=1,2$, the functions $\varpi_k(z)$ give solutions to the equation (\ref{MAPFeq}). 
After some simple manipulations, it is easy to see that the contribution to $\CJ_0(\kappa)$ of the residue at $s=-2n$, $n\not=0$, is given by 
\begin{equation}
-{1\over 8 \pi^2} {\rm Res}_{\epsilon=0} {  \Gamma \left(1-\frac{\epsilon }{2}\right)^4 \Gamma \left(\frac{\epsilon }{2}\right)^4  \over  \Gamma (1-\epsilon ) \Gamma (\epsilon )}{\Gamma
   \left(n-\frac{\epsilon }{2}\right) \Gamma \left(n-\frac{\epsilon }{2}+\frac{1}{2}\right)^2  \over \Gamma
   \left(n-\frac{\epsilon }{2}+1\right)^3}(4/\kappa)^{2 n-\epsilon }.
   \end{equation}
 By setting $\rho=-\epsilon/2$ and comparing to (\ref{MAanrho}), we find
\begin{equation}
\label{MAj0-omegas}
\CJ_0(\mu)= -{1\over 12 \pi^2} \varpi_3(z) - {1 \over 6} \varpi_1(z)+ {2 \zeta(3) \over  \pi^2}, 
\end{equation}
where
\begin{equation}
z=\MAre^{-2 \mu}. 
\end{equation}
We also have
 \begin{equation}\label{MAlist-rels}
 \begin{aligned}
 \sum_{\ell=1}^{\infty} a_{0,\ell}z^\ell& =-{1\over \pi^2} \widetilde \varpi_1(z), \\
  \sum_{\ell=1}^{\infty} b_{0,\ell}z^\ell& ={1\over 2 \pi^2} \widetilde \varpi_2(z), \\
   \sum_{\ell=1}^{\infty} c_{0,\ell}z^\ell& =-{1\over 12 \pi^2} \widetilde \varpi_3(z) -{1\over 6}\widetilde \varpi_1(z). 
   \end{aligned}
   \end{equation}

The structure above indicates a connection to topological string theory. The differential operator (\ref{MApfop}) 
is the Picard--Fuchs operator describing the genus zero topological string 
on local $\IP^1 \times \IP^1$ (see for example \cite{MAkz}). In this context, it is useful to define a so-called flat coordinate $t$ 
and a genus zero free energy $F_0(t)$ by the equations, 
\begin{equation}
\begin{aligned}
t&=-\varpi_1(z), \\
{\partial F_0 \over \partial t}&={1\over 2} \varpi_2(z) -{\pi^2\over 3}, 
\end{aligned}
\end{equation}
so that
\begin{equation}
F_0(t)={t^3 \over 6}- {\pi^2 t\over 3}-2 \zeta(3) - 4 \MAre^{-t} -{9 \over 2} \MAre^{-2t}-\cdots. 
\end{equation}
In terms of these quantities, one finds
\begin{equation}
\label{MAj0-f0}
\CJ_0(\mu)={1\over 2\pi^2} \left( t {\partial F_0 \over \partial t} - 2 F_0\right). 
\end{equation}

We would like to understand now the higher order WKB corrections to the grand potential in the context of 
topological string theory, in line with what we have done for the leading, semiclassical function $\CJ_0(\mu)$. To do this, 
and following \cite{MAkm}, we will look at the WKB expansion of the 
energy levels, i.e. we will  consider the spectral problem defined by (\ref{MAspec}), (\ref{MAspec-2}). 
The first step is to reformulate (\ref{MAspec}) as a spectral problem for a difference equation. Let us define 
\begin{equation}
|\psi \rangle =\MAre^{{1\over 2} U(\mathsf{x})} |\phi \rangle.
\end{equation}
It follows from (\ref{MArhopq}) that (\ref{MAspec}) can be written as (we remove the indices for the discrete energies)
\begin{equation}
\MAre^{ U(\mathsf{x})} \MAre^{T(\mathsf{p})}|\psi\rangle = \MAre^{E} |\psi\rangle,
\end{equation}
or, equivalently, in the coordinate representation,
\begin{equation}
\label{MAdiff-eq}
\psi\left( x +\ri \pi k\right) +  \psi \left( x-\ri \pi k\right) = {\MAre^{E} \over 2 \cosh \left( {x\over 2} \right)} \psi(x). 
\end{equation}
This difference equation is equivalent to the original problem (\ref{MAspec}) provided some analyticity and boundary conditions are imposed on the function $\psi(x)$. 
Let us denote by $\CS_a$ the strip in the complex $x$-plane defined by 
\begin{equation}
\left|{\rm Im}(x)\right|<a. 
\end{equation}
Let us also denote by $A\left(\CS_{a} \right)$ those functions $g$ which are bounded and analytic in the strip, continuous on its closure, and for which 
$g(x+ \ri y) \to 0$ as $x\rightarrow \pm \infty$ through real values, when $y \in \IR$ is fixed and satisfies $|y|<a$. 
It can be seen, by using for example the results in \cite{MAtw}, that the equivalence of (\ref{MAdiff-eq}) and (\ref{MAspec}) requires that $\psi(x)$ belongs to the space 
$A\left(\CS_{\pi k} \right)$. 

The difference equation (\ref{MAdiff-eq}) can be solved in the WKB approximation, just as the Schr\"odinger equation (see for example \cite{MAdingle}). One introduces a WKB ansatz, 
\begin{equation}
\label{MAWKBpsi}
\psi(x, \hbar) = \exp \left( {1\over \hbar} S(x, \hbar)\right), 
\end{equation}
where
\begin{equation}
\label{MAWKBS}
S(x,\hbar)= \sum_{n\ge 0} S_n(x) \hbar^{n}, 
\end{equation}
and we remember that $\hbar$ is given by (\ref{MAhache}). The leading order approximation gives
\begin{equation}
S_0'(x)= p(x). 
\end{equation}
This defines a curve in phase space 
\begin{equation}
\label{MAclass-curve}
y=p(x), 
\end{equation}
as well as a differential on that curve
\begin{equation}
\lambda(x)= p(x) \rd x. 
\end{equation}
In the case of the difference equation (\ref{MAdiff-eq}), the curve (\ref{MAclass-curve}) is nothing but the equation for the Fermi surface (\ref{MAclfermi}). Geometrically, this is a curve of genus one, 
with two one-cycles $A$ and $B$. The $B$ period of the differential $\lambda$ gives the classical volume of phase space,  
\begin{equation}
{\rm vol}_0(E)= \oint_B \lambda, 
\end{equation}
and the Bohr--Sommerfeld quantization condition says that this volume is quantized as
\begin{equation}
\label{MAbs}
{\rm vol}_0 (E) = 2 \pi \hbar \left( n+{1\over 2}\right), \qquad n=0, 1,2, \cdots. 
\end{equation}
The quantum corrections in (\ref{MAWKBS}) can be also interpreted geometrically: we introduce a  ``quantum" differential 
\begin{equation}
\lambda(x;\hbar)=\partial_x S(x, \hbar) \rd x, 
\end{equation}
and the perturbative, ``quantum" volume of phase space is defined as  
\begin{equation}
\label{MAp-vol}
{\rm vol}_{\rm p} (E; \hbar)= \oint_B \lambda(x;\hbar)=\sum_{n\geq0} {\rm vol}_n(E)\hbar^{2n}. 
\end{equation}
As it is well-known since the work of Dunham \cite{MAdunham}, this leads to a quantum-corrected quantization condition of the form 
\begin{equation}
\label{MAvolbs}
{\rm vol}_{\rm p} (E) = 2 \pi \hbar \left( n+{1\over 2}\right), \qquad n=0, 1,2, \cdots. 
\end{equation}
It is straightforward to do an analysis of this problem order by order in $\hbar$, but exact as a function of $E$. The classical volume is given essentially 
by a Meijer function \cite{MAmp}, 
\begin{equation}
 \label{MAvolexact}
{\rm vol}_0(E)=\frac{\MAre^E}{\pi}  G_{3,3}^{2,3}\left(\frac{\MAre^{2E}}{16}\left|
\begin{array}{c}
 \frac{1}{2},\frac{1}{2},\frac{1}{2} \\
 0,0,-\frac{1}{2}
\end{array}
\right.\right)-4\pi^2, 
\end{equation}
which has the following behavior at large $E$, 
\begin{equation}
{\rm vol}_0(E) = 8 E^2 -{4  \pi^2 \over 3} + \CO\left( E\, \MAre^{-2E} \right).  
\end{equation}
The leading term is nothing but the area of the polygon (\ref{MApolygon}). The corrections 
incorporate the difference between the volume, as computed by the exact Fermi surface, and the volume as computed in the polygonal approximation. One can also 
find \cite{MAkm}, 
\begin{equation}
\label{MAvol01}
{\rm vol}_1(E)=\frac{\MAre^{-E} \left((32 \, \MAre^{-2E}-1) E(k_E)-K(k_E)\right)}{6 \left(16 \,  \MAre^{-2E}-1 \right)},
\end{equation}
where $K(k_E)$, $E(k_E)$ are elliptic integrals of the first and second kind, respectively, with modulus 
\begin{equation}
k_E^2=1-\frac{\MAre^{2E}}{16 }~.
\end{equation}
As we will explain in a moment, this calculation is the counterpart of the perturbative calculation of $\CJ_n(\mu)$ 
that we considered in the previous section. What we really need, in order to understand the 
theory in the M-theory regime, is an approach which is exact in $k$ (i.e. in $\hbar$) but leads to 
an expansion at large $E$, since this corresponds to large $N$. In the case of the WKB problem 
we are analyzing here, we need to resum the WKB expansion of the perturbative volume at all orders in $\hbar$, 
but order by order in $\MAre^{-2 E}$. To do this, we will relate the spectral problem (\ref{MAspec}) to another 
one, which makes contact with the refined topological string \cite{MAkm}. Let us consider
\begin{equation}
\label{MAqsc}
\left( \MAre^{\mathsf{u}} + z_1 \MAre^{-\mathsf{u}}+ \MAre^{\mathsf{v}} + z_2 \MAre^{-\mathsf{v}}- 1\right) |\psi \rangle=0,
\end{equation}
where $\mathsf{u}$, $\mathsf{v}$ are operators satisfying the commutation relation, 
\begin{equation}
[\mathsf{v}, \mathsf{u}]= {\ri \hbar \over 2}, 
\end{equation}
and $z_1$, $z_2$ are complex parameters. If we do the following change of variables, 
\begin{equation}
\label{MAch-v}
\mathsf{v}= {\mathsf{x}+\mathsf{p} \over 2}+{\ri \pi k \over 4}  -E, \qquad \mathsf{v}={\mathsf{x}-\mathsf{p} \over 2}-{\ri \pi k \over 4} -E, 
\end{equation}
and consider the specialization
\begin{equation}
\label{MAzz-q}
z_1=q^{1/2}z \qquad z_2=q^{-1/2}z, 
\end{equation}
where
\begin{equation}
z=\MAre^{-2E}
\end{equation}
and
\begin{equation}
\label{MAqh}
q=\MAre^{\ri \hbar \over 2}= \MAre^{\pi \ri k}, 
\end{equation}
the equation (\ref{MAqsc}) becomes the difference equation (\ref{MAdiff-eq}). Note that the change of variables (\ref{MAch-v}) is a canonical transformation, i.e. its classical version 
preserves the volume element of phase space, up to an overall factor. It turns out that the difference equation (\ref{MAqsc}) appears in the context of refined topological 
string theory \cite{MAacdkv} (see also \cite{MAmirmor,MAnps}). It implements the ``quantization" of the mirror curve of local $\IP^1 \times \IP^1$, 
and it leads to the NS limit of the refined topological string. 
In particular, the periods of the exact quantum differential (or {\it quantum periods}) can be calculated as an expansion at small $z_1$, $z_2$ 
but exactly in $k$ \cite{MAacdkv, MAhmmo}. Note that, in this context, the variables $z_1$, $z_2$ 
are interpreted as complex deformation parameters of the mirror curve. In the 
case at hand, we have two quantum $A$-periods and two quantum $B$-periods, denoted by 
$\Pi_{A_I}(z_1, z_2; \hbar)$, $\Pi_{B_I}(z_1, z_2; \hbar)$, $I=1,2$. The $A$-periods have the structure
\begin{equation}
\Pi_{A_I}(z_1,z_2;\hbar)=\log z_I +\widetilde{\Pi}_{A}(z_1,z_2;\hbar), \quad I=1,2. 
\end{equation}
The two quantum $B$-periods are related by the exchange of the moduli, 
\begin{equation}
\Pi_{B_2}(z_1,z_2;\hbar)=\Pi_{B_1}(z_2,z_1;\hbar), 
\end{equation}
and they have the structure
\begin{equation}
\begin{aligned}
\label{MAq-B}
\Pi_{B_1}(z_1, z_2;\hbar)&=-{1\over 8}\left( \log^2 z_1  -2 \log z_1\log z_2 -\log^2 z_2 \right) +{1\over 2} \log z_2\, \widetilde \Pi_A (z_1, z_2;\hbar)\\
&+ {1\over 4} \widetilde \Pi_B (z_1, z_2;\hbar).
\end{aligned}
\end{equation}
The expansion of the periods around $z_1=z_2=0$ is given, to the very first orders, by, 
\begin{equation}
\label{MAq-ABper}
\begin{aligned}
\widetilde{\Pi}_{A}(z_1,z_2;\hbar)&=2(z_1+z_2)+3(z_1^2+z_2^2)+2(4+q+q^{-1})z_1z_2+\frac{20}{3}(z_1^3+z_2^3) \\
&\quad +2(16+6q+6q^{-1}+q^2+q^{-2})z_1 z_2(z_1+z_2)+{\cal O}(z_i^4), \\
\widetilde{\Pi}_{B}(z_1, z_2;\hbar)&=8 \left[\frac{q+1}{2(q-1)} \log q \right] z_1+4 \left[ 1+\frac{5 q^2 + 8 q + 5}{2(q^2-1)}\log q \right] z_1^2 
\\ 
& +8 \left[ 1+\frac{(1+q)^3}{2 q(q-1)}\log q \right]z_1z_2
+4 z_2^2+{\cal O}(z_i^3),
\end{aligned}
\end{equation}
where $q$ is given in (\ref{MAqh}). In general, on a local CY manifold with 
$n$ moduli, the quantum $A$ periods define a  ``quantum" mirror map \cite{MAacdkv}, relating the flat coordinates $t_I$ to the complex moduli $z_I$, $I=1, \cdots, n$, 
while the quantum $B$ periods define the NS free energy, $F^{\rm NS}$, 
\begin{equation}
\label{MAq-pers}
-t_I(\hbar)= \Pi_{A_I}(z_I;\hbar), \qquad {\partial F^{\rm NS}\over \partial t_I}={1\over \hbar} \Pi_{B_I}(z_I;\hbar), \qquad I=1, \cdots, n. 
\end{equation}
The NS free energy has an asymptotic expansion around $\hbar=0$, of the form
\begin{equation}
\label{MANSh}
F^{\rm NS}(t_I;\hbar)= \sum_{n\ge 0} F^{\rm NS}_n (t_I) \hbar^{2n-1}, 
\end{equation}
and the leading order term 
\begin{equation}
F^{\rm NS}_0(t_I)= F_0(t_I)
\end{equation}
is the standard prepotential of the local CY manifold. The knowledge of the quantum mirror map and of the NS free energy, 
as a function of the flat coordinates $t_I$, is equivalent to knowing both periods.

Let us now come back to the problem of calculating (\ref{MAp-vol}). This is a quantum period for the spectral curve defined by (\ref{MAdiff-eq}), 
but this curve is just a specialization of (\ref{MAqsc}) with the dictionary 
(\ref{MAzz-q}) and after a canonical transformation. Therefore, (\ref{MAp-vol}) should be a combination 
of the quantum periods of local $\IP^1 \times \IP^1$, specialized to the ``slice" (\ref{MAzz-q}). Let us denote
\begin{equation}\label{MAq-ABperrelab}
\begin{aligned}
\widetilde \Pi_{A}(z; 
\hbar)& \equiv \widetilde \Pi_A(q^{1/2}z,q^{-1/2}z; \hbar)=\sum_{\ell \ge 1}  \widehat a_\ell (\hbar) z^\ell, \\ 
\widetilde \Pi_{B}(z;\hbar)& \equiv {1\over 2} \left( \widetilde \Pi_B (q^{1/2}z,q^{-1/2}z; \hbar)+ \widetilde \Pi_B (q^{-1/2}z,q^{1/2}z; \hbar)\right) =\sum_{\ell \ge 1} \widehat b_\ell (\hbar)  z^\ell,
\end{aligned}
\end{equation}
where $\widehat a_\ell (\hbar)$, $\widehat b_\ell (\hbar)$ have the $\hbar$-expansion, 
\begin{equation}
\widehat a_\ell (\hbar)= \sum_{n=0}^\infty \widehat a_\ell ^{(n)} \hbar^{2n}, \qquad \widehat b_\ell (\hbar)= \sum_{n=0}^\infty \widehat b_\ell ^{(n)} \hbar^{2n}.
\end{equation}
Note that the classical limits of $\widetilde \Pi_{A, B}(z; \hbar)$ are the power series $\widetilde \varpi_{1,2}(z)$ written down in (\ref{MAom-pers}).  
Requiring the combination of quantum periods to have the correct classical limit, and that only even powers of $\hbar$ appear, we find, 
\begin{equation}
\begin{aligned}
\label{MApvol}
{\rm vol}_{\rm p} (E; \hbar)&= 4 \Pi_{B_1} ( q^{1/2} z, q^{-1/2} z; \hbar) + 4 \Pi_{B_2} ( q^{1/2} z, q^{-1/2} z; \hbar)  -{4 \pi^2 \over 3} -{\hbar^2 \over 12}\\
&= 8 E^2 -{4 \pi^2 \over 3} +{\hbar^2 \over 24} - 8 E \sum_{\ell\ge 1}  \widehat a_\ell (\hbar) \MAre^{-2 \ell E} + 2 \sum_{\ell\ge1} \widehat b_{\ell } (\hbar) \MAre^{-2 \ell E}. 
\end{aligned}
\end{equation}
This is the resummation we were looking for, and solves the problem of computing the perturbative, quantum volume of phase space by re-expressing it 
in terms of quantities associated to the quantum mirror symmetry of local $\IP^1 \times \IP^1$. 

Let us make contact with the grand potential. Of course, this is not independent of the quantum volume of phase space, since in an ideal gas the spectrum of the Hamiltonian 
determines the thermodynamics. From the expression (\ref{MAfred}) as a Fredholm determinant, we have
\begin{equation}
\CJ(\mu, k)= \sum_{n \ge 0} \log \left(1 + \MAre^{\mu-E_n} \right).  
\end{equation}
For the moment being we will restrict ourselves to the perturbative regime, but at all orders in the WKB expansion (as we will see,
there are important non-perturbative corrections in $\hbar$). The perturbative energies will be denoted by $E_n^{\rm p}$, 
and they are determined by the WKB quantization condition (\ref{MAvolbs}), which defines in fact an implicit function $E^{\rm p}(n)$ for arbitrary 
values of $n$. In order to perform the sum over discrete energy levels, we will use 
the Euler--Maclaurin formula, which reads
\begin{equation}
\label{MAemc}
\sum_{n\geq 0}f(n)=\int_{0}^\infty f(n) \rd n+\frac{1}{2}\left(f(0)+f(\infty)\right)+\sum_{r\geq 1} \frac{B_{2r}}{(2r)!}\left(f^{(2r-1)}(\infty)-f^{(2r-1)}(0)\right). 
\end{equation}
Notice that, in general, this formula gives an asymptotic expansion for the sum. In our case, the function $f(n)$ is given by 
\begin{equation}
f(n)= \log{\left(1+ \MAre^{\mu-E^{\rm p}(n)}\right)}. 
\end{equation}
Since $E^{\rm p}(n)\rightarrow \infty$ as $n\rightarrow\infty$, we have $f(\infty)=0$, $f^{(2r-1)}(\infty)=0$ for all $r \ge 1$. The first terms of (\ref{MAemc}) give, 
\begin{equation}
\label{MAfirst-qv}
\int_{E^{\rm p}_0}^\infty \frac{\rd n(E)}{\rd E}\log{\left(1+\MAre^{\mu-E}\right)}\rd E+\frac{1}{2}f(0)=\frac{1}{2\pi\hbar}\int_{E^{\rm p}_0}^\infty \frac{ {\rm vol}_{\rm p}(E)}{1+\MAre^{E-\mu}}\rd E. 
\end{equation}
In deriving this equation, we first changed variables from $n$ to $E$, by using 
\begin{equation}
n(E)=\frac{{ \rm vol}_{\rm p} (E)}{2\pi\hbar}-\frac{1}{2},
\end{equation}
we integrated by parts, and we took into account that 
\begin{equation}
\frac{{\rm vol}_{\rm p} (E^{\rm p}_0)}{2\pi\hbar}=\frac{1}{2}, 
\end{equation}
as well as the asymptotic behavior 
\begin{equation}
{\rm vol}(E)\approx E^2, \qquad E\rightarrow \infty.
\end{equation}
We conclude that the WKB expansion of the grand potential is related to the perturbative quantum volume as 
\begin{equation}
\label{MAj-int}
\CJ^{\rm WKB}(\mu, k)={1\over 2 \pi \hbar}  \int_{E^{\rm p}_0}^\infty {{\rm vol}_{\rm p} (E) \rd E \over \MAre^{E-\mu}+1}-\sum_{r\geq 1} \frac{B_{2r}}{(2r)!}f^{(2r-1)}(0).
\end{equation}
A clarification is needed concerning the above derivation. The first term in the l.h.s. of (\ref{MAfirst-qv}) looks identical to the standard formula 
(\ref{MAst-form}) which one finds in textbooks. However, (\ref{MAfirst-qv}) 
has an infinite number of corrections due to the Euler--Maclaurin formula. How is this compatible with (\ref{MAst-form})? 
The answer is that in (\ref{MAst-form}) the function 
$\rho(E)$ is not really smooth, but rather a sum of delta functions. In contrast, in (\ref{MAfirst-qv}) we use a smooth function, 
the perturbative quantum volume. The price to pay for using a smooth 
function, in a situation in which one has a discrete set eigenvalues, is precisely including the corrections to 
the Euler--Maclaurin formula (see for example \cite{MAbb-book} for a discussion on the discrete versus the 
smooth density of eigenvalues). 

We can now plug the expansion (\ref{MApvol}) in the r.h.s. of (\ref{MAj-int}) and integrate term by term. The resulting integrals can be done by using the Mellin transform, see \cite{MAkm} for the 
details, and one finds in the end
\begin{equation}
\label{MAwkb-j}
\begin{aligned}
\CJ^{\rm WKB}(\mu, k)&= {2 \mu^3 \over 3 k \pi^2} + \left( {1\over 3k} + {k \over 24}\right) \mu+ \widehat A(\hbar) \\
&\, \, + \sum_{\ell \ge 1} \left( -{\widehat a_\ell (\hbar) \over \pi^2 k} \mu^2  + {\widehat b_\ell (\hbar) \over 2 \pi^2 k} \mu  + { \widehat c_\ell (\hbar) 
\over 2 \pi^2 k}  \right)\MAre^{-2 \ell  \mu}- \sum_{\ell \ge 0}  {  \widehat{d}_\ell (\hbar) 
\over 2 \pi^2 k} \MAre^{-(2 \ell+1) \mu}. 
\end{aligned}
\end{equation}
Here, $A(\hbar)$, $\widehat c_\ell (\hbar)$ and $\widehat{d}_\ell (\hbar)$ have complicated expressions which can be found in \cite{MAkm}. They 
involve $E^{\rm p}(n)$ and its derivatives, evaluated at $n=0$, as well as the coefficients $\widehat a_\ell (\hbar)$, $\widehat b_\ell (\hbar)$. 
By comparing (\ref{MAwkb-j}) with (\ref{MApert-gp}), we find that the coefficients 
appearing in (\ref{MAjm2k}) are related to the coefficients of the quantum periods introduced before by
\begin{equation}
\label{MAabrels}
a_\ell (k)=-{ \widehat a_\ell (\hbar)\over \pi^2 k}~, \qquad b_\ell(k)= {\widehat b_\ell (\hbar)\over 2 \pi^2 k}. 
 \end{equation}
 Note that, in the classical limit $k \rightarrow 0$, we recover the first two equations in (\ref{MAlist-rels}). 
This relationship between the WKB expansion of the grand potential and the quantum periods of local $\IP^1 \times \IP^1$ 
was first conjectured in \cite{MAhmmo}, and it can be checked against explicit calculations of both sets of coefficients. The above argument explains 
why this relation holds: it is due to the 
fact that the WKB solution of the spectral problem of the ABJM Fermi gas can be mapped to the problem of computing these quantum periods. 
This method also provides explicit, but complicated expressions for the remaining coefficients, $\widehat A(\hbar)$, $\widehat c_\ell (\hbar)$ and $ \widehat{d}_\ell (\hbar)$. 
In order to have a result consistent with the perturbative results for $\CJ_n(\mu)$, one should have $ \widehat{d}_\ell (\hbar)=0$ for all $\ell \ge 0$. This can be verified 
in the very first orders of a power series expansion around $k=0$ \cite{MAkm}. 

We can also give a more conceptual understanding of the 
quantum corrections to the grand potential. We have shown that the coefficients $a_\ell (k)$ and $b_\ell(k)$ in 
$\CJ^{\rm WKB}(\mu, k)$ can be obtained by promoting the 
periods $\varpi_{1,2}(z)$ (which encode their classical limit) to quantum periods. However, $\CJ_0(\mu)$ is itself a classical period, 
as we showed in (\ref{MAj0-omegas}) and (\ref{MAj0-f0}). Therefore, the full 
function $\CJ^{\rm WKB}(\mu, k)$ should be obtained by promoting this period to its quantum counterpart. 
For example, one can write down a quantum version of (\ref{MAj0-f0}) by using the 
NS free energy defined in (\ref{MAq-pers}). To do this, we focus on the period appearing in (\ref{MAj0-f0}), which in a general CY with $n$ moduli is given by
\begin{equation}
\varpi_3= 2 F_0 - \sum_{i=1}^n t_i {\partial F_0 \over \partial t_i}. 
\end{equation}
 This can be written in a more natural way by introducing the so-called 
homogeneous coordinates 
\begin{equation}
t_i = {X_i \over X_0}, \qquad i=1, \cdots, n. 
\end{equation}
Here, $X_0$ plays the r\^ole of the inverse string coupling constant, which in our case is 
$1/\hbar$. Let us define the homogeneous prepotential $\CF_0(X_0, X_i)$ as 
\begin{equation}
\CF_0 (X_0, X_i)=X_0^2 F_0 \left( {X_i \over X_0} \right). 
\end{equation}
Then, one has (see for example \cite{MAhkty})
\begin{equation}
{1\over X_0} {\partial \CF_0 \over \partial X_0}= 2 F_0 - \sum_{i=1}^n t_i {\partial F_0 \over \partial t_i}. 
\end{equation}
If follows from (\ref{MAj0-f0}) that
\begin{equation}
\label{MAlead-NS}
{1\over k} \CJ_0(\mu)= -{1\over \pi} {\partial \CF_0 \over \partial X_0}. 
\end{equation}
The natural generalization of the homogeneous prepotential, including 
all the corrections in $\hbar$, involves the NS free energy introduced in (\ref{MAq-pers}). 
\begin{equation}
\CF(X_0, X_i)= X_0 F^{\rm NS}\left( {X_i \over X_0}; {1\over X_0} \right)= X_0^2 F_0\left( {X_i \over X_0} \right) +\cdots
\end{equation}
Therefore, the generalization of (\ref{MAlead-NS}) to an all-orders formula is 
\begin{equation}
\label{MAcomplete-jwkb}
\CJ^{\rm WKB}(\mu, k)=-{1\over \pi} {\partial \over \partial X_0} \left[ X_0 F^{\rm NS}\left( {X\over X_0}; {1\over X_0} \right) \right],  
\end{equation}
where, after performing the derivative, we set $X_0=1/\hbar$, $X= X_0 t$ and $\hbar=2 \pi k$. This formula for the 
all-orders modified grand potential in the WKB approximation, 
$\CJ^{\rm WKB}(\mu, k)$, was first proposed in \cite{MAhmmo} based on detailed calculations of the WKB expansion. 
We can now interpret it as the quantum 
version of the period $\varpi_3$ of special geometry. 

The above considerations suggest that we write the WKB grand potential in the way first proposed in \cite{MAhmo3}. 
In quantum mirror symmetry, one introduces 
a quantum mirror map depending on $\hbar$. In terms of the variables of the grand potential, this amounts to introducing an ``effective" 
chemical potential $\mu_{\rm eff}$ through the equation, 
\begin{equation}
\label{MAmueff}
\mu_{\rm eff}= \mu+{1\over C(k)} \sum_{\ell \ge 1} a_\ell(k) \MAre^{-2\ell \mu}. 
\end{equation}
In the equation (\ref{MAj0-f0}), we re-expressed the semiclassical grand potential in terms of the ``flat" coordinate $t$. In the all-orders WKB expansion, one should 
re-express it in terms of the effective chemical potential introduced before. After doing this, one obtains 
\begin{equation}
\label{MAgpmueff}
\CJ^{\rm WKB}(\mu, k)=J^{(\rm p)}(\mu,k)+ J^{\rm M2}(\mu, k)
=J^{(\rm p)}(\mu_{\rm eff},k)+\mu_{\rm eff} \widetilde{J}_b(\mu_{\rm eff},k)+\widetilde{J}_c(\mu_{\rm eff},k).
\end{equation}
The two functions $\widetilde{J}_b(\mu_{\rm eff},k)$ and
$\widetilde{J}_c(\mu_{\rm eff},k)$, when expanded at large $\mu_{\rm eff}$, 
define the coefficients $\widetilde{b}_\ell(k)$, $\widetilde{c}_\ell(k)$:
\begin{align}
 \widetilde{J}_b(\mu_{\rm eff},k)=\sum_{\ell=1}^\infty\widetilde{b}_\ell(k)\MAre^{-2\ell\mu_{\rm eff}}, \qquad \widetilde{J}_c(\mu_{\rm eff},k)
=\sum_{\ell=1}^\infty\widetilde{c}_\ell(k)\MAre^{-2\ell\mu_{\rm eff}}. 
\end{align}
In terms of generating functionals for the three sets of coefficients appearing in $J^{\rm M2}(\mu, k)$,   
\begin{align}
\label{MAJabc}
J_a(\mu,k)=\sum_{\ell=1}^\infty a_\ell(k)\MAre^{-2\ell \mu},\quad
J_b(\mu, k)=\sum_{\ell=1}^\infty b_\ell(k)\MAre^{-2\ell \mu},\quad
J_c(\mu, k)=\sum_{\ell=1}^\infty c_\ell(k)\MAre^{-2\ell \mu},
\end{align}
we have
\begin{align}
\label{MAwidetildeJ}
\widetilde{J}_b(\mu_{\rm eff},k)
&
=J_b(\mu,k)-\frac{J_a(\mu,k)^2}{C(k)},\\
\widetilde{J}_c(\mu_{\rm eff},k)
&
=J_c(k,\mu)-\frac{J_a(\mu, k) J_b(\mu, k)}{C(k)}
-\frac{B(k)}{C(k)}J_a(\mu, k)+\frac{2J_a(\mu,k)^3}{3C(k)^2}.\notag
\end{align}
As a final step, we can put together the decomposition (\ref{MAgpmueff}) with the formula (\ref{MAcomplete-jwkb}). It was conjectured in \cite{MAacdkv} (and confirmed in many examples \cite{MAhkrs})
that the NS free energy of a general, toric CY manifold, defined in terms of quantum periods, agrees with the NS limit of the refined topological string energy, which can be 
computed with many other methods (the refined topological vertex \cite{MAikv}, the refined holomorphic anomaly equations of \cite{MAhk-di}, and the geometric perspective on 
refined BPS invariants in \cite{MAckk}). In particular, the NS free energy can be 
expressed in terms of {\it refined BPS invariants} \cite{MAikv,MAckk}, which are integer invariants encoding enumerative information 
on the local CY manifold. One has the following formula for the instanton part of the NS free energy, i.e. for the part involving exponentially small corrections at large $t_I$: 
\begin{equation}
\label{MANS-j}
F^{\rm inst}_{\rm NS}(t, \hbar) =\sum_{j_L, j_R} \sum_{w, {\bf d} } 
N^{{\bf d}}_{j_L, j_R}  \frac{\sin\frac{\hbar w}{2}(2j_L+1)\sin\frac{\hbar w}{2}(2j_R+1)}{2 w^2 \sin^3\frac{\hbar w}{2}} \MAre^{-w {\bf d}\cdot{\bf  t}}.
\end{equation}
In this formula, ${\bf t}$ is the vector of K\"ahler parameters $t_I$, and ${\bf d}$ is the vector of degrees. The refined BPS invariants, denoted by $N^{{\bf d}}_{j_L, j_R}$, depend 
on the degrees and on two half-integer spins, $j_L$ and $j_R$. Note that, when expressed in terms of the $z_I$s, the 
$t_I$ depend as well on $\hbar$, as explained in (\ref{MAq-pers}). It follows from (\ref{MANS-j}) and (\ref{MAgpmueff}) that the coefficient  $\widetilde{b}_\ell(k)$ has the 
following expression \cite{MAhmmo} 
\begin{equation}
\label{MAblj}
\widetilde{b}_\ell(k)=-\frac{\ell}{2\pi}\sum_{j_L,j_R}\sum_{\ell=dw}\sum_{d_1+d_2=d}N^{d_1,d_2}_{j_L,j_R}q^{\frac{w}{2}(d_1-d_2)}
\frac{\sin\frac{\pi kw}{2}(2j_L+1)\sin\frac{\pi kw}{2}(2j_R+1)}{w^2\sin^3\frac{\pi kw}{2}}. 
\end{equation}
In this equation, $N^{d_1, d_2}_{j_L, j_R}$ are the refined BPS invariants of the CY manifold local $\IP^1 \times \IP^1$. The extra factor of 
$q$ appearing in this formula is due to the choice of $z_{1,2}$ in (\ref{MAzz-q}). 
In addition, one finds the following formula for the coefficient $\widetilde c_\ell(k)$, 
\begin{equation}
\widetilde{c}_\ell(k)=- k^2 \frac{\partial}{\partial k} 
\left(\frac{\widetilde{b}_\ell(k)}{2\ell k}\right).
\label{MAbcrel}
\end{equation}
This relationship was first conjectured in \cite{MAhmo3}, and verified in many examples. As explained above, this relation follows from the fact that the full WKB grand potential 
can be regarded as the quantum version of the period given in (\ref{MAj0-f0}). This interpretation of the WKB grand potential also 
explains the fact that the quantum corrections $\CJ_n^{\rm WKB}(\mu)$ can be obtained by acting with differential 
operators on $\CJ_0^{\rm WKB}(\mu)$ \cite{MAmoriyaman,MAhatsuda-spectral}, since this 
is a known property of quantum periods \cite{MAhuang}.

\subsubsection{The TBA approach}

The ABJM Fermi gas can be studied with a different tool, which in principle is exact: a TBA-like system which determines 
the grand potential through a set of coupled non-linear integral equations. This TBA-like system was first considered in \cite{MAcfiv}, in the context of $\CN=2$ 
models in two dimensions, but its relevance to the study of integral kernels was first noticed by Al. Zamolodchikov in \cite{MAzamo}. His results were further elaborated and 
justified in the work of Tracy and Widom \cite{MAtw}. Although TBA-like systems have been very useful in the study of integrable models in QFT, in the case of the 
ABJM Fermi gas this approach has had a limited use, as we will eventually explain. 
Nevertheless, it makes it possible to compute in an efficient way the partition functions $Z(N,k)$ for finite $N$ \cite{MApy,MAhmo}, and is well suited for a WKB analysis \cite{MAcm}.

In \cite{MAzamo}, the following type of integral kernel is considered: 
\begin{equation}
\label{MAkernel}
\rho(\theta, \theta')={1\over 2 \pi} {\exp\left(-u(\theta) - u (\theta')\right) \over 2 \cosh{ \theta-\theta'\over 2}}. 
\end{equation}
Under suitable assumptions on the potential function $u(\theta)$, this kernel is of trace class. The spectral problem 
\begin{equation}
\label{MAeigen-K}
\int_{-\infty}^{\infty} \rho(\theta, \theta') f(\theta') = \lambda f(\theta) 
\end{equation}
leads to a discrete set of eigenvalues $\lambda_n$, $n=0,1,\cdots$. The Fredholm determinant of the operator $\rho$ is defined by 
\begin{equation}
\label{MAfredholm}
\Xi (\kappa) =\prod_{n\ge 0} (1+ \kappa \lambda_n)
\end{equation}
and it is an entire function of $\kappa$. We will regard $\Xi$ as a grand canonical partition function, as in (\ref{MAfred}). The grand potential is then given by 
\begin{equation}
\CJ(z)=\log \, \Xi(z),
\end{equation}
and it has the expansion (\ref{MAjsum}). Let us now introduce the iterated integral 
 \begin{equation}
 \label{MAit-int}
 R_\ell (\theta)=\MAre^{-2 u(\theta)} \int_{-\infty}^{\infty} {\MAre^{-2u (\theta_1)-\cdots -2u(\theta_\ell) } \over  \cosh {\theta-\theta_{1} \over 2}   \cosh {\theta_1-\theta_2 \over 2} \cdots 
  \cosh {\theta_\ell-\theta \over 2}}\rd \theta_1 \cdots \rd \theta_\ell, \qquad \ell \ge 1,
 \end{equation}
 and
 \begin{equation}
 R_0(\theta)=\MAre^{-2 u(\theta)}.
 \end{equation}
Notice that 
\begin{equation}
\label{MARZ}
\int_{-\infty}^{\infty} \rd \theta \, R_\ell(\theta)= (4 \pi)^{\ell+1} Z_{\ell+1}, 
\end{equation}
where $Z_\ell$ is the coefficient appearing in (\ref{MAjsum}). The generating series 
\begin{equation}
R(\theta|\kappa)= \sum_{\ell \ge 0} \left( -{\kappa \over 4 \pi} \right)^\ell R_\ell (\theta)
\end{equation}
satisfies
\begin{equation}
\label{MAzamoJ}
\int_{-\infty}^{\infty}  {\rd \theta \over 4 \pi} R(\theta|\kappa)=\sum_{\ell \ge 0} \left( -\kappa\right)^\ell Z_{\ell+1} ={\partial \CJ \over \partial \kappa}.
\end{equation}

It was conjectured in \cite{MAzamo} and proved in \cite{MAtw} that the function $R(\theta|\kappa)$ can be obtained by using 
TBA-like equations. We first define
\begin{equation}
\label{MArs}
\begin{aligned}
R_+(\theta|\kappa)&={1\over 2} \left( R(\theta|\kappa)+ R(\theta|-\kappa) \right), \\
R_-(\theta|\kappa)&={1\over 2} \left( R(\theta|\kappa)- R(\theta|-\kappa) \right).
\end{aligned}
\end{equation}
Let us now consider the TBA-like system
\begin{equation}
\label{MAtba}
\begin{aligned}
2u(\theta)&= \epsilon(\theta) + \int_{-\infty}^{\infty} {\rd \theta' \over 2\pi} {\log \left(1+\eta^2(\theta')\right) \over \cosh(\theta-\theta')}, \\
\eta(\theta)&=-\kappa \int_{-\infty}^{\infty} {\rd \theta'  \over  2\pi} {\MAre^{-\epsilon(\theta')}\over \cosh(\theta-\theta')}. 
\end{aligned}
\end{equation}
Then, one has that
\begin{equation}
\begin{aligned}
R_+(\theta|\kappa)&= \MAre^{-\epsilon(\theta)},\\
R_-(\theta|\kappa)&=R_+(\theta|\kappa) \int_{-\infty}^{\infty} {\rd \theta' \over \pi} {\arctan\, \eta(\theta') \over \cosh^2(\theta-\theta')}.
\end{aligned}
\end{equation}
This conjecture has been proved in \cite{MAtw} for general $u(\theta)$. In general, the system (\ref{MAtba}) has to be solved numerically, although an exact solution exists for 
$u(\theta) =\MAre^{\theta}$ in terms of Airy functions \cite{MAfendley}.

It is obvious that the ABJM Fermi gas is a particular case of the above formalism, up to a simple change of variables. Indeed, if we set
\begin{equation}
x=k \theta,
\end{equation}
and compare with (\ref{MAit-int}), we find 
\begin{equation}
\rho^{\ell +1}(x,x)= \langle x| \mathsf{\rho}^{\ell+1} |x \rangle={1\over (4 \pi)^{\ell+1} k } R_\ell(x),
 \end{equation}
where we have denoted 
\begin{equation}
R_\ell (x)\equiv R_\ell \left( \theta ={x\over k} \right).
\end{equation}
The function $R_\ell(\theta)$ is calculated with the TBA system (\ref{MAtba}) and the potential 
\begin{equation}
\label{MAutheta}
u(\theta)={1\over2} \log \left(2 \cosh{ k \theta \over 2} \right),
\end{equation}
which depends explicitly on $k$. The grand potential is given by 
\begin{equation}
\label{MAgrand-k}
{\partial \CJ \over \partial \kappa}= {1\over 4 \pi k} \int_{-\infty}^{\infty} \rd x R\left( x |\kappa\right). 
\end{equation}
Notice that the function $R(x|\kappa)$ can be written as 
\begin{equation}
\label{MArxz}
R\left( x |\kappa\right)= {4 \pi k \over \kappa} \left \langle x \left| {1\over \MAre^{\mathsf{H}-\mu} +1}\right| x \right\rangle, 
\end{equation}
where the Hamiltonian $\mathsf{H}$ is defined by an equation similar to (\ref{MAhamil}), 
\begin{equation}
\label{MAgen-hamil}
\langle \theta | \MAre^{-\mathsf{H}}|\theta'\rangle = \rho(\theta, \theta'), 
\end{equation}
and $\kappa$ is the fugacity (\ref{fuga}). The expression (\ref{MArxz}), is up to an overall factor, 
the diagonal value of the number of particles in an ideal Fermi gas with Hamiltonian $\mathsf{H}$. 

There is an important property of the TBA system of \cite{MAzamo} which is worth discussing in some detail: the functions $\epsilon(\theta)$, $\eta(\theta)$ make it possible to calculate {\it both}
 $R(x|\kappa)$ and $R(x|-\kappa)$. The last quantity is given by  
\begin{equation}
R\left( x|-\kappa\right)= {4 \pi k \over \kappa}  \left \langle x \left| {1\over \MAre^{\mathsf{H}-\mu} -1}\right| x \right\rangle, 
\end{equation}
and it corresponds to the same one-particle Hamiltonian (\ref{MAgen-hamil}) but with {\it Bose--Einstein} statistics. If we now take into account the expression (\ref{MAfredholm}), 
we deduce that for Bose--Einstein statistics there is a physical singularity at 
\begin{equation}
\kappa=\lambda_0^{-1}>0
\end{equation}
where $\lambda_0$ is the largest eigenvalue of the operator $\rho(x_1, x_2)$. This singularity corresponds of course to the onset of Bose--Einstein condensation 
in the gas, and as a consequence the functions $R_\pm (x|\kappa)$ will have singularities in the $x$-plane for $\kappa\ge  \lambda_0^{-1}$. But this is precisely the regime in which we are interested in the 
case of the ABJM Fermi gas, since large $N $ corresponds to $\mu \gg 1$. Of course, the singularity at large and positive $\kappa$ {\it cancels}, once one adds up $R_+$ and $R_-$, 
but it appears in intermediate steps and leads to technical problems. For example, it is very difficult to use the TBA approach presented in this section 
to obtain numerical information on the grand potential at large fugacity $\kappa$, since the 
standard iteration of the integral equation does not converge when $\kappa$ is moderately large. 

One can then try the opposite regime and perform an expansion around $\kappa=0$ of all the quantities involved in the TBA system \cite{MAhmo,MApy}. 
By doing this, one can compute the coefficients $Z_\ell$ recursively, 
up to very large $\ell$, and obtain in this way exact results for the partition functions $Z(N,k)$, for $N=1,2,\cdots$. In practice, $k$ is taken to be a small integer, and no surprisingly the cases $k=1,2$ are the 
easiest ones: for these values of $k$, ABJM theory has extended $\CN=8$ supersymmetry \cite{MAbkk, MAgr,MAnotes, MAkapustin}, and one would expect additional simplifications. 
For example, for $k=2$, one finds, for the very first values of $N$, 
\begin{equation}
\label{MAzn1}
Z(1,2)= {1\over 8 }, \qquad Z(2,2)={1\over 32 \pi^2}, \qquad Z(3,2)={10-\pi^2  \over 512 \pi^2}.
\end{equation}
 Another use of the TBA approach was proposed in \cite{MAcm}, where it was noticed that, for $k\rightarrow 0$, 
 the kernel of the Fermi gas becomes a delta function, and the TBA system becomes algebraic. Exploiting 
 this observation, one can perform a systematic expansion of all the quantities around $k=0$, and calculate the functions $\CJ_n(\mu)$ directly to high order in $n$. 

\subsubsection{A conjecture for the exact grand potential}

So far we have obtained two different pieces of information on the matrix model of ABJM theory: on the one hand, 
the full 't Hooft expansion of the partition function, and on the other hand, the full WKB 
expansion of the grand potential. Can we put these two pieces of information together? It turns out that the 't Hooft expansion can be incorporated in the 
grand potential, but this requires a subtle handling of 
the relationship between the canonical and the grand-canonical ensemble. In the standard thermodynamic 
relationship, the canonical partition function is given by the integral (\ref{MAfinite-n}). As we have seen in the 
derivation of (\ref{MAairy}), it is very convenient to extend the integration contour to infinity, along the Airy contour $\CC$ shown in Fig. \ref{MAairy-c}. However, 
this cannot be done without changing the value of the integral: As already noted in \cite{MAmp}, if we extend the contour in (\ref{MAfinite-n}) to infinity, we will add to 
the partition function non-perturbative terms of order 
\begin{equation}
\label{MAnp-dif}
\sim  \MAre^{- \mu/k}, 
\end{equation}
If we want to understand the structure of the non-perturbative effects in the ABJM partition function, we have to handle 
this issue with care. A clever way of proceeding was found by Hatsuda, Moriyama and Okuyama in \cite{MAhmo2}. Following their work, we will introduce an auxiliary object, 
which we will call the {\it modified} grand potential, denoted it by $J(\mu, k)$. The modified grand potential 
is {\it defined} by the equality 
\begin{equation}
\label{MAnaif}
Z(N,k)= \int_\CC  {\rd \mu \over 2 \pi \ri} \MAre^{J(\mu,k) - N \mu}, 
\end{equation}
where $\CC$ is the Airy contour shown in Fig. \ref{MAairy-c}. As it was noticed in \cite{MAhmo2}, if we know $J(\mu, k)$, we can recover the conventional 
grand potential $\CJ(\mu,k)$ by the relation
\begin{equation}
\label{MAnaif-gp}
\MAre^{\CJ(\mu, k)} = \sum_{n \in \IZ} \MAre^{J(\mu+2 \pi \ri n , k)}. 
\end{equation}
Indeed, if we plug this in (\ref{MAfinite-n}), we can use the sum over $n$ to extend the integration region
from $[-\pi \ri, \pi \ri]$ to the full imaginary axis. If we then deform the contour 
to $\CC$, we obtain (\ref{MAnaif}). Note that the difference between $J(\mu, k)$ and $\CJ(\mu, k)$ 
involves non-perturbative terms of the form (\ref{MAnp-dif}), and it is not seen in a perturbative 
calculation around $k=0$. Therefore the WKB calculation of 
the full grand potential still gives the perturbative expansion of the modified grand potential, and we have
\begin{equation}
J(\mu, k) = \CJ^{\rm WKB}(\mu, k) + \CO\left(  \MAre^{- \mu/k} \right). 
\end{equation}

Let us now try to understand how to incorporate the information of the 't Hooft expansion in the grand potential. It is much better to use the modified grand potential, due to the fact that 
the relationship (\ref{MAnaif}) involves an integration going to infinity. Let us denote the 't Hooft contribution to the modified grand potential by $J^{\text{ 't Hooft}}(\mu, k)$. 
If we plug this function in the r.h.s. of (\ref{MAnaif}), we should obtain the 't Hooft expansion of the standard free energy. This requires doing the integral by a saddle point calculation at $k \rightarrow \infty$, and it also requires the following scaling for the modified grand potential (see \cite{MAkkn} for a related discussion), 
\begin{equation}
\label{MAws-ge}
J^{\text {'t Hooft}}(\mu, k)=\sum_{g=0}^\infty k^{2-2g} J_g \left({\mu \over k} \right). 
\end{equation}
This can be regarded as the 't Hooft expansion of the modified grand potential, and contains exactly the same information than the 't Hooft expansion of the 
canonical partition function. As usual in the saddle-point expansion of a Laplace transform, the leading terms, which 
are the genus zero pieces $J_0$ and $-F_0/(4 \pi^2)$, are related by a Legendre transform: we first solve for the 't Hooft parameter $\lambda$, in terms of $\mu/k$, through the 
equation 
\begin{equation}\label{MAchemicalpot}
{\mu \over k}={1\over 4 \pi^2} {\rd F_0 \over \rd \lambda}, 
\end{equation}
and then we have, 
\begin{equation}
\label{MAjf}
J_0 \left({\mu \over k} \right)= -{1\over 4 \pi^2} \left( F_0(\lambda) - \lambda {\rd F_0 \over \rd \lambda} \right).
\end{equation}
Similarly, the genus one grand potential $J_1$  is related to the genus one free energy $F_1$ through the equation
\begin{equation} 
J_1\left({\mu \over k}\right)=F_1(\lambda)+{1\over 2} \log \left( 2 \pi k^2 {\partial^2 _ {\mu }  }J_0\left( {\mu \over k} \right)\right), 
\end{equation}
which takes into account the one-loop correction to the saddle-point. Since the integration contour in (\ref{MAnaif}) goes to infinity, 
doing the saddle-point expansion with Gaussian integrals is fully 
justified and no error terms of the form (\ref{MAnp-dif}) are introduced in this way. This is clearly one of the advantages of using the modified grand potential, 
instead of the standard grand potential. 

We should recall now that the genus $g$ free energies $F_g$ are given by the topological 
string free energies of local $\IP^1 \times \IP^1$, in the so-called orbifold frame. It turns out that the Laplace transform which takes us to the grand potential has an interpretation 
in topological string theory: as shown in \cite{MAmp} by using the general theory of \cite{MAabk}, 
it is the transformation that takes us from the orbifold frame to the so-called large radius frame. Therefore, we can interpret $J^{\text {'t Hooft}}(\mu,k)$ as the 
total free energy of the topological string on local $\IP^1 \times \IP^1$ at large radius. 
This free energy has a polynomial piece, which reproduces precisely the perturbative piece (\ref{MAjpmu}), and then an infinite series of corrections of the form, 
\begin{equation}
\MAre^{-4 \mu/k}. 
\end{equation}
These corrections, after the inverse Legendre transform, give back the worldsheet instanton 
corrections that we found in the 't Hooft expansion of the free energy. Notice however that, from 
the point of view of the Fermi gas approach, these are non-perturbative in $\hbar$, and correspond to 
instanton-type corrections in the spectral problem (\ref{MAspec}) \cite{MAmp,MAkm}. 

The large radius free energy of local $\IP^1 \times \IP^1$ is a well studied quantity (in fact, it has been much more studied than the orbifold free energies). 
In particular, there is a surprising result of Gopakumar and Vafa \cite{MAgv}, valid for any CY manifold, which makes it possible to resum the genus expansion in (\ref{MAws-ge}). In our 
case, this means that we can resum the genus expansion, order by order in $\exp(-4 \mu/k)$ \cite{MAhmo2}. The result 
can be written as, 
\begin{equation}
\label{MAthooft-j}
J^{\text{'t Hooft}}(\mu, k)= {C(k) \over 3} \mu^3 + B(k) \mu + A(k)+ J^{\rm WS}(\mu, k), 
\end{equation}
where 
\begin{equation}
\label{MAjws}
J^{\rm WS}(\mu, k)= \sum_{m \ge 1} (-1)^m d_m(k) {\rm e}^{-{4 m \mu \over k}}, 
\end{equation}
and the coefficients $d_m(k)$ are given by 
\begin{equation}\label{MAgvone}
d_m(k)=\sum_{g\ge 0} \sum_{m=wd} n^d_g \left( 2 \sin {2 \pi w \over k} \right)^{2g-2}. 
\end{equation}
In this equation, we sum over the positive integers $w$, $d$ satisfying the constraint $wd=m$. The quantities $n^d_g$ are integer numbers 
called Gopakumar--Vafa (GV) invariants. One should note that, for any given $d$, the $n_d^g$ are different from zero only for a finite number of $g$, therefore (\ref{MAgvone})
is well-defined as a formal power series in $\exp(-4 \mu/k)$. The GV invariants can be computed by various techniques, and in the case of non-compact CY manifolds, 
there are algorithms to determine them for all possible values of $d$ and $g$ (like for example the theory of the topological vertex \cite{MAakmv-tv}.) It is important to note 
that it is only when we use the modified grand potential that we 
obtain the results (\ref{MAthooft-j}), (\ref{MAjws}) for the 't Hooft expansion. If we use the standard grand potential, 
there are additional contributions coming from the ``images" of the modified grand potential in the sum (\ref{MAnaif-gp}). Note also that the resummation of 
(\ref{MAws-ge}) leads to a resummation of the genus $g$ free energies $F_g(\lambda)$, i.e. the terms of the same order in the expansion parameters 
\begin{equation}
\exp\left(-2 \pi {\sqrt{\hat \lambda}} \right), \qquad {1\over {\sqrt{\lambda}}}
\end{equation}
can be resummed to all genera. 

We have now the most important pieces of the total grand potential, $J^{\rm WS}(\mu, k)$, given in (\ref{MAjws}), and $J^{\rm M2}(\mu, k)$, given in (\ref{MAgpmueff}). 
From the point of view of the 't Hooft expansion, $J^{\rm WS}(\mu, k)$ is 
a resummation of a perturbative series, while $J^{\rm M2}(\mu, k)$ contains non-perturbative information. Conversely, from the 
point of view of the WKB expansion, $J^{\rm M2}(\mu, k)$ is the resummation 
of a perturbative expansion, while $J^{\rm WS}(\mu, k)$ is non-perturbative. 
One would be tempted to conclude that the total, modified grand potential is given by 
\begin{equation}
J^{({\rm p})}(\mu,k)+  J^{\rm M2}(\mu,k)+J^{\rm WS}(\mu, k). 
\end{equation}
However, it can be seen that this is not the case: there is a ``mixing" of the ``membrane" and ``worldsheet" contributions, which 
was found experimentally in \cite{MAhmo3}. It was noted in that reference that agreement with the calculations done at low orders in the expansion was 
achieved by promoting 
\begin{equation}
\label{MAmu-mueff}
\mu \rightarrow \mu_{\rm eff}
\end{equation}
in $J^{\rm WS}(\mu, k)$. From the point of view of the 't Hooft expansion, the corrections added in this form are again non-perturbative. It was noted in \cite{MAkm} 
that the prescription (\ref{MAmu-mueff}) is natural from the point of view of the spectral problem: in calculating instanton corrections to the WKB result for 
the volume (\ref{MApvol}), the weight of an instanton should involve the quantum A-period, which means that one should use $\mu_{\rm eff}$ rather than $\mu$. 
 
Then, the final proposal for the modified grand potential, putting 
together all the pieces from \cite{MAmpabjm, MAdmp, MAmp, MAcm, MAhmo2, MAhmo3, MAhmmo}, is the following: 
\begin{equation}
\label{MAfinal-proposal}
J(\mu, k)= J^{(p)}(\mu,k)+ J^{\rm WS}(\mu_{\rm eff}, k)+ J^{\rm M2}(\mu,k). 
\end{equation}
When the function (\ref{MAfinal-proposal}) is expanded at large $\mu$, one finds 
exponentially small corrections in $\mu$ of the form, 
\begin{equation}
\exp\left\{ -\left( {4 n \over k} + 2 \ell\right) \mu\right\}. 
\end{equation}
In \cite{MAcm} these mixed terms were interpreted as bound states of worldsheet instantons and membrane instantons in the M-theory dual. 

One of the most important properties of the proposal (\ref{MAfinal-proposal}) is the following. It is easy to see, by looking at the 
explicit expressions (\ref{MAjws}) and (\ref{MAgvone}), that $J^{\rm WS}(\mu, k)$ is singular 
for any rational $k$. This is a puzzling result, since it implies that the genus resummation of the free energies 
$F_g$ is also singular for infinitely many values of $k$, including the integer values 
for which the theory is in principle well-defined non-perturbatively. It is however clear that this divergence is an artifact 
of the genus expansion, since the original matrix integral (\ref{MAabjm-mm}), 
as well as its Fermi gas form (\ref{MAtanh-form}), are perfectly well-defined for any real value of $k$. What is going on? 

It was first proposed in \cite{MAhmo2} that the divergences in the resummation of 
the genus expansion of $J^{\rm WS}(\mu, k)$ should be cured by other terms in the modified grand potential, 
in such a way that the total result is finite. It turns out that $J^{\rm M2}(\mu, k)$ is also 
singular, and its singularities cancel those of $J^{\rm WS}(\mu, k)$ in such a way that the total $J(\mu, k)$ is finite. 
This remarkable property of $J(\mu, k)$ was called in \cite{MAcm} the HMO cancellation 
mechanism. Originally, this mechanism was used as a way to understand the structure of $J^{\rm M2}(\mu, k)$ 
at finite $k$. In \cite{MAhmmo} it was shown that this cancellation is a consequence of the 
structure of the the modified grand potential, and it can be proved by using the underlying geometric structure of 
$J^{\rm M2}(\mu, k)$ and $J^{\rm WS}(\mu, k)$. Let us review this proof here.

The poles in $J^{\rm M2}(\mu, k)$ are due to simple poles appearing in the coefficients $\widetilde b_\ell(k)$ in (\ref{MAblj}), and to double poles in 
the coefficients $\widetilde c_\ell(k)$ given in (\ref{MAbcrel}). On the other hand, the coefficients $d_m(k)$ are singular due to the double poles with $g=0$ in the formula (\ref{MAgvone}).
The Gopakumar--Vafa invariants appearing in (\ref{MAgvone}) can be also expressed in terms of the refined 
BPS invariants appearing in (\ref{MAblj}), and one has
 \begin{equation}
 \label{MAdmk}
 d_m(k)=\sum_{j_L,j_R}\sum_{m=dn}\sum_{d_1+d_2=d}N^{d_1,d_2}_{j_L,j_R}
\frac{2j_R+1}{\left(2\sin\frac{2\pi n}{k}\right)^2}\frac{ \sin\left( \frac{4\pi n}{k}(2j_L+1) \right)}{n\sin\frac{4\pi n}{k}}. 
\end{equation}
Using these results, it is easy to see that the poles cancel. The coefficient \eqref{MAdmk} has double poles when 
$k \in 2n/\IN$. The coefficient (\ref{MAblj}) has 
a simple pole when $k \in 2 \IN/w$, and the coefficient $\tilde c_\ell (k)$ has 
a double pole at the same values of $k$. 
These poles contribute to terms of the same order in $\MAre^{-\mu_{\rm eff}}$ when $k$ is of the form
\begin{equation}
\label{MAksing}
k={2n \over w}={2m \over \ell}.
\end{equation}
 We have then to examine the pole structure of (\ref{MAgpmueff}) at these values of $k$. Since both (\ref{MAdmk}) and (\ref{MAblj}) involve a sum over refined BPS invariants, 
  we can look at the contribution to the pole structure of each of these invariants. In the worldsheet instanton contribution, the singular part around $k=2n/w$ associated to 
  the refined BPS invariant $N^{d_1,d_2}_{j_L,j_R}$ is 
 \begin{equation}
 \label{MAws-pole}
{(-1)^m \over \pi^2}  \left[  {n\over  w^4 \left( k -{2n \over w}\right)^2 } + {1\over    k -{2n \over w}}  \left( {1\over w^3} +   {m \over  n w^2}   \mu_{\rm eff}   \right) \right] (1+2 j_L)(1+ 2j_R) 
N^{d_1,d_2}_{j_L,j_R} \MAre^{-{2 mw\over n}  \mu_{\rm eff} }.
 \end{equation}
The singular part in $\mu_{\rm eff} \widetilde J_b (\mu_{\rm eff},k) $ associated to the same invariant is given by 
\begin{equation}
\label{MAb-pole}
-{  \MAre^{\pi \ri k w(d_1-d_2)/2} \over \pi^2}  {\ell \over w^3 \left( k -{2n \over w} \right)} (-1)^{n(2j_L+2j_R-1)} (1+2 j_L)(1+ 2j_R)  N^{d_1,d_2}_{j_L,j_R}\mu_{\rm eff}\MAre^{-2\ell\mu_{\rm eff}}. 
\end{equation}
Using (\ref{MAbcrel}), we find that the corresponding singular part in $\widetilde J_c (\mu_{\rm eff},k) $ is given by 
\begin{equation}
\label{MAc-pole}
-{\MAre^{\pi \ri k w(d_1-d_2)/2} \over \pi^2} \left[  {n\over  w^4 \left( k -{2n \over w}\right)^2 } +
 {1\over    w^3 \left( k -{2n \over w}  \right)} \right] (-1)^{n(2j_L+2j_R-1)} (1+2 j_L)(1+ 2j_R)  N^{d_1,d_2}_{j_L,j_R}\MAre^{-2\ell\mu_{\rm eff}}
\end{equation}
By using (\ref{MAksing}), one notices that 
\begin{equation}
 \MAre^{\pi \ri k w(d_1-d_2)/2}=(-1)^m
\end{equation}
and it is easy to see that all poles in (\ref{MAws-pole}) cancel against the poles in (\ref{MAb-pole}) and (\ref{MAc-pole}), for any value of $\mu_{\rm eff}$, provided that
\begin{equation}
(-1)^{n(2j_L+2j_R-1)}=1.
\end{equation}
However, this is the case, since for local $\IP^1 \times \IP^1$, the only non-vanishing BPS indices $N^{d_1, d_2}_{j_L, j_R}$ have 
\begin{equation}
\label{MA2jzero}
2j_L + 2j_R-1\equiv 0 \quad {\rm mod}\, 2. 
\end{equation}
This can be justified with a geometric argument explained in \cite{MAhmmo}. 

The HMO cancellation mechanism is conceptually important for a deeper understanding of the non-perturbative structure of M-theory. In the M-theory expansion, 
we have to resum the worldsheet instanton contributions to the free energy at fixed $k$ and large $N$, which is precisely what the Gopakumar--Vafa representation (\ref{MAjws}) 
does for us. However, after this resummation, the resulting expression is singular and displays an infinite number of poles. We obtain a finite result only when the contribution 
of membrane instantons has been added. This shows very clearly that a theory based solely on fundamental strings is fundamentally incomplete, and that additional 
extended objects are needed in M-theory. Of course, this has been clear since the advent of M-theory, 
but the above calculation shows that the contribution of membranes is not just a correction to 
the contribution of fundamental strings; it is crucial to remove the poles and to make the amplitude well-defined. Conversely, a theory based only on membrane instantons 
will be also incomplete and will require fundamental strings in order to make sense. 

The result (\ref{MAfinal-proposal}) is the current proposal for the grand potential of the ABJM matrix model. It is a remarkable exact result. From the point of view of gauge theory, it encodes 
the full $1/N$ expansion, at all genera, as well as non-perturbative corrections at large $N$ (due presumably to some form of large $N$ instanton). From the point of view of M-theory, 
it incorporates both membrane instantons and worldsheet instantons. This can be seen very clearly by writing
\begin{equation}
\MAre^{J(\mu, k)} = \MAre^{J^{({\rm p})} (\mu,  k)} \sum_{l, n} a_{l,n} (k)\mu^n \MAre^{- l \mu}, 
\end{equation}
where $l$ is of the form $2p + 4 q/k$, and $p$, $q$ are non-negative integers. Plugging this in (\ref{MAnaif}), we find
\begin{equation}
\label{MAz-exp}
Z(N, k) =\frac{\MAre^{A(k)}}{\left(C(k) \right)^{1/3}}
\sum_{l,n} a_{l,n} \left(-\frac{\partial}{\partial N}\right)^n \mathrm{Ai}
\left(\frac{N+l -B(k)}{\left(C(k)\right)^{1/3}}\right), 
\end{equation}
where ${\rm Ai}(z)$ is again the Airy function. The first term in this expansion is of course (\ref{MAzpnk}), while the remaining terms are non-perturbative 
corrections, exponentially suppressed as $N$ becomes large. The conjectural formula 
(\ref{MAz-exp}) provides an {\it exact} and {\it convergent} series for the partition function $Z(N,k)$, in the M-theory expansion. It is remarkable that, in the framework of 
M-theory, the asymptotic and divergent expansions in the string coupling constant are promoted to convergent series. 

It is instructive to 
compare these results for $Z(N,k)$ with the well-known number theory results for the number of partitions $p(N)$ of an integer $N$. For $N$ large, the leading asymptotic behavior of $p(N)$ is given by Hardy--Ramanujan formula, 
\begin{equation}
p(N) \sim \exp\left(  \pi {\sqrt{2 N \over 3}} \right), \qquad N \gg 1. 
\end{equation}
This leading asymptotics was dramatically improved by Rademacher into a convergent series expression. In our case, the analogue of the Hardy--Ramanujan formula is the SUGRA result (\ref{MAmsugra}), while (\ref{MAz-exp}) is the analogue of the Rademacher expansion. 

As noted in \cite{MAcgm}, one interesting aspect of the formulae (\ref{MAnaif}) and (\ref{MAz-exp}) is the following. 
In principle, the partition function $Z(N,k)$ is only defined for integer $N$, since 
it is given by a matrix integral for $N$ variables. However, using (\ref{MAnaif}) or (\ref{MAz-exp}), 
one can define $Z(N,k)$ for arbitrary complex $N$, and the resulting function seems to be 
entire. It is not clear how natural is this promotion of a function defined on integer 
values of $N$, to a function on the complex plane. This is similar to the promotion 
of the factorial to the Gamma function. It is known that this problem has many solutions (for example, 
Hadamard's Gamma function also does the job). 
There is in fact another way to promote $Z(N,k)$ to a function defined on a larger domain, 
which is by performing a Borel resummation of the $1/N$ expansion. This type of promotion, in the context of 
of matrix models, was first worked out in \cite{MAmmnp} for a unitary matrix model, and it has been recently explored in much detail in \cite{MArrr}, in the case of 
the quartic matrix model. In general, in order to recover the exact non-perturbative answer, even for 
integer $N$, one has to add (Borel-resummed) instanton contributions to the resummed $1/N$ series. In the case of ABJM theory, the results of \cite{MAgmz} 
indicate that such contributions must be present, although their precise form is not known yet. 
It would be very interesting to see if these two different ways of extrapolating 
$Z(N,k)$ to non-integer $N$ (namely, the exact formulae (\ref{MAnaif}), (\ref{MAz-exp}), and the Borel 
resummation of the $1/N$ expansion and its instanton corrections) lead to the same function. If so, this would be a strong indication 
that this is the natural function picked up by the underlying mathematical structures of the problem.

The result (\ref{MAfinal-proposal}) is relatively complicated, since it involves an enormous amount of information, 
including the all-genus Gopakumar--Vafa invariants of local $\IP^1 \times \IP^1$, and 
the quantum periods of this manifold to all orders. However, there are certain cases in which (\ref{MAfinal-proposal}) 
can be simplified, as first noted in \cite{MAcgm}. This happens when $k=1$ or $k=2$ and ABJM theory 
has enhanced $\CN=8$ supersymmetry. In the underlying 
topological string theory, only the genus zero and genus one free energies contribute, 
so we have a sort of ``non-renormalization theorem." In these cases, as shown in \cite{MAcgm}, one can 
write down closed formulae for the modified grand potential and for the grand canonical partition function. The simplest case is $k=2$, and one finds
\begin{equation} 
\label{MAstd2}
\Xi(\mu, 2)=  \exp \left\{ {\mu \over 4}+F_1+F_1^{\rm NS} -{1\over  \pi^2}  \left( F_0(\lambda)-\lambda \partial_{\lambda}F_0(\lambda)+{\lambda^2 \over 2} 
\partial^2_{\lambda}F_0(\lambda)\right) \right\}
\vartheta_3(\bar \xi, \bar \tau). 
\end{equation}
In this equation, $F_0(\lambda)$ is the planar free energy of ABJM theory defined by (\ref{MAcomf}), and $\lambda$ is given, as a function of $\kappa$ (or $\mu$) by 
(\ref{MAlamkap}). In (\ref{MAstd2}), $\vartheta_3(\bar \xi, \bar \tau)$ is a Jacobi theta function 
with arguments
\begin{equation}
\begin{aligned}
\bar \xi&= {\ri \over 4 \pi^3} \left( \lambda \partial_\lambda^2 F_0 (\lambda) - \partial_\lambda F_0 (\lambda)\right), \\
\bar \tau&={\ri \over 8 \pi^3} \partial_\lambda^2 F_0 (\lambda). 
\end{aligned} 
\end{equation}
Note that $\bar \tau$ is related to the parameter $\tau$ in (\ref{MAtauABJM}) by the linear relation $\bar \tau = (\tau-1)/2$. $F_1$ is the genus one free energy 
introduced in (\ref{MAgenus-one}). Finally, $F_1^{\rm NS}$ is the next-to-leading term in the $\hbar$ expansion (\ref{MANSh}), and it has the explicit expression
\begin{equation}
F^{\rm NS}_1 (t)= -{1\over 6} \mu-{1\over 24} \log(1+16\,  \MAre^{-2 \mu}). 
\end{equation}
It can be checked \cite{MAcgm} that the small $\kappa$ expansion of (\ref{MAstd2}) reproduces the values of the partition function (\ref{MAzn1}) and all the values 
computed in \cite{MAhmo2}. In addition, from the study of the zeroes of the theta function in (\ref{MAstd2}), 
one finds an exact quantization condition determining the energy levels in (\ref{MAspec}) 
for $k=2$. This quantization condition takes the form of a Bohr--Sommerfeld formula, 
\begin{equation}
{\rm vol}(E,2)= 8 \pi^2 \left( n+{1\over 2} \right), \qquad  n=0,1,2, \cdots, 
\end{equation}
where the exact quantum volume function has the form
\begin{equation}
\label{MAeqv}
\begin{aligned}
{\rm vol}(E, 2) &= 8\pi {K'( 4 \ri \, \MAre^{-E} ) \over K(4 \ri \, \MAre^{-E} )} \left( E + 2  \MAre^{-2E} \, _4F_3\left(1,1,\frac{3}{2},\frac{3}{2};2,2,2;-16 \,  \MAre^{-2E}\right) \right)
\\ & -{4\over  \pi} 
G_{3,3}^{3,2}\left(-16 \,  \MAre^{-2E}\left|
\begin{array}{c}
 \frac{1}{2},\frac{1}{2},1 \\
 0,0,0
\end{array}
\right.\right).
\end{aligned}
\end{equation}
In this equation, $K(k)$ is the elliptic integral of the first kind, and $G_{3,3}^{3,2}$ is a Meijer function. The spectrum obtained from the above quantization condition 
has been checked in detail against numerical calculations starting from (\ref{MAspec}) \cite{MAkm,MAcgm}. Similar considerations can be made for $k=1$, see \cite{MAcgm} for detailed 
formulae. The exact quantum volume (\ref{MAeqv}) was first proposed in \cite{MAkm}, 
based on some simplifying assumptions which turn out to hold only 
in the maximally supersymmetric cases of ABJM theory. In the general case, the quantization condition proposed in \cite{MAkm} receives corrections, 
as pointed out in \cite{MAfhw} based on a numerical analysis of the spectrum. These corrections 
can be derived analytically within the framework proposed in \cite{MAghm-abjm}, which provides an exact quantization condition for arbitrary $k$. As in the 
maximally supersymmetric cases, the quantization condition in \cite{MAghm-abjm} is based on finding the zeroes of the spectral determinant, 
which is in turn determined by (\ref{MAnaif-gp}) and (\ref{MAfinal-proposal}).

\section{Generalizations}

In the previous section, we have focused on ABJM theory, for simplicity. However, many of the above results can be extended, in one way or another, to other Chern--Simons--matter theories. 
Not surprisingly, and as a general rule, the more supersymmetry the theory has, 
the more explicit are the results that one can obtain. In this section we will make a brief overview of these generalizations, which should mostly 
serve as a guide to the growing literature on the subject. 

\subsection{The 't Hooft expansion of Chern--Simons--matter theories}

Localization can be applied to a large range of Chern--Simons--matter theories with $\CN\ge 2$ supersymmetry \cite{MAkwy,MAhama,MAjafferis}. The partition function of such theories is reduced in this way 
to a matrix integral, which can then be studied in the 't Hooft expansion with conventional techniques. However, for most of these theories, obtaining even the planar free 
energy has been a very difficult task, due to the sheer complexity of the resulting matrix integrals. Some results exist for ABJM theory with fundamental matter \cite{MAcmp}, 
for the generalization of ABJM theory in which the levels are different \cite{MAsuyama-gt} (also known as Gaiotto--Tomasiello theory \cite{MAgt}), and for a single Chern--Simons theory coupled 
to adjoint hypermultiplets \cite{MAsuyama-adjoint}. In addition, T. Suyama has provided some general results for a class of $\CN\ge 3$ Chern--Simons--matter theories \cite{MAsuyama-1,MAsuyama-2}. 

One case where we have much information is ABJ theory \cite{MAabj}. As shown in \cite{MAmpabjm,MAdmp}, the 't Hooft expansion of this theory is still completely 
controlled by topological string theory on local $\IP^1 \times \IP^1$. Since this is now a two-parameter problem, computing the genus $g$ free energies is technically more 
complicated, but one can determine the analogue of the function $J^{\rm WS}(\mu, k)$ for ABJ theory in terms of Gopakumar--Vafa invariants in a completely straightforward way. 

Perhaps the simplest theory, beyond ABJ(M) theory, where the planar limit of the free energy is known, is the superconformal field theory in three dimensions 
consisting of supersymmetric $U(N)$ Yang--Mills theory, coupled to a single 
adjoint hypermultiplet and to $N_f$ fundamental hypermultiplets. When $N_f=1$, this theory is related by mirror symmetry to $\CN=8$ super Yang--Mills theory, therefore 
to ABJM theory with $k=1$ \cite{MAkwytwo}. From the point of view of M-theory, this gauge theory 
is supposed to describe $N$ M2 branes probing the space \cite{MAkapustin, MAbcc}
\begin{equation}
\IC^2 \times \left( \IC^2/\IZ_{N_f} \right), 
\end{equation}
where $\IZ_{N_f}$ acts on $\IC^2$ as
\begin{equation}
\MAre^{2 \pi \ri/N_f}\cdot(a, b)= \left( \MAre^{2 \pi \ri /N_f} a, \MAre^{-2 \pi \ri /N_f} b\right). 
\end{equation}
The corresponding quotient is an $A_{N_f-1}$ singularity, which can be resolved to give a multi-Taub-NUT space, as expected from the engineering of the theory in terms of D6 branes. 
The large $N$ dual description of this theory is in terms of M-theory on AdS$_4\times \IS^7/\IZ_{N_f}$, where the action of $\IZ_{N_f}$ is the one inherited by the action on $\IC^2 \times \IC^2$. 
  
The rules for localization of Chern--Simons--matter theories obtained in \cite{MAkwy,MAhama,MAjafferis} (see \volcite{WI}) imply 
that the partition function on the three-sphere $\IS^3$ is given by the matrix integral  
\begin{equation}
\label{MApf}
Z(N, N_f)={1\over N!}  \int \prod_{i=1}^N {\rd x_i \over 4 \pi}  {1\over \left( 2 \cosh {x_i \over 2} \right)^{N_f}} \prod_{i<j} \left( \tanh \left( {x_i - x_j \over 2} \right) \right)^2. 
\end{equation}
The $\tanh$ interaction between the eigenvalues includes both a $\sinh$ factor due to the Yang--Mills vector multiplet, and a $1/\cosh$ due to the hypermultiplet in the 
adjoint representation. Notice that, when $N_f=1$, this model leads to the same matrix integral than ABJM theory with $k=1$, in the representation (\ref{MAtanh-form}). The natural 
't Hooft parameter for this model is 
\begin{equation}
\label{MAthooft-nf}
\lambda={N\over N_f},
\end{equation}
where $N_f$ is the number of flavours (in fact, we should rather call this parameter the Veneziano parameter, since it takes into account the scaling of the number of flavors 
with the rank of the gauge group). The matrix model (\ref{MApf}) was called the $N_f$ matrix model in \cite{MAgm}, where it was analyzed 
in detail. It turns out that this model can be mapped to an $O(2)$ matrix model \cite{MAkostov-on} 
and studied with the techniques of \cite{MAek,MAek2} (similarly to what was done in \cite{MAsuyama-adjoint}). 
The planar solution turns out to be very simple. Let us write the large $N$ expansion of the free energy as 
\begin{equation}
F(N, N_f)= \sum_{g \ge 0} N_f^{2-2g}F_g(\lambda). 
\end{equation}
As in the ABJM matrix model, in order to write the planar solution, we need to introduce an auxiliary parameter $k$, related to the 't Hooft parameter (\ref{MAthooft-nf}) by  
 \begin{equation}
 \label{MAtex}
 \lambda= -{1\over 8} + {(1+k)^2 \over 8 \pi^2} K'(k)^2. 
 \end{equation}
Then, the planar free energy is given, as a function of $k$, by 
  \begin{equation}
  \label{MAfree2}
  {\rd^2 F_0 \over \rd \lambda^2}= -2 \pi { K(k) \over K'(k)}.
  \end{equation}
The integration constants can be fixed by a perturbative calculation in the matrix 
model (\ref{MApf}). The planar free energy can be expanded at strong 't Hooft coupling, and it has the behavior 
\begin{equation}
\label{MApf-nf}
F_0(\lambda)= -{ \pi \sqrt{2}\over 3} \hat \lambda^{3/2}+ {1\over 8} \left(\log (2)-{\zeta(3) \over \pi^2} \right) + F^{\rm WS}_0 (\lambda),
\end{equation}
where
\begin{equation}
\label{MAws-f0} \begin{aligned}
F^{\rm WS}_0(\lambda)&=-{\MAre^{- 2 \pi {\sqrt{2 \hat\lambda}}} \over 4 \pi^2} \left( 1+ 2 \pi {\sqrt{2\hat \lambda}} \right)-{\MAre^{- 4 \pi {\sqrt{2 \hat\lambda}}} 
\over 32 \pi^2} \left( 7+ 28 \pi {\sqrt{2 \hat\lambda}} + 64 \pi^2\hat \lambda \right)+\cdots
\end{aligned} \end{equation}
and the variable $\hat \lambda$ is defined by 
\begin{equation}
\hat \lambda =\lambda+{1\over 8}. 
\end{equation}
This shift is reminiscent of the shift in $-1/24$ which appears in the planar solution of ABJM theory, in (\ref{MAshift-lam}). The constant appearing in the second term of (\ref{MApf-nf}) was 
determined numerically in \cite{MAgm}, and then conjectured analytically in \cite{MAho}. 

The expansion of the planar free energy found above is conceptually very similar to what was found in ABJM theory: 
the leading term displays the expected $N^{3/2}$ behavior, and the coefficient 
is in agreement with the expectations from the M-theory dual (this was already checked in the strict large $N$ limit in \cite{MAjkps}). The subleading, 
exponentially small corrections appearing in (\ref{MAws-f0}) have the expected form of worldsheet instanton corrections. However, they do {\it not} 
have the structure that one would expect from a 
topological string theory on a CY manifold. This is easier to see by considering the grand potential of the theory in 
the 't Hooft expansion, i.e. the analogue of the quantity (\ref{MAjf}) in ABJM theory. Its structure 
turns out to be more complicated than the one in ABJM theory, and in particular it does not have the simpler structure appearing in topological string theory. 

Unfortunately, it is hard to calculate $1/N$ corrections to the above result. The genus one free energy was determined in 
\cite{MAgm} by using matrix model techniques, but 
beyond that nothing is known. In \cite{MAho}, some all-genus results for the worldsheet instantons were guessed up 
to third order in the instanton expansion, but much work 
remains to be done if we want to understand the $N_f$ matrix model at the level of precision that we have achieved for the ABJM matrix model.

\subsection{The Fermi gas approach to Chern--Simons--matter theories}

As we have seen, the Fermi gas approach to ABJM theory is very powerful, and it can be applied to 
other Chern--Simons--matter theories. However, it also has some limitations. It relies on algebraic identities (such as (\ref{MAcauchy})) 
which make it possible to rewrite the relevant matrix models, in the form of the canonical partition function of an ideal Fermi gas. 
This can be done for a large class of $\CN\ge 3$ theories, 
but in the case of $\CN=2$ theories the resulting gas is interacting \cite{MAmp-inter} and much harder to analyze. 

Let us mention some generalizations of the Fermi gas approach. The $\CN=3$ models where it can be applied more successfully are 
the necklace quiver gauge theories constructed in \cite{MAjt,MAik}. These theories are given by a Chern--Simons quiver with gauge groups and levels,
\begin{equation}
U(N)_{k_1} \times U(N)_{k_2} \times \cdots U(N)_{k_r}.
\end{equation}
Each node is labelled with the letter $a=1, \cdots, r$. There are bifundamental chiral superfields $A_{a a+1}$, $B_{a a-1}$ connecting 
adjacent nodes, and in addition there can be $N_{f_a}$ matter superfields $(Q_a, \tilde Q_a)$ in each node, in the fundamental representation. We will write
\begin{equation}
k_a=n_a k, 
\end{equation}
and we will assume that 
\begin{equation}
\label{MAadd0}
\sum_{a=1}^r n_a=0. 
\end{equation}
The matrix model computing the $\IS^3$ partition function of such a necklace quiver gauge theory is given by 
\begin{equation}
\label{MAnecklace}
Z\left(N, n_a, N_{f_a}, k \right)={1\over (N!)^r} \int  \prod_{a,i} {\rd \lambda_{a,i} \over 2 \pi}  
{\exp \left[ {\ri n_a k\over 4 \pi}\lambda_{a,i}^2 \right] \over \left( 2 \cosh{\lambda_{a,i} \over 2}\right)^{N_{f_a}} } \prod_{a=1}^r  {\prod_{i<j} \left[ 2 \sinh \left( {\lambda_{a,i} -\lambda_{a,j} \over 2} \right)\right]^2 \over \prod_{i,j} 2 \cosh \left( {\lambda_{a,i} -\lambda_{a+1,j} \over 2} \right)}.
\end{equation}
To construct the corresponding Fermi gas, we define a kernel corresponding to a pair of connected nodes $(a,b)$ by, 
\begin{equation}
\label{MAkerquiv}
 K_{ab}(x',x)=\frac{1}{2\pi k}\frac{\exp\left\{\ri\frac{n_b x^2}{4\pi k}\right\}}{2\cosh\left(\frac{x'-x}{2k}\right)}\,\left[2\cosh\frac{x}{2k}\right]^{-N_{f_b}},
 \end{equation}
 where we set $x=\lambda/k$. If we use the Cauchy identity (\ref{MAcauchy}), it is easy to see that we can write the grand canonical partition function for this theory in the form (\ref{MAfred}), 
 where now \cite{MAkostov}
 \begin{equation}
 \label{MArhoquiv}
\mathsf{\rho}=\hat K_{r1}\hat K_{12}\cdots \hat K_{r-1,r}
\end{equation}
is the product of the kernels (\ref{MAkerquiv}) around the quiver. Therefore, we still have a Fermi gas, albeit the Hamiltonian is 
quite complicated, and we can apply the same techniques that were used for ABJM theory in the previous section. For example, 
it is possible to analyze the gas in the thermodynamic limit \cite{MAmp}. One can show 
that, at large $\mu$, the grand potential of the theory is still of the form 
\begin{equation}
\label{MAj-quiver}
\CJ(\mu, k) \approx {C \mu^3 \over 3} + B \mu, \qquad \mu \gg 1. 
\end{equation}
The coefficient $C$ for a general quiver is also determined, as in ABJM theory, by the volume of the Fermi surface at large energy. This limit is a polygon 
and one finds, 
\begin{equation}
\pi^2 C=\,{\rm vol}\left\{(x,y)\, : \,\sum_{j=1}^r\left|y-\left(\sum_{i=1}^{j-1}k_i\right) x\right|+\left(\sum_{j=1}^r N_{f_j}\right)|x|< 1\right\}, 
 \label{MAnecklacevolume}
\end{equation}
which can be computed in closed form \cite{MAherzog}. The $B$ coefficient can be obtained in a case by case basis, although no general formula is known 
for all $\CN=3$ quivers (a general formula is however known for a class of special quivers which preserve $\CN=4$ supersymmetry, see \cite{MAmoriyaman}.)

The result (\ref{MAj-quiver}) has two important consequences. First, the free energy at large $N$ has the behavior 
\begin{equation}
 F(N)\approx -\frac{2}{3}C^{-1/2}N^{3/2}.
\end{equation} 
It can be seen \cite{MAmp}, by using the results of \cite{MAherzog}, that this agrees with the prediction from the M-theory dual (this was also 
verified in \cite{MAhkpt,MAherzog} with the techniques 
that we reviewed in section \ref{MAhkpt-review}). Second, by using the same techniques as in ABJM theory, we conclude that the M-theory expansion of 
the partition function is also given, at leading order, by an Airy function:
\begin{equation}
\label{MAai-gen}
Z(N) \approx C^{-1/3}(k) {\rm Ai} \left[ C^{-1/3} \left( N- B\right) \right]. 
\end{equation}
A Fermi gas formulation also exists for $\CN=3$ theories with other 
gauge groups \cite{MAm-pufu,MAadf,MAmori-so}, and one finds again the Airy behavior (\ref{MAai-gen}). 
This behavior is arguably one of the most important results obtained from the Fermi gas approach, and it has been conjectured in \cite{MAmp} that it is {\it universal}. More precisely, 
it was conjectured that, in any Chern--Simons--matter theory which displays 
the $N^{3/2}$ behavior for the strict large $N$ limit of its partition function, the leading term of the M-theory expansion will have the form (\ref{MAai-gen}). 
The subleading, universal logarithmic prediction for $Z(N)$ implied by (\ref{MAai-gen}) has been also tested by \cite{MAbgms}, and it has been argued that 
the full Airy function can be obtained from a localization computation in AdS supergravity \cite{MAddg}.

Of course, one of the most important advantages of the Fermi gas approach is that it makes it possible to calculate non-perturbative corrections to the 't Hooft expansion. 
However, for theories other than ABJM theory, this has been in general difficult. An exception is again ABJ theory. The Fermi gas formulation of this theory requires some 
work, but it has been achieved in the papers \cite{MAahs,MAhonda} (a different formulation has been proposed in \cite{MAmatsumori}.) A conjectural formula for the modified 
grand potential of this theory has been finally proposed in \cite{MAmatsumori,MAhonda-o}. It has exactly the same properties that we found in ABJM theory: poles appear in the resummation 
of the genus expansion, and they are cancelled by non-perturbative contributions (presumably coming from membrane instantons), just as in the HMO mechanism of \cite{MAhmo2}. 

For other theories which admit a Fermi gas description, the computation of non-perturbative contributions at the level of detail that was done in ABJ(M) theory remains 
largely an open problem. However, all the existing results indicate that generically the 't Hooft expansion of these theories is radically {\it insufficient}: not only there are non-perturbative 
corrections to it, but in addition the resummation of the genus expansion has poles which have to be cured by these non-perturbative effects through a 
generalization of the HMO mechanism. This pattern has been observed in the $N_f$ matrix model \cite{MAgm, MAho}, in some $\CN=4$ 
quiver theories \cite{MAmoriyaman, MAmoriyaman-cancel, MAmoriyaman-2,MAhho}, and in theories involving orientifolds, where new types of instanton effects appear \cite{MAmori-or,MAkazumi-or,MAhonda-or}. 

The Fermi gas approach has had other applications. For example, in \cite{MAdr-fe} it is shown that the mirror symmetry relating some 
Chern--Simons--matter theories can be understood as a canonical transformation 
of the one-dimensional fermions. 

\subsection{Topological strings}
\label{MAts-sec}

One of the key ingredients for the exact determination of the grand potential of ABJM theory has been the relationship to topological string theory. Conversely, one might hope that 
the structures emerging in the context of Chern--Simons--matter theories are relevant for a better understanding of topological strings. 
In particular, as already pointed out in \cite{MAmm-talk}, the results of \cite{MAmp} suggest a Fermi gas approach to topological string theory on toric CY manifolds
which has been pursued in \cite{MAhmmo,MAkm,MAhw}. When the CY $X$ is given by the canonical 
bundle 
\begin{equation}
\CO(K_S) \rightarrow S
\end{equation}
over a del Pezzo surface $S$, a detailed proposal has been presented in \cite{MAghm} and further developed in \cite{MAkas-mar,MAmz,MAgkmr,MAwzh,MAcgm-2,MAoz, MAkp, MAgrassi, MAeigen} (see \cite{MAmm-stms} for a review.) The proposal of 
\cite{MAghm} gives a correspondence between topological string theory on $X$, and the spectral properties of a quantum-mechanical operator  $\mathsf{\rho}_S$ 
obtained by quantizing the mirror curve to $X$. This correspondence shares 
many ingredients of the gauge theory/string theory correspondence, since the topological string emerges in the large $N$ limit of the quantum-mechanical system. In addition, 
it provides a non-perturbative definition of (closed) topological string theory on $X$. 
The resulting picture is also very similar to the theory of the ABJM partition function reviewed in this article, since the operator $\mathsf{\rho}_S$ plays the r\^ole of the canonical 
density matrix of the Fermi gas (\ref{MAdensitymat}). For example, in the case of local $\IP^2$, the mirror curve is given by
\begin{equation}
W(x,y)=\MAre^{x}+ \MAre^{y}+ \MAre^{- x - y} = u, 
\end{equation}
where $u$ is the complex deformation moduli and we have chosen appropriate coordinates. The relevant operator is 
\begin{equation}
\label{MAp2-op}
\mathsf{\rho}_{\IP^2}= \left( \MAre^{\mathsf{x}}+ \MAre^{\mathsf{y}}+ \MAre^{-\mathsf{x} - \mathsf{y}} \right)^{-1}, 
\end{equation}
where $\mathsf{x}$, $\mathsf{y}$ are canonically conjugate Heisenberg operators, as in (\ref{MAcc-operators}). In contrast to what happens in ABJM theory, the kernel 
of the operator $\mathsf{\rho}_S$ is not elementary. However, in some cases (like the operator for local $\IP^2$), it can be calculated explicitly in 
terms of Faddeev's quantum dilogarithm \cite{MAkas-mar}. 

\begin{figure}
\center
\includegraphics[height=6cm]{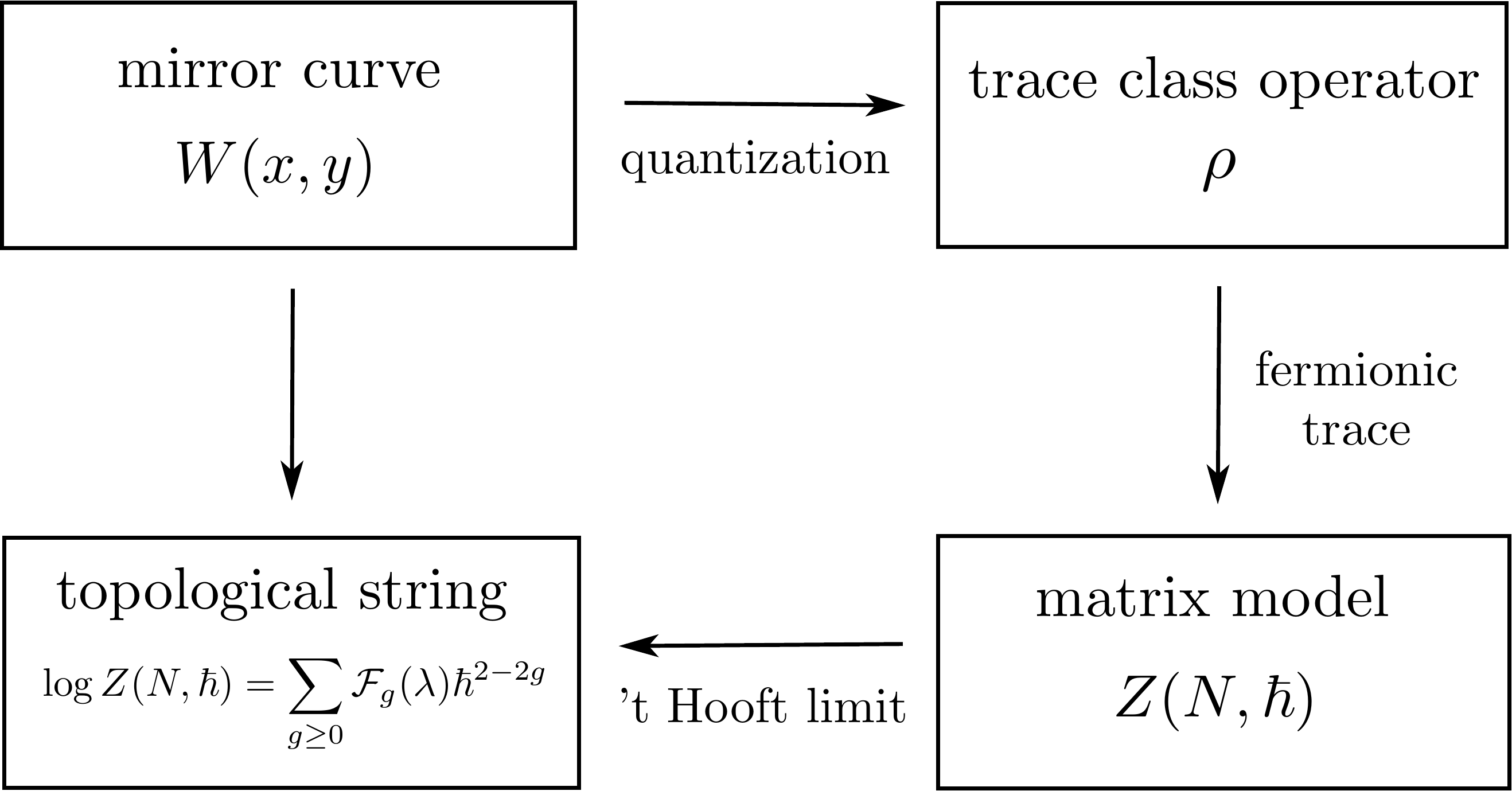} 
\caption{Given a toric CY threefold, the quantization of its mirror curve leads to a trace class operator $\rho$. 
The standard topological string free energy is obtained as the 't Hooft limit of its fermionic traces $Z(N, \hbar)$.}
\label{MArels}
\end{figure}

We can now consider a Fermi gas of $N$ particles whose canonical density matrix is given by $\mathsf{\rho}_S$. Its canonical partition function $Z(N, \hbar)$ 
has a matrix model representation similar to (\ref{MAfgasform}),  
\begin{equation}
\label{MAznmm}
Z_S (N, \hbar)= {1 \over N!} \sum_{\sigma  \in S_N} (-1)^{\epsilon(\sigma)}  \int  \rd ^N x \prod_i \rho_S(x_i, x_{\sigma(i)})
= {1 \over N!}  \int  \rd ^N x \, {\rm det}\left( \rho_S(x_i, x_j) \right), 
\end{equation}
where
\begin{equation}
\rho_S(x, x')=\langle x|\mathsf{\rho}_S |x' \rangle
\end{equation}
is the kernel of $\mathsf{\rho}_S$. The $1/N$ expansion of this matrix model, in the 't Hooft limit 
\begin{equation}
N \rightarrow \infty, \qquad {N \over \hbar}=\lambda=\text{fixed}, 
\end{equation}
gives the all-genus topological string free energies, 
\begin{equation}
\log Z(N, \hbar)= \sum_{g\ge 0} \CF^S_g(\lambda) \hbar^{2-2g}, 
\end{equation}
where $\lambda$ parametrizes the moduli space of the local del Pezzo. Therefore, the genus expansion of the topological string on $X$ is 
realized as the asymptotic expansion of a well-defined quantity, $Z_S (N, \hbar)$. In addition, there are non-perturbative corrections to this expansion which are captured by 
the refined topological string free energy of the CY manifold, in the NS limit. The structure of the correspondence is summarized in Fig. \ref{MArels}. 

This picture can be made very concrete in many examples. For example, for local $\IP^2$, the matrix model takes the form \cite{MAmz}, 
\begin{equation}
\label{MAp2mm}
Z_{\IP^2}(N,\hbar)=\frac{1}{N!}  \int_{\IR^N}  { \rd^N u \over (2 \pi)^N}  
\prod_{i=1}^N \MAre^{- V_{\IP^2}(u_i, \hbar)}  \frac{\prod_{i<j} 4 \sinh \left( {u_i-u_j \over 2} \right)^2}{\prod_{i,j} 2 \cosh \left( {u_i -u_j \over 2} + {\ri \pi \over 6} \right)}, 
\end{equation}
where
\begin{equation}
V_{\IP^2}(u, \hbar)={\hbar u \over 2 \pi} + \log \left[ {\Phi_b \left( {b u \over 2 \pi} + {\ri b \over 3} \right) \over \Phi_b \left( {b u \over 2 \pi} - {\ri b \over 3} \right)} \right]. 
\end{equation}
In this equation, $\Phi_b(x)$ is Faddeev's quantum dilogarithm (where we use the conventions of \cite{MAkas-mar,MAkz}) and $b$ is related to $\hbar$ by
\begin{equation}
b^2={3 \hbar \over 2 \pi}. 
\end{equation}
The function $V_{\IP^2}(u, \hbar)$ appearing here is relatively complicated. However, it is 
completely determined by the operator (\ref{MAp2-op}). It has an asymptotic expansion at large $\hbar$ which 
can be used to compute the asymptotic expansion of the matrix integral (\ref{MAp2mm}) in the 't Hooft regime (\ref{MAthooft-regime}), 
as explained in \cite{MAmz}. In fact, this matrix integral is structurally very similar 
to the generalizations of the $O(2)$ matrix model considered in \cite{MAkostov-beta}, and might be 
exactly solvable in the planar limit. 

Another geometry where the integral kernel of the operator can be written in detail is local $\IP^1 \times \IP^1$ \cite{MAkmz}. 
The expression (\ref{MAznmm}) leads to an $O(2)$ matrix model, 
which can be regarded as a generalization of the ABJ matrix model of \cite{MAahs,MAhonda}. In \cite{MAbgt}, the field theory limit of topological string 
theory on local $\IP^1 \times \IP^1$ was implemented directly in the integral kernel found in \cite{MAkmz}. In this limit, one recovers the integral 
kernel studied in \cite{MAzamo} in connection with the Painlev\'e III equation. On the topological 
string side, the field theory limit leads to Seiberg--Witten theory \cite{MAkkv}. Based on this, \cite{MAbgt} provided a Fermi gas picture of Seiberg--Witten theory, as well as a 
proof of the conjecture of \cite{MAghm} for local $\IP^1 \times \IP^1$, in the field theory limit.

\section{Conclusions and outlook}

In this paper we have reviewed in detail the large $N$ solution to the matrix model calculating the partition function of ABJM theory and other 
Chern--Simons--matter theories, beyond the strict large $N$ limit. This analysis has given us unexpected and valuable information. First of all, it has made possible 
to derive the famous $N^{3/2}$ scaling of M2 branes theories from the gauge theory point of view. This result turns out to be just the tip of the iceberg: a detailed treatment of the 
ABJM matrix integral has led to many new exact results on the $1/N$ expansion of this interacting QFT. We have now, thanks to the localized matrix model, 
the full 't Hooft expansion of the free energy, at all orders in the genus expansion. We also have a conjectural exact result for the partition function 
which includes all non-perturbative corrections at large $N$. 

This non-perturbative result shows that the 't Hooft expansion is fundamentally incomplete, and 
instanton effects are crucial to obtain a meaningful answer. From the point of 
view of the dual superstring theory, it shows in a quantitative and precise way 
that the genus expansion is incomplete, and one needs membrane instantons coming from M-theory 
in order to have a consistent answer. The results obtained for ABJM theory can be generalized, to some extent, 
to other Chern--Simons--matter theories. Although much more work is needed for generic 
Chern--Simons--matter theories, the same pattern emerges, and non-perturbative effects are crucial in order to understand the theory.

The work done so far in this field leads to many interesting and important open problems for the future. Let us mention four of them. 

First of all, although many of the ingredients of the 
conjectural solution for the exact grand potential have been justified analytically, we are still lacking a 
complete derivation of the result (even by physics standards.) Mathematically, this is a challenging problem, since it 
amounts to computing the exact spectral determinant of the operator (\ref{MAdensitymat}). At some point, these loopholes should be addressed seriously. 
Such a derivation might also be useful in the second problem: in order to have a complete 
picture of the non-perturbative aspects of more general Chern--Simons--matter theories, one should develop techniques to analyze the 
resulting matrix models. Even in the case of $\CN=3$ theories, 
where a Fermi gas formulation is possible, we lack the necessary technology to analyze their spectrum in full detail. 
Concrete progress in this field will depend much 
on our ability to develop this technology. It is likely that the same tools which will fill out the loopholes in 
the derivation of (\ref{MAfinal-proposal}) will make it also possible 
to obtain analytic results in other theories. 

It is clear, in view of the AdS/CFT correspondence, that all these analytic results on 
Chern--Simons--matter theories give us precise predictions on the structure of the M-theory partition function on AdS$_4$ backgrounds. In particular, they provide many 
interesting predictions for the counting on worldsheet instantons and membrane instantons (in a sense, these results can be regarded as a generalization of mirror 
symmetry, in which the matrix model computes instanton corrections due to extended configurations in superstring theory and M-theory.) 
It would be very interesting to see to which extent these predictions can be tested with the current technology. 

Finally, as we pointed out briefly in section \ref{MAts-sec}, the mathematical structures appearing in the study of the partition function of Chern--Simons--matter theories provide a non-perturbative 
formulation of the topological string on toric CY manifolds. It has been speculated in \cite{MAghm} that this formulation might be understood in terms of a theory of M2 branes. For example, 
the partition function (\ref{MAznmm}) has, at large $N$ and fixed $\hbar$, the $N^{3/2}$ behavior typical of these theories. Such a formulation would clearly shed new light 
on topological string theory and lead to new perspectives in enumerative geometry. In relation to this, it has been recently noted that the matrix model for topological 
strings on local $\IP^2$, when $\hbar=2 \pi$, computes as well the partition function 
of ABJM theory with $k=1$ and on an ellipsoid with deformation parameter $b^2=3$ \cite{MAhatsuda-ellipsoid}, making in this way an explicit connection between M2 brane theories and the 
non-perturbative formulation of topological strings proposed in \cite{MAghm,MAmz}, for this particular case. It would be interesting to understand whether this is a happy coincidence or there is a more systematic relation 
between theories of M2 branes and topological strings.

\section*{Acknowledgements}
I would like to thank Sayantani Bhattacharyya, Flavio Calvo, Santiago Codesido, Ricardo Couso, Nadav Drukker, Alba Grassi, Yasuyuki Hatsuda, 
Johan  K\"all\'en,  Rinat Kashaev, Albrecht Klemm, Sanefumi Moriyama, Kazumi Okuyama, Pavel Putrov, Marc Schiereck, Ashoke Sen, Masoud Soroush and Szabolcs Zakany
for the collaborations which led to most of the results  presented in this review paper. Special thanks to Alba Grassi, Yasuyuki Hatsuda and Kazuo Hosomichi for a careful reading of the manuscript. 
I would also like to thank the coordinators of this project, Vasily Pestun and Maxim Zabzine, for inviting me to write a contribution. 
This work is supported in part by the Fonds National Suisse, subsidies 200021-156995 and 200020-141329, and by the 
NCCR 51NF40-141869 ``The Mathematics of Physics" (SwissMAP).

\documentfinishBBL